\documentclass[12pt]{article}
\usepackage{xr-hyper}
\usepackage[top=1in, bottom=1in, left=1in, right=1in]{geometry}
\usepackage{setspace}
\usepackage{lmodern}
\usepackage{booktabs}
\usepackage{multirow}
\usepackage{rotating}
\usepackage{graphicx}
\usepackage[section]{placeins} 
\usepackage[small, bf]{caption}
\usepackage{subcaption}
\usepackage{pdfpages}
\usepackage{pdflscape}
\usepackage{textcomp}
\usepackage{longtable}
\usepackage{nicefrac}
\usepackage{adjustbox}	
\usepackage[hyphens]{url}
\usepackage{bbm}
\usepackage{floatrow}
\usepackage{wrapfig}

\usepackage{natbib}
\usepackage{bibunits}
\bibpunct{(}{)}{;}{a}{,}{,}

\usepackage{amsmath, amsfonts, amssymb, amsthm}
\usepackage[pdftex]{hyperref}
\hypersetup{
    colorlinks=true,
    linkcolor=red,
    citecolor=blue,
    filecolor=magenta,      
    urlcolor=cyan,
}

\newtheorem{lemma}{Lemma}
\newtheorem{proposition}{Proposition}

\theoremstyle{definition}
\newtheorem{definition}{Definition}
\newtheorem{assumption}{Assumption}

\usepackage{titlesec}
\titlespacing*{\section}{0pt}{1.5ex plus 1ex minus .2ex}{0.8ex plus .2ex}
\titlespacing*{\subsection}{0pt}{1.2ex plus 1ex minus .2ex}{0.8ex plus .2ex}

\usepackage{dcolumn}
\newcolumntype{d}[0]{D{.}{.}{5}}

\usepackage{graphicx}
\usepackage[space]{grffile}
\usepackage{colortbl}

\usepackage{amssymb} 

\usepackage{tikz}
\usepackage{tikzpeople}
\usetikzlibrary{positioning}
\usetikzlibrary{snakes}
\usetikzlibrary{calc}
\usetikzlibrary{arrows}
\usetikzlibrary{decorations.markings}
\usetikzlibrary{shapes.misc}
\usetikzlibrary{matrix,shapes,arrows,fit,tikzmark}

\usepackage[linesnumbered,ruled,vlined]{algorithm2e}
\usepackage{algpseudocode}
\SetKwInput{KwInput}{Input}                
\SetKwInput{KwOutput}{Output}              

\usepackage[most]{tcolorbox}

\usepackage{subfiles} 

\title{Large Language Models: \\ An Applied Econometric Framework\thanks{This paper was supported by the Center for Applied Artificial Intelligence at the University of Chicago and the Altman Family Fund at MIT. 
We thank Haya Alsharif and Janani Sekar for excellent research assistance. We thank Peter Bergman, Tim Christensen, Oeindrila Dube, Larry Katz, Lindsey Raymond, Suproteem Sarkar, Andrei Shleifer, and David Yanagizawa-Drott as well as numerous seminar audiences and conference participants for helpful comments.
We also thank the editor and three reviewers for their valuable feedback. Wharton Research Data Services (WRDS) was used in preparing this paper. This service and the data available constitute valuable intellectual property and trade secrets of WRDS and/or its third-party suppliers. All interpretations and any errors are our own.}}
\author{Jens Ludwig \and Sendhil Mullainathan \and Ashesh Rambachan\thanks{Ludwig: University of Chicago and NBER. 
Mullainathan: Massachusetts Institute of Technology and NBER. Rambachan: Massachusetts Institute of Technology.}}	
\date{\today}

\begin{document}

{\singlespacing \maketitle}
\thispagestyle{empty} 
\setcounter{page}{0}

{\singlespacing
\begin{abstract}
Large language models (LLMs) enable researchers to analyze text at unprecedented scale and minimal cost.
Researchers can now revisit old questions and tackle novel ones with rich data.  
We provide an econometric framework for realizing this potential in two empirical uses.
For prediction problems -- forecasting outcomes from text -- valid conclusions require ``no training leakage'' between the LLM's training data and the researcher's sample, which can be enforced through careful model choice and research design. 
For estimation problems -- automating the measurement of economic concepts for downstream analysis -- valid downstream inference requires combining LLM outputs with a small validation sample to deliver consistent and precise estimates.
Absent a validation sample, researchers cannot assess possible errors in LLM outputs, and consequently seemingly innocuous choices (which model, which prompt) can produce dramatically different parameter estimates.
When used appropriately, LLMs are powerful tools that can expand the frontier of empirical economics.
\end{abstract}}

\newpage 
\section{Introduction}

Large language models (LLMs) enable economists to process text at unprecedented scale and minimal cost, unlocking questions previously out of reach: predicting stock returns from earnings call transcripts, measuring partisan polarization in social media posts, backcasting historical consumer sentiment from newspapers, or simulating survey responses cheaply. By making these questions -- and countless others -- tractable, LLMs offer transformative potential for empirical research.

To realize this potential, we must answer a practical question: how should LLM outputs be incorporated into empirical workflows? Can they be plugged in ``as-is,'' or do our estimation strategies require adjustment to use these powerful tools?

This question is challenging because LLMs resist traditional statistical analysis. 
LLMs are complex, often proprietary, constantly evolving, and trained on sprawling, heterogeneous corpora that defy tractable modeling -- extraordinary engineering achievements accomplished without our usual statistical foundations. 
Existing evaluation methods have been remarkably effective for developing better models but offer limited guidance for how any given LLM will perform on a new task.

We develop an econometric framework that addresses these complexities and provides practical guidance for empirical research using LLM outputs, focusing on two empirical uses: prediction problems and estimation problems.
For each use, we clarify what assumptions and practices ensure valid conclusions.

In prediction problems, researchers use collected text to predict some economic outcome (e.g., predicting stock prices from corporate earnings calls).
Extracting meaning from text requires modeling the complex structure of language.
LLMs, having already learned this structure from enormous training corpora, can possibly serve as the foundation upon which economists can build in prediction problems --- either by directly prompting an LLM to make a prediction or by using its representations as features in a predictor.

Using LLMs in prediction problems requires one condition: ``no training leakage.''
If a researcher forms predictions using an LLM and evaluates those predictions on an evaluation dataset, this reflects the LLM's out-of-sample performance if and only if there is no overlap between the LLM's training dataset and the researcher's dataset. 
This can be violated because many LLMs are trained on intentionally obscured datasets.
Computer scientists have found that LLMs are often trained on common benchmark evaluations.
We find that LLMs appear to be trained on economic datasets; see also \citet{glasserman2023assessinglookaheadbiasstock, sarkar2024lookahead, lopezlira2025memorizationproblemtrustllms} among others.

We provide guidance on enforcing no training leakage, clarifying that it requires careful attention to both the choice of model and research design.
For example, when the goal is to predict future, unseen documents (such as tomorrow's financial news), open-source LLMs with fixed, published weights \citep[e.g.,][]{touvron2023llama2openfoundation, dubey2024llama3herdmodels} or time-stamped training data \citep[e.g.,][]{Sarkar(24), he2025chronologicallyconsistentlargelanguage} can be paired with evaluation samples constructed only from documents published after the model's publication date or training cutoff.

In estimation problems, researchers estimate relationships between economic concepts expressed in text (e.g., partisanship in social media posts) and downstream parameters (e.g., the causal effect of a policy change).
There is \textit{some} resource-intensive procedure for measuring the economic concept; if it could be scaled, we would use these measurements and report the resulting estimates.
Often, this would involve the researcher carefully reading each text piece themselves.
We would like to instead use an LLM to economize on measurement costs.
How can we use LLM outputs for valid downstream inference?

We highlight a constructive solution borrowing from an old idea in econometrics: collect a small validation sample and use it to \textit{empirically} correct for possible LLM errors.
We illustrate how researchers can incorporate validation data in the context of linear regression. 
Debiasing LLM outputs preserves our usual econometric guarantees of consistency and asymptotic normality.
This has been well-studied in econometrics \citep[e.g.,][]{LeeSepanski(95)-ErrorsInVariable, chen2005measurement, schennach2016recent} and extended in machine learning, such as \citet[][]{WangMcCormickLeek(20)}, \citet[][]{AngelopoulosEtAl(23)-PPI}, \citet[][]{EgamiEtAl(24)-ImperfectSurrogates}, \citet[][]{BattagliaEtAl(24)}, and \citet[][]{carlson2025unifyingframeworkrobustefficient}.
We illustrate that incorporating debiased LLM outputs can substantially improve the precision of downstream estimates – resulting in estimates that are often more precise than using validation data alone. 
Used correctly, imperfect LLM outputs serve not as substitutes but as amplifiers of validation samples, allowing researchers to achieve tighter standard errors.

Absent a validation sample, researchers cannot assess the magnitude or pattern of errors in LLM outputs—and therefore cannot evaluate their impact on downstream parameter estimates.
We demonstrate this problem empirically: absent a validation sample, seemingly innocuous choices---which LLM to use, how to phrase the prompt---lead to dramatically different parameter estimates in applications to finance and political economy, with coefficients varying in magnitude, sign, and significance. 
In estimation problems, researchers have no choice but to invest effort and collect validation samples when using LLMs.

Finally, this framework is flexible enough to account for creative uses of LLMs in economics. 
We argue that using LLMs for hypothesis generation \citep[e.g.,][]{ludwig2024machine} can be cast as a form of prediction problem, so that the key assumption required is again no training leakage. 
LLM simulation of human subjects in surveys or experiments \citep[e.g.,][]{park2023generativeagentsinteractivesimulacra, horton2023large, manning2024automatedsocialsciencelanguage, park2024generativeagentsimulations1000} can be thought of as an estimation problem, and so in-silico subjects can amplify -- but not fully replace -- a small validation sample of actual human responses.

Our framework clarifies when and how researchers can harness LLMs in empirical research. 
The requirements are straightforward: ensure no training leakage for prediction and collect small validation samples for estimation. 
We provide a checklist based on our framework in Section \ref{section: checklist for research}. 
These simple practices unlock the transformative potential of LLMs for empirical research.

\section{An Econometric Framework for Large Language Models}

LLMs are complex architectures trained on vast corpora through multi-stage processes: pretraining (learning to predict next tokens), instruction fine-tuning, reinforcement learning from human feedback (RLHF), and reinforcement learning from verifiable rewards (RLVR). 
Reasoning models employ test-time computation to further improve performance.
The field evolves rapidly, with new architectures, training procedures, and capabilities emerging regularly.\footnote{Many resources are available about the technical foundations of large language models.
See, for example, \citet[][]{chang2024_llmevals}, \citet[][]{minaee2025largelanguagemodelssurvey}, and \citet[][]{zhao2025surveylargelanguagemodels}. 
For overviews aimed at economists, see \citet[][]{Korinek(23)}, \citet[][]{dell2024deeplearningeconomists}, \citet[][]{Korinek(24)}, and \citet[][]{ash2024llm}.}

This complexity creates a fundamental challenge for econometric analysis. 
We typically study statistical procedures by articulating assumptions about the data-generating process and combining them with a computational understanding of the procedure.
This approach is at present intractable for LLMs. 
They have billions of parameters, proprietary training datasets, and algorithms that resist formal characterization.

We adopt a different strategy based on how economists actually use these tools.
We treat LLMs as black boxes -- prompting them or extracting embeddings without characterizing internal mechanisms -- and identify conditions they must satisfy for valid empirical use. 
We focus on two applications: prediction problems and estimation problems. 
By black-boxing their inner workings, our approach provides guidance that is robust to inevitable changes in architectures and training procedures.

\subsection{Setting and the Researcher's Dataset}

Let $\Sigma^*$ denote a collection of strings (up to some finite length) in an alphabet with elements $\sigma \in \Sigma^*$. 
A training dataset is any collection of strings, summarized by the vector $t$ whose elements $t_{\sigma}$ are sampling indicators for whether string $\sigma$ was collected. 

For any empirical question, only some strings are economically relevant. 
Denote these as $\mathcal{R} \subseteq \Sigma^*$ with elements $r \in \mathcal{R}$ that we refer to as text pieces. 
The researcher's dataset is summarized by the vector $d$ whose elements $d_{\sigma}$ are sampling indicators for whether the researcher collected string $\sigma$. 
The researcher only collects economically relevant text pieces, and so $d_{\sigma} = 0$ for all $\sigma \in \Sigma^* \setminus \mathcal{R}$. 

Each text piece $r$ can be linked to observable economic variables $(Y_r, W_r)$, which are economic outcomes $Y_r$ that might be influenced by the text or candidate covariates $W_r$ that might influence the text. 

\vspace{-1em}

\paragraph{Example: Congressional legislation} Consider descriptions of bills introduced in the United States Congress. Each text piece $r \in \mathcal{R}$ is a bill's description such as ``\texttt{A bill to revise the boundary of Crater Lake National Park in the State of Oregon}.'' The economic outcome $Y_r$ might be whether the bill passed its originating chamber. The covariate $W_r$ might be the party affiliation or roll-call voting score of the bill's sponsor. $\blacktriangle$

\vspace{-1em}

\paragraph{Example: Financial news headlines} Consider financial news headlines about publicly traded companies. Each text piece $r \in \mathcal{R}$ is a financial news headline such as ``\texttt{Bank of New York Mellon Q1 EPS \$0.94 Misses \$0.97 Estimate, Sales \$3.9B Misses \$4.01B Estimate}.'' The economic outcome $Y_r$ might be the company's realized return in some window after the headline's publication date, while the covariate $W_r$ could be the company’s past fundamentals. $\blacktriangle$

\vspace{1em}

Each text piece $r$ expresses some economic concept $V_r := f^*(r)$, where $f^*(\cdot)$ is the measurement procedure the researcher would use if time and resources were no constraint.
Typically these measurements are what the researcher themselves or an appropriate domain expert would produce by carefully reading each text piece.\footnote{For example, \cite{AshHansen(23)} write, ``The most accurate approach to concept detection is perhaps direct human reading with appropriate domain expertise'' (p. 672). \cite{HansenEtAl(23)-RemoteWork} write, ``The most precise way of classifying [text pieces] is arguably via direct human reading'' (p. 6).} 
In defining the economic concept, researchers must answer: what exactly am I measuring from the text, and how would I label it if resources were no constraint?
Moving forward, we assume a measurement procedure $f^*(\cdot)$ exists that the researcher would be satisfied to use \textit{if} it could be scaled.

This creates a text processing problem: measuring the economic concept requires processing each text piece $r$, which may be prohibitively costly at scale. 
Absent a solution to this text processing problem, the researcher cannot collect $V_r$ on all text pieces.

In settings like this, researchers would like use the collected text pieces to tackle two types of empirical analyses. 
The first is a \textit{prediction problem}: predict the linked variable $Y_r$ using the associated text piece $r$. 
For example, can we predict stock returns from news headlines? 
The second is an \textit{estimation problem}: estimate some parameter that relates the economic concept $V_r$ to the linked variables $(Y_r, W_r)$. 
If bill descriptions $r$ express the policy topic $V_r$ of the bill (e.g., whether it is related to foreign affairs or health policy), how do policy topics relate to the party affiliation of the bill's sponsor $W_r$? 

LLMs are general-purpose and easy-to-use models for processing text, so researchers would like to use them in prediction and estimation problems.
We next introduce LLMs into our framework.

\subsection{Incorporating Large Language Models in Empirical Research}

To capture how we often interact with LLMs as black boxes --- generating responses from prompts without knowing their design nor training data --- we define a \textit{large language model} as any mapping from possible training datasets $t$ to mappings between strings, where $\widehat{m}(\cdot; t) \colon \Sigma^* \rightarrow \Sigma^*$ is its text generator when trained on dataset $t$ and $\widehat{m}(\sigma; t)$ is the LLM's response when prompted by string $\sigma$.

This definition has a key implication: the LLM is actually two distinct algorithms. 
First, a training algorithm that takes any dataset $t$ and learns the mapping between strings. 
Second, the text generator $\widehat{m}(\cdot; t)$ is the output of the training algorithm and what users interact with. 

Our analysis will not depend on how exactly these algorithms work. 
Since the state of the art is constantly evolving, we should expect the exact implementation of training algorithms and text generators to change.  
Furthermore, alternative LLMs may differ in their implementations. 
Studying LLMs at this level of abstraction will provide interpretable conditions for empirical research and ensures the durability of our analysis.
Researchers will have conditions under which any model's output can be incorporated into empirical research.

Our framework captures all key LLM design choices --- some made by researchers, others by algorithm builders (often invisibly to researchers).

\vspace{-1em}
\paragraph{Interpreting the Text Generator} The text generator $\widehat{m}(\cdot; t)$ captures all choices that influence how an LLM generates responses once trained.
This includes prompt engineering strategies that materially affect the quality of responses \citep[e.g.,][]{LiuEtAl(23), wei2023chainofthoughtpromptingelicitsreasoning, white2023promptpatterncatalogenhance, chen2024unleashingpotentialpromptengineering}. 
Alternative prompt engineering strategies can be cast as alternative specifications of the text generator $\widehat{m}(\cdot; t)$.

The text generator captures other important choices governing generation. 
Parameters like temperature, top-$p$ sampling, or top-$k$ sampling control randomness in sampling from the LLM's probability distribution over next tokens. 
Reasoning models employ test-time computation, generating intermediate ``thought tokens'' before producing final outputs to enhance performance on complex tasks.\footnote{For many reasoning models, users cannot control generation parameters like temperature. Consequently, there is inherent randomness that cannot be eliminated.}
By defining the text generator as a deterministic mapping from prompts to responses, our framework can be interpreted as focusing on the case in which these parameters are such that the LLM greedily generates its most likely token.
Our results extend naturally to stochastic text generators as well, but at the expense of more cumbersome notation.\footnote{A practical challenge is ensuring LLM outputs are formatted appropriately for empirical analysis. Researchers often prompt LLMs to return structured outputs, but LLMs may not reliably comply. Our text generator abstraction takes further post-processing steps as given, and $\widehat{m}(\cdot; t)$ represents the resulting final output.}

\vspace{-1em} 
\paragraph{Interpreting the Training Algorithm} The training algorithm captures all aspects of design and training that produce the text generator. 
This includes architectural choices (e.g., parameter count, attention layers, context window), the pretraining objective (typically next-token prediction), and optimization details.
LLMs undergo multiple training stages beyond pretraining. 
Instruction fine-tuning trains models to follow user instructions, and reinforcement learning from human feedback (RLHF) aligns output with human preferences.
More recent developments include reinforcement learning from verifiable rewards (RLVR), where models are trained on tasks with objectively correct answers.
Crucially, the training dataset $t$ in our framework encompasses all strings used in pretraining \emph{and} post-training stages. 

\vspace{1em}

So how do researchers use LLMs in empirical research? 
In prediction problems, researchers often prompt an LLM with each text piece $r$ to generate a prediction $\hat{Y}_r = \widehat{m}(r; t)$.
For instance, they might prompt a model to predict whether a stock return will be positive.
In estimation problems, researchers prompt an LLM to generate labels $\hat{V}_r =  \widehat{m}(r; t)$ for the economic concept. 
For example, they might prompt a model to classify a bill's policy topic.

\subsection{The Challenge of Evaluating Large Language Models}\label{section: challenge of LLM evaluation}

At some level, the viability of using LLMs for prediction and estimation problems hinges on the quality of LLM outputs: how large are possible errors $Y_r - \widehat{Y}_r$ and $V_r - \widehat{V}_r$ in any application? 
A natural starting point is to understand how computer science approaches this evaluation problem. 

Computer scientists adapted methods from supervised learning.
In supervised learning, the ``common task framework'' \citep[][]{Donoho2024Data} builds benchmark datasets like ImageNet for image classification and the Netflix Prize for recommender systems and evaluates competing algorithms on these benchmarks. 
Since LLMs aspire to be general-purpose technologies useful across all tasks, computer scientists have built increasingly diverse benchmarks.
For example, the "Beyond-the-Imitation-Game benchmark" (BIG-bench) collects problems across 204 tasks \citep[][]{srivastava2022beyond}, while the "Massive Multitask Language Understanding" (MMLU) benchmark spans 57 categories from logic to social sciences \citep[][]{hendrycks2020measuring}. 
Other benchmarks assess specialized capabilities: SWE-bench for coding ability \citep[][]{jimenez2024swebenchlanguagemodelsresolve}, GSM8K for mathematical reasoning \citep[][]{cobbe2021trainingverifierssolvemath}, and standardized exams such as the SAT, GRE, and AP tests.
Modern LLMs perform remarkably well on these benchmarks, and this impressive performance forms much of the quantitative basis for current enthusiasm. 

Can economists use these benchmark evaluations to reason about how LLMs will perform on specific empirical tasks? 
The answer is more complicated than it might initially appear.

\subsubsection{The Limits of Benchmarks}

We do not intrinsically care about benchmark performance itself --- after all, who deploys LLMs to solve SAT problems? 
We hope to \textit{generalize} from benchmarks to new economically relevant tasks.

This generalization happens intuitively. We assume an LLM that aces AP chemistry must, like a person, handle many chemistry-related tasks. 
Evidence for ``anthropomorphic generalization'' suggests that people apply similar generalization heuristics to LLMs as they do to humans when predicting performance across tasks \citep[][]{vafa2024humangeneralization, DreyfussRaux(24)}. 

Yet this intuition is misleading. 
LLMs exhibit what \citet[][]{mancoridis2025potemkin} call ``potemkin understanding'' --- performing well on benchmarks without grasping underlying concepts. 
This manifests as remarkable brittleness: performance is sensitive to seemingly minor details that would not affect human performance. 
Consider several examples from an accumulating body of evidence. 
While humans who master one math problem typically solve easier variations, LLMs have not readily generalized this way. 
An LLM that could reliably solve $(9/5) x + 32$ could not solve $(7/5) x + 31$ \citep{mccoy2023embersautoregressionunderstandinglarge}. 
An LLM correctly defines an ABAB rhyming scheme but then generates poems violating its own definition \citep[][]{mancoridis2025potemkin}. 
An LLM trained on ``A is B" will not know ``B is A" \citep{berglund2023reversal} and LLMs struggle with logical puzzle variations \citep[][]{nezhurina2024alicewonderlandsimpletasks}. 
Brittleness extends to presentation: LLMs answer multiple-choice questions correctly but fail when answer order is permuted \citep{zong2024foolvisionandlanguage}, succeed at programming with 0-based indexing but fail with 1-based indexing \citep[][]{wu2024reasoningrecitingexploringcapabilities}, and show inconsistent performance on similar spatial reasoning tasks \citep{MelanieMitchell(23)-Brittleness}.

LLM errors align poorly with human intuitions precisely because we expect them to understand and misunderstand as humans do.
\citet[][]{Dellacqua(23)} aptly characterize this as AI's ``jagged frontier'' -- a landscape where performance on similar tasks varies unpredictably.  
This jagged frontier captures why benchmark performance cannot be intuitively generalized to specific economic applications.

\subsubsection{Do Large Language Models have World Models?}

Perhaps this is too pessimistic, and this brittleness is superficial -- noise around a deeper structural understanding embedded within LLMs. 
LLMs are trained on massive corpora containing rich information about reality, followed by extensive reinforcement learning. 
Perhaps LLMs have successfully learned ``world models'' -- internal, structured representations of how the world works -- that would enable reliable generalization despite occasional errors. 

This hypothesis is an active research area focused on settings where researchers can both infer the LLM's implicit world model and compare it to the truth \citep[e.g.,][]{li2024emergentworldrepresentationsexploring, nanda2023emergentlinearrepresentationsworld, jylin04_2024_othellogpt, nikankin2025arithmeticalgorithmslanguagemodels}. 
The evidence so far is at best mixed and at worst suggests LLMs may not learn generalizable world models. 
For example, \citet[][]{vafa2024worldmodel} trained a generative sequence model on turn-by-turn driving data from 12.6 million NYC taxi rides. 
While it predicts the next turn between locations with high accuracy, the authors reverse-engineered the model's implicit map of Manhattan's street grid: the implicit map bears no resemblance to the actual grid. 
High predictive accuracy did not require learning the underlying spatial structure.
\citet[][]{vafa2025has} trained a generative sequence model on planetary orbits. 
Despite accurately predicting planetary movements, the model reveals no understanding of Newtonian physics. 
Rather than learning the gravitational law, it relies on task-specific heuristics that fail to generalize beyond its training distribution. 
Even advanced reasoning models exhibited similar failures.

Taken together, the evidence suggests that LLMs not only make errors, but these errors cannot be reliably predicted.
They fail in ways misaligned with how humans generalize across tasks, nor can we assume these errors are noise around accurate world models.
This creates our challenge. Since LLMs are impressive yet brittle tools, how should researchers incorporate their outputs into empirical research?

\section{Prediction with Large Language Models}\label{section: prediction with large language models}

We begin with prediction problems -- using text to predict economic outcomes. 
LLMs' extensive training makes them natural candidates for this task, having already learned rich representations of language.
Valid inference in prediction problems hinges on one condition: no leakage between the model's training data and the researcher's evaluation sample.
While training leakage plagues benchmark evaluations in computer science, economists can manage it through careful model choice and research design. 

\subsection{The Researcher's Prediction Problem}

Suppose the researcher prompts an LLM to predict the linked variable from text pieces, $\widehat{Y}_r = \widehat{m}(r; t)$.\footnote{Researchers may use an LLM to construct embeddings for each text piece $r$, and the resulting embeddings may then be used as features by a supervised machine learning algorithm to predict $Y_r$. Our analysis equally applies to evaluating the performance of a prediction function using LLM embeddings as features.}
For some loss function $\ell(y, \tilde{y})$, the researcher calculates the sample average loss
\begin{equation}\label{equation: sample average prediction loss}
    \frac{1}{N} \sum_{r \in \mathcal{R}} D_r \ell(Y_r, \widehat{m}(r; t)),  
\end{equation}
where $N = \sum_{r} D_r$ is the number of text pieces collected.
We would like to draw conclusions about the predictability of $Y_r$ from the text piece $r$ based on the LLM's sample average loss. 
When is this valid?

Researchers face two sources of uncertainty about LLMs: What data was the LLM trained on? How does it generate output from text?
We introduce two objects to formalize these uncertainties. 
The research context formalizes what we know (and do not know) about the LLM's training data.
The LLM's guarantee summarizes high-level properties about how the model behaves without requiring exact knowledge of its internal workings.
Together, these will allow us to derive interpretable conditions for using LLMs in prediction problems.

The \textit{research context} $Q(\cdot) \in \mathcal{Q}$ is a joint distribution over the sampling indicators $(D, T)$.
It encodes two distinct features. 
The sampling distribution over $D$ defines the researcher's out-of-sample prediction problem -- the collection of text pieces over which they want to evaluate the LLM's predictive performance.
The sampling distribution over $T$ captures the researcher's uncertainty about the LLM's training data. 
We make a technical assumption about the collection of research contexts $\mathcal{Q}$.

\begin{assumption}\label{asm: research contexts}
Letting $t = (t_{\sigma_1}, \hdots, t_{\sigma_{|\Sigma^*|}} )^\prime$ denote the sampling indicators summarizing the LLM's realized training dataset, all research contexts $Q(\cdot) \in \mathcal{Q}$ satisfy: (i) For all values $d$, $Q(D = d, T = t) = \Pi_{\sigma \in \Sigma^*} Q(D_{\sigma} = d_{\sigma}, T_{\sigma} = t_{\sigma})$; and (ii) $\mathbb{E}_{Q}[\sum_{r \in \mathcal{R}} D_r] = \mathbb{E}_{Q}[\sum_{r \in \mathcal{R}} D_r \mid T = t]$.
\end{assumption}

\noindent Assumption \ref{asm: research contexts}(i) states that sampling across strings is independent but not necessarily identically distributed. 
Assumption \ref{asm: research contexts}(ii) states that the researcher's expected sample size does not depend on the LLM's training corpus. 
Let $q^{T \mid D}_{\sigma}(t_\sigma) = Q(T_{\sigma} = t_{\sigma} \mid D_{\sigma} = 1)$ denote the conditional probability the string is sampled by the LLM's training dataset given that it is sampled by the researcher. 
The marginal probabilities are $q^{T}_{\sigma}(t_{\sigma}) = Q(T_{\sigma} = t_{\sigma})$ and $q^{D}_{\sigma} = Q(D_{\sigma} = 1)$.

Our goal is to assess the LLM's predictive performance over the population of text pieces defined by the research context $Q(\cdot)$:
\begin{equation}\label{eq: prediction target}
\mathbb{E}_{Q}\left[\sum_{r \in \mathcal{R}} D_r \ell(Y_r, \widehat{m}(r; t))\right].
\end{equation}
This is summarizes whether the model's predictive performance on the broader target population.
As the number of economically relevant text pieces grows large (see Appendix \ref{section: prediction and estimation, large sample}), we can characterize what the sample average loss converges to. 
Conditional on the LLM's realized training dataset,
\begin{equation}\label{equation: convergence of sample avg loss}
\frac{1}{N} \sum_{r \in \mathcal{R}} D_{r} \ell(Y_r, \widehat{m}(r; t)) - \frac{1}{\mathbb{E}_{Q}[\sum_{r \in \mathcal{R}} D_r \mid T = t]} \mathbb{E}_{Q}\left[ \sum_{r \in \mathcal{R}} D_{r} \ell(Y_r, \widehat{m}(r; t)) \mid T = t \right] \xrightarrow{p} 0.
\end{equation}
Conditioning on the training dataset $t$ treats the LLM $\widehat{m}(\cdot; t)$ as a fixed mapping, avoiding strong assumptions about how the LLM was trained (e.g., how would it behave over counterfactual training datasets?).
The sample average loss recovers our target quantity in Equation \eqref{eq: prediction target} if $\mathbb{E}_{Q}\left[\sum_{r \in \mathcal{R}} D_r \ell(Y_r, \widehat{m}(r; t)) \mid T = t\right] = \mathbb{E}_{Q}\left[\sum_{r \in \mathcal{R}} D_r \ell(Y_r, \widehat{m}(r; t))\right].$

Recall the second source of uncertainty: researchers do not know precisely how the LLM generates outputs from text. 
Yet researchers often observe other high-level information about the LLM's behavior, such as performance on benchmarks (e.g., "achieves 88.7\% on MMLU" or "scores 1520 on the SAT").
We formalize this through a \textit{guarantee} $\mathcal{M}$, a collection of possible text generators that captures what the researcher knows about the LLM's behavior. 
The researcher only knows that $\widehat{m}(\cdot; t) \in \mathcal{M}$. 

Given a guarantee $\mathcal{M}$ and a research context $Q(\cdot)$, we define when this workflow is valid for prediction problems.

\begin{definition}[Prediction problem]\label{defn: generalization and gpt for prediction}
The LLM $\widehat{m}(\cdot; t)$ with guarantee $\mathcal{M}$ \textit{generalizes} in research context $Q(\cdot)$ if, for all text generators satisfying the guarantee $\widehat{m}(\cdot) \in \mathcal{M}$, 
$$
    \mathbb{E}_{Q}\left[ \sum_{r \in \mathcal{R}} D_r \ell(\widehat{m}(r), Y_r) \mid T = t \right] = \mathbb{E}_{Q}\left[ \sum_{r \in \mathcal{R}} D_r \ell(\widehat{m}(r), Y_r) \right].
$$
\end{definition}

\noindent This definition formalizes an out-of-sample prediction goal. 
The LLM generalizes if its sample average loss reflects its predictive performance on the target population. 
Under Definition \ref{defn: generalization and gpt for prediction}, the researcher's workflow in the prediction problem is justified for \textit{any} LLM satisfying the guarantee $\mathcal{M}$. 
The researcher can draw conclusions based on the sample average loss knowing only the guarantee $\mathcal{M}$ is satisfied.

\vspace{-1em}
\paragraph{Example: Congressional legislation}
Consider researchers predicting whether a bill passes either house of Congress $Y_r$ using only its text description $r$.
The LLM generates predictions $\widehat{m}(r; t)$ using a specific prompting strategy.
Each researcher calculates the sample average loss of the LLM's predictions on their own collected sample of Congressional bills.
Different researchers face different prediction problems, formalized by alternative research contexts $Q(\cdot)$. 
One researcher might assess whether the LLM can predict outcomes for future legislation by sampling bills after a cutoff date, such as all bills from 2025 onward. 
Another might evaluate whether the LLM's predictions generalize to novel policy domains, sampling bills on as cryptocurrency regulation or artificial intelligence policy. $\blacktriangle$

\vspace{-1em}
\paragraph{Example: Financial news headlines}
Consider researchers predicting a company's realized returns $Y_r$ from a financial news headline $r$.
The LLM generates predictions $\widehat{m}(r; t)$ using a specific prompting strategy.
Each researcher calculates the sample average loss on their collected headlines.
Different researchers again face different out-of-sample prediction problems, formalized by alternative research contexts $Q(\cdot)$.
One researcher might assess whether the LLM predicts returns during future market conditions, sampling headlines from 2025 onward.
Another might evaluate whether predictions generalize across company types, focusing on small-cap technology firms or international equities.
$\blacktriangle$

\subsection{Training Leakage as a Threat to Prediction}\label{section: training leakage as threat to prediction}

Using LLMs in prediction problems requires one condition: no training leakage between the LLM's training data and the researcher's evaluation sample.

\begin{lemma}\label{lemma: leakage decomposition}
Under Assumption \ref{asm: research contexts}, for any research context $Q(\cdot) \in \mathcal{Q}$ and text generator $\widehat{m}(\cdot)$,
$$
\mathbb{E}_{Q}\left[ \sum_{r \in \mathcal{R}} D_r \ell(Y_r, \widehat{m}(r)) \right] = \mathbb{E}_{Q}\left[\sum_{r \in \mathcal{R}} D_r \ell(Y_r, \widehat{m}(r)) \mid T = t\right] - \mathbb{E}_{Q}\left[ \sum_{r \in \mathcal{R}} D_r \left( \frac{q_r^{T \mid D}(t_r)}{q^{T}_r(t_r)} - 1 \right) \ell(Y_r, \widehat{m}(r)) \right].
$$
\end{lemma}

\begin{proposition}\label{proposition: prediction gpts iff no leakage}
The LLM $\widehat{m}(\cdot; t)$ generalizes for research context $Q(\cdot) \in \mathcal{Q}$ if and only if it satisfies the guarantee $\mathcal{M}(Q)$ for
\begin{equation}\label{equation: no leakage condition}
   \mathcal{M}(Q) = \left\{ \widehat{m}(\cdot) \colon -\mathbb{E}_{Q}\left[ \sum_{r \in \mathcal{R}} D_r \left( \frac{q^{T \mid D}_r(t_r)}{q^{T}_r(t_r)} - 1 \right) \ell(Y_r, \widehat{m}(r)) \right] = 0 \right\}.
\end{equation}
\end{proposition}

\noindent The no-training leakage condition (Equation \ref{equation: no leakage condition}) captures the extent to which overlap between the LLM's training data and the researcher's sample covaries with the LLM's predictive performance. 
It has an intuitive interpretation as omitted variable bias.
The term $\frac{q^{T \mid D}_r(t_r)}{q^{T}_r(t_r)} - 1$ measures how learning that the researcher sampled text piece $r$ updates beliefs about whether $r$ appeared in the LLM's training data.
When this correlation is positive -- text pieces in the researcher's sample are more likely to have been in the training data -- and the LLM performs well on such pieces, the overall bias term is positive and the sample average loss overstates true predictive performance.
Uncertainty over what entered into the LLM's training data acts like an omitted variables bias in the prediction problem.

This bias arises from both the researcher's choice of out-of-sample prediction and our beliefs about what entered into the LLM's training data. 
Section \ref{section: prediction problem guidance} discusses how researchers can prevent training leakage through careful choices of model and prediction exercise.

\subsubsection{Some Evidence of Training Leakage}\label{section: evidence of training leakage}

Training leakage is a well-documented problem in computer science. LLMs' training datasets frequently contain examples from popular benchmark evaluations \citep[][]{sainz2023nlpevaluationtroubleneed, golchin2024timetravelllmstracing, lm_index}. 
This has generated skepticism about evaluating LLMs on any publicly available data \citep[e.g.,][]{ravaut2024largelanguagemodelscontaminated}.
But this evidence focuses on CS applications. 
Might economic prediction problems face similar risks? 

\vspace{-1em}
\paragraph{Assessing Training Leakage in Congressional Legislation} We assess training leakage in an empirical setting relevant for economists studying politics and political economy: congressional legislation. 
The Congressional Bills Project \citep[][]{CAP, CBP} contains descriptions $r$ for over 400,000 bills proposed in Congress, along with whether each passed the House or Senate $Y_r$. 
We sample $10,000$ bills introduced from 1973 to 2016 and test whether passage can be predicted from text descriptions alone --- a potentially challenging task given the strategic dynamics of congressional voting. 
Among these bills, only 7.4\% pass the House and 6.0\% pass the Senate.

We generate predictions $\widehat{Y}_{r} = \widehat{m}(r; t)$ based on each bill's description $r$ by prompting GPT-4o (see Appendix Figure \ref{fig: congressional bills prediction, prompts} for the specific prompt).
GPT-4o correctly predicts the bill's outcome 91.2\% of the time in the House and 92.5\% of the time in the Senate (left panel of Appendix Table \ref{tab: congressional bills prediction, performance of GPT-4o}).
What drives the model's ability to accurately predict whether a Congressional bill will pass the House or the Senate based only on its description?
The answer is that the text of congressional legislation is likely included in its training dataset. 

To evaluate training leakage, we prompt the model to complete each bill's text description based on only the first half of its text, following research in computer science such as \cite{golchin2024timetravelllmstracing} (see Appendix Figure \ref{fig: congressional bills completion, prompts} for the specific prompt).  
If a model reproduces the text exactly, it has likely seen it during training. 
This is not a necessary condition for training leakage; models may benefit from exposure to a text piece without memorization. Perfect reproduction nonetheless offers compelling evidence.

On 344 bills, GPT-4o completes the bill's text description \textit{exactly} as it is written, indicating that not only was GPT-4o likely trained on these text pieces but it appears to have memorized them. 
Appendix Figure \ref{figure: congressional bills prediction, no date restriction, memorization examples} provides two examples of successful completions. 
On the other bills, GPT-4o's completed bill descriptions are close to the original bill descriptions.
Word embeddings of GPT-4o's descriptions are far closer to those of the originals than the average distance between the word embeddings of two randomly selected bills (left panel of Appendix Table \ref{table: congressional bills prediction, embedding closeness}).

One might wonder whether this training leakage could be addressed through prompt engineering --- for example, by commanding the models to not pay attention to any information past a certain date.
Applying this prompt engineering strategy to our sample of Congressional bills (see Appendix Figure \ref{fig: congressional bills prediction, prompts} and \ref{fig: congressional bills completion, prompts} for associated prompts), we still find substantial evidence of training leakage.
Even when explicitly told to not consider any information past the bill's introduction date in Congress, GPT-4o can still accurately predict its outcome in the House and the Senate based on these small snippets of text (right panel of Appendix Table \ref{tab: congressional bills prediction, performance of GPT-4o}). GPT-4o still exactly completes nearly the same number of bill descriptions as without the prompt engineering (330 versus 344), and the word embeddings of its completed descriptions remain quite close on average to those of the originals (right panel of Appendix Table \ref{table: congressional bills prediction, embedding closeness}). 

\vspace{-1em}
\paragraph{Assessing Training Leakage in Financial News Headlines} We consider another domain relevant for economists: financial markets.
Existing work \citep[e.g.,][]{glasserman2023assessinglookaheadbiasstock, lopezlira2024chatgptforecaststockprice} found that LLMs predict stock returns accurately from news headlines. 
We test whether this reflects training leakage using publicly available headline data \citep[][]{Headlines} covering nearly 4 million headlines for 6,000 publicly traded companies from 2009-2020.

We sampled 10,000 financial news headlines from 2019, and we prompt GPT-4o to complete each financial news headline based on only 50\% of its text (see Appendix Figure \ref{fig: financial news headlines completion, prompts} for the specific prompt). 
GPT-4o reproduces 60 financial news headlines exactly as they were written in the publicly available dataset, indicating that GPT-4o was likely trained on these headlines and memorized them.
Appendix Figure \ref{figure: financial news headlines prediction, no date restriction, memorization examples} provides two examples of successfully completions.
On all other headlines, word embeddings of the model's completions are close to those of the original headlines (left panel of Appendix Table \ref{table: financial news headlines prediction, embedding closeness}).

We again explore whether explicitly incorporating date restrictions into the LLM prompt moderate this evidence of training leakage (see Appendix Figure \ref{fig: financial news headlines completion, prompts} for the associated prompts).\footnote{See also \citet[][]{wongchamcharoen2025largelanguagemodelsllms}. Another prompting strategy -- entity masking -- aims to prevent memorized information by masking identifiers in prompts \citep[][]{glasserman2023assessinglookaheadbiasstock, sarkar2024lookahead, EnglebergEtAl(25)-EntityNeutering}. However, implementation details matter, and its wider applicability is unclear. In our congressional legislation example, it is not obvious what should be masked.}
Surprisingly, this appears to make the problem \textit{worse}; GPT-4o now reproduces 73 headlines exactly, and word embeddings of GPT-4o's completions with the date restriction are still on average close to those of the original headlines.
 
\cite{sarkar2024lookahead} provide additional evidence of training leakage in finance: prompting Llama 2 to predict risks from September-November 2019 earnings calls, the LLM mentions Covid-19 in over 25\% of cases—a form of "look-ahead bias" from training on future information. Combined with our findings and other evidence \citep[][]{glasserman2023assessinglookaheadbiasstock, lopezlira2025memorizationproblemtrustllms}, this indicates substantial risk of training leakage in finance applications.

\subsection{Practical Guidance for Prediction Problems}\label{section: prediction problem guidance}

The no-training-leakage condition (Equation \ref{equation: no leakage condition}) may seem abstract, but it provides concrete guidance for empirical practice. 
Researchers must consider what population they are predicting on, how their evaluation sample relates to it, and how their evaluation sample relates to the LLM's training corpus.

To make this concrete, recall that the bias term in Proposition \ref{proposition: prediction gpts iff no leakage} depends on $\frac{q^{T \mid D}(t_r)}{q^{T}(t_r)} - 1$ -- how much does learning that the researcher sampled text piece $r$ update our beliefs about whether $r$ was in the training data.
No training leakage requires this term to be zero, or at least uncorrelated with the LLM's performance. 
Different prediction problems and model choices achieve this in different ways. 
We illustrate this by considering the threat of training leakage and how to manage it across several examples.

\vspace{-1em}
\paragraph{Example: Lookahead Bias and Time-Stamped Models}
Consider a researcher who wants to predict stock returns from future financial news headlines. 
If the researcher evaluates on recent headlines from 2024 using an LLM trained on data through 2025, the leakage term is almost certainly positive: both the researcher and the LLM builders had access to 2024 headlines. 
Any predictive success may reflect training leakage.

The solution in this case is to use open-source LLMs with fixed published weights, such as the Llama family \citep[][]{touvron2023llama2openfoundation, dubey2024llama3herdmodels} and others (e.g., BLOOM, OLMo, etc.), or time-stamped training data \citep[e.g.,][]{Sarkar(24), he2025chronologicallyconsistentlargelanguage}. 
If the researcher uses a model with weights published on date $\tau$ and constructs an evaluation sample using documents published after $\tau$, then $q^{T}_r(t_r) = 0$ mechanically since they could not have been in the training data. This eliminates training leakage by design.

Closed models like the GPT family from OpenAI or the Claude family from Anthropic do not permit this solution.
Their training data is undisclosed, and they may be continuously fine-tuned.  
Researchers in natural language processing now regularly caution against sending test data to the APIs or chat interfaces of closed LLMs since these data may be used in further fine-tuning or the development of new models \citep[][]{jacovi-etal-2023-stop}.\footnote{\cite{balloccu2024leakcheatrepeatdata} estimates that 263 benchmarks may have been inadvertently leaked to OpenAI through use of the chat interface and API, and \cite{cheng2024dateddatatracingknowledge} questions the validity of publicly stated ``knowledge cutoffs'' in closed LLMs.} 
Most worryingly, \cite{BarrieEtAl(24)-LLMReplication} illustrates that the results of submitting the same prompt to GPT-4o yields results that change month-to-month, despite there being no publicly announced changes to the underlying model. $\blacktriangle$

\vspace{-1em}
\paragraph{Example: Prediction on Confidential Documents}  Suppose a researcher wants to predict, for example, whether administrative case notes are predictive of some decision, like pretrial release. 
In examples like this, the target population consists of confidential documents that were never publicly released. 
Since these documents never entered any public corpus, we can again credibly argue that $q^{T}_r(t_r) = 0$ for all documents since they were never available for training.
This reasoning hinges on credibly claiming the documents' provenance. $\blacktriangle$

\vspace{-1em}
\paragraph{Example: Random Sampling from a Known Corpus} Consider a researcher with a complete corpus of economics papers published between 2000 and 2020 who wants to predict citation counts from abstracts. She draws a random sample to form her evaluation set. 
In this case, the researcher's sampling process is independent of the LLM's training data: her random number generator knows nothing about which papers OpenAI included in their training data. 
This implies $q^{T \mid D}_r(t_r) = q^{T}_r(t_r)$ -- learning that a paper was randomly selected provides no information about whether it was trained on --- so the leakage term equals zero.

The logic is familiar from causal inference: randomization eliminates omitted variable bias by making treatment assignment independent of potential outcomes. Here, random sampling eliminates training leakage by making the researcher's evaluation sample independent of the LLM's training data. 
This can be applied whenever the researcher draws a genuinely random sample from a well-defined corpus before using an LLM. 
The converse is also important: just as non-random treatment assignment resurrects possible confounding in causal inference, non-random sampling may resurrect training leakage. $\blacktriangle$

\vspace{-1em}
\paragraph{Example: Prediction on the Complete Population} As a final example, suppose the researcher collected all Congressional bill descriptions from 1973–2016 and wants to understand what textual features predict passage. 
The researcher seeks to descriptively characterize patterns in this specific historical corpus, not to make predictions on some population of legislation.
When the researcher observes the entire population, the prediction problem changes character. 
There is no sampling, and hence no inference to a broader population.
In this case, $q^{D}_r = 1$ and the leakage condition is satisfied trivially: we are not asking whether performance generalizes beyond what we observe.
The complete-population case licenses only descriptive conclusions about the collected corpus. $\blacktriangle$

\vspace{1em}

More broadly, these examples illustrate a general principle: no training leakage is ultimately about \textit{both} research design and model properties. 
Researchers must be explicit about their target population, their sampling procedure, and how that procedure relates to possible training corpora. 
When in doubt, the safest approach combines open-source models with published weights or documented training data and evaluation samples constructed to mechanically exclude the model's training corpus.

\section{Estimation with Large Language Models}\label{section: estimation with LLMs}

In estimation problems, the researcher measures economic concepts $V_r$ expressed in text pieces $r$ to estimate downstream parameters.
The measurement $f^*(\cdot)$ is costly to scale across thousands or millions of text pieces.
This is where LLMs enter as potential substitutes.

Using LLM outputs in plug-in estimation requires a strong assumption: the LLM must reproduce the existing measurement everywhere. As Section \ref{section: challenge of LLM evaluation} discussed, LLM performance varies across similar tasks, and benchmark evaluations provide little guidance on new applications.
We demonstrate this empirically: seemingly minor choices -- which LLM, which prompt -- substantially affect downstream parameter estimates in applications to finance and political economy. 
These choices change the magnitude, significance, and even sign of estimated parameters.
The solution: collect a small validation sample and use it to debias the LLM's outputs. 

\subsection{The Researcher's Estimation Problem}

The researcher specifies a parameter $\theta \in \Theta$ and a moment condition $g(\cdot)$ that identifies this parameter. 
If she collected the economic concept $V_r$ on each text piece $r$, she would calculate
\begin{equation}\label{eqn: target sample parameter}
\widehat{\theta}^* = \arg \min_{\theta \in \Theta} \frac{1}{N} \sum_{r \in \mathcal{R}} D_r g(V_r, W_r; \theta).
\end{equation}
For example, letting $g(V_r, W_r; \theta) = (V_r - W_r^\prime \theta)^2$, the researcher studies how $V_r$ relates to the linked variables $W_r$, and $\widehat{\theta}^*$ is the sample regression coefficient. 
Due to the text processing problem, however, the researcher cannot calculate $\widehat{\theta}^*$ directly.

The researcher prompts an LLM to measure the economic concept on each text piece, $\widehat{V}_r := \widehat{m}(r; t)$, and plugs in the LLM's labels 
\begin{equation}\label{eqn: plug in sample parameter}
\widehat{\theta} = \arg \min_{\theta \in \Theta} \frac{1}{N} \sum_{r \in \mathcal{R}} D_r g(\widehat{m}(r; t), W_r; \theta).
\end{equation}
\noindent When can we draw conclusions about $\widehat{\theta}^*$ based on $\widehat{\theta}$? 

We associate the researcher with a research context $Q(\cdot) \in \mathcal{Q}$ satisfying Assumption \ref{asm: research contexts}. 
We assume that the researcher could study any moment condition $g(\cdot) \in \mathcal{G}$ satisfying the following assumption. 

\begin{assumption}\label{asm: moment conditions}
For all $g(\cdot) \in \mathcal{G}$, $g(\cdot)$ is differentiable and there exists some $\overline{G} > 0$ such that $\left| \frac{\partial g(v, W_r; \theta)}{\partial \theta} \right| \leq \overline{G}$ for all $r \in \mathcal{R}$, $\theta \in \Theta$, and values of the economic concept $v$.
\end{assumption}

As the number of economically relevant text pieces grows large (see Appendix \ref{section: prediction and estimation, large sample}), we can characterize the behavior of both estimators. 
The target moment condition converges to, at any parameter value $\theta \in \Theta$, 
\begin{equation}
    \frac{1}{N} \sum_{r \in \mathcal{R}} D_r g(V_r, W_r; \theta) - \frac{1}{\mathbb{E}_{Q}[\sum_{r \in \mathcal{R}} D_r ]} \mathbb{E}_{Q}\left[ \sum_{r \in \mathcal{R}} D_r g(V_r, W_r; \theta) \right] \xrightarrow{p} 0.
\end{equation}
Conditional on the LLM's realized training dataset, the plug-in moment condition converges to, at any parameter value $\theta \in \Theta$, 
\begin{equation}
    \frac{1}{N} \sum_{r \in \mathcal{R}} D_r g(V_r, W_r; \theta) - \frac{1}{\mathbb{E}_{Q}[\sum_{r \in \mathcal{R}} D_r \mid T = t ]} \mathbb{E}_{Q}\left[ \sum_{r \in \mathcal{R}} D_r g(\widehat{m}(r; t), W_r; \theta) \mid T = t \right] \xrightarrow{p} 0.
\end{equation}
As in the prediction problem, conditioning on the training dataset simplifies analysis by treating the LLM as a fixed mapping.
 
Given an LLM with guarantee $\mathcal{M}$, the researcher would like to recover the moment condition defined using the economic concept. 

\begin{definition}\label{definition: validity and GPT for estimation}
The LLM $\widehat{m}(\cdot; t)$ with guarantee $\mathcal{M}$ \textit{automates} the existing measurement $f^*(\cdot)$ for the moment condition $g(\cdot)$ in research context $Q(\cdot)$ if, for all models satisfying the guarantee $\widehat{m}(\cdot) \in \mathcal{M}$ and all $\theta \in \Theta$, 
$$
\mathbb{E}_{Q}\left[ \sum_{r \in \mathcal{R}} D_r g(\widehat{m}(r), W_r; \theta) \mid T = t \right] = \mathbb{E}_{Q}\left[ \sum_{r \in \mathcal{R}} D_r g(V_r, W_r; \theta) \right].
$$
The LLM $\widehat{m}(\cdot; t)$ with guarantee $\mathcal{M}$ is a \textit{general-purpose technology for estimation} if it automates the researcher's measurement process for all $g(\cdot) \in \mathcal{G}$ and $Q(\cdot) \in \mathcal{Q}$. 
\end{definition}

\noindent The excitement around LLMs stems from their potential as ``general-purpose technologies'' -- tools deployable across diverse applications without task-specific engineering \citep[e.g.,][]{ElondouEtAl(24)}. 
For estimation problems, this would mean reliably substituting for existing measurements across different economic concepts and research contexts. 
Definition \ref{definition: validity and GPT for estimation} formalizes what this requires: the LLM must produce moment conditions matching the existing measurement procedure, regardless of which moment condition or population the researcher studies.
If an LLM satisfies this property, researchers could confidently use its outputs for plug-in estimation knowing only the guarantee $\mathcal{M}$.

\subsection{Measurement Error as a Threat to Estimation}\label{section: measurement error as a threat}

We clarify what guarantee $\mathcal{M}$ is necessary and sufficient for an LLM to automate the existing measurement. 
We decompose the difference between the plug-in moment condition and the target moment condition into two terms. 

\begin{lemma}\label{lemma: training leakage decomp for estimation}
Under Assumption \ref{asm: research contexts}, for any research context $Q(\cdot) \in \mathcal{Q}$, moment condition $g(\cdot) \in \mathcal{G}$ and text generator $\widehat{m}(\cdot)$, $\mathbb{E}_{Q}[\sum_{r \in \mathcal{R}} D_r g(\widehat{m}(r), W_r; \theta) \mid T = t] - \mathbb{E}_{Q}[\sum_{r \in \mathcal{R}} D_r g(V_r, W_r; \theta)]$ equals
\begin{equation}\label{eqn: training leakage decomp for estimation}
\small
\begin{aligned}
    \left(\mathbb{E}_{Q}\left[ \sum_{r \in \mathcal{R}} D_r g(\widehat{m}(r), W_r; \theta) \right] - \mathbb{E}_{Q}\left[ \sum_{r \in \mathcal{R}} D_r g(V_r, W_r; \theta) \right]\right) + \mathbb{E}_{Q}\left[ \sum_{r \in \mathcal{R}} D_r \left( \frac{q_{r}^{T \mid D}(t_r)}{q_{r}^{T}(t_r)} - 1 \right) g(\widehat{m}(r), W_r; \theta) \right].
\end{aligned}
\end{equation}
\end{lemma}

\noindent The second term captures training leakage -- potential overlap between the LLM's training dataset and the researcher's evaluation sample. 
As discussed in Section \ref{section: prediction problem guidance}, training leakage can be controlled through appropriate choice of model and research context.
For example, if the researcher collects all text pieces in their research context or randomly samples text pieces, then training leakage is mechanically zero.

We therefore focus on the first term, which captures possible LLM errors. 
Intuitively, LLM errors $\Delta_r = \widehat{m}(r; t) - V_r$ can bias parameter estimates if they correlate with the economic variables $W_r$.
A natural question arises: what if we know the LLM is ``pretty good''  -- say, within some $\delta$ of the true measurement everywhere?  
This is what we might be tempted to conclude from impressive benchmark evaluations and demonstrations of LLM capabilities.
More precisely, suppose the LLM satisfies the guarantee $\mathcal{M}(Q, \delta)$ --- the collection of text generators satisfying $\left\| \widehat{m}(\cdot) - f^*(\cdot) \right\|_{\infty, Q} = \max_{r \in \mathcal{R} \colon q_{r}^D > 0} | \widehat{m}(r; t) - f^*(r)| \leq \delta$.
Does this ensure valid plug-in estimation? 

Unfortunately, knowing the guarantee $\mathcal{M}(Q, \delta)$ does not tell us about the exact pattern of errors --- whether they relate to economic variables in ways that bias our estimates. 
We say that a moment condition $g(\cdot)$ is \textit{sensitive} to the economic concept $V_r$ in research context $Q(\cdot)$ if $q_r^{D} > 0$ and there exists some $\underline{G} > 0$ such that $| \frac{\partial g(v, W_r; \theta)}{\partial v} | \geq \underline{G}$ for all $v, \theta$. 
Let $\mathcal{R}(g, Q)$ denote the collection of sensitive text pieces.

\begin{lemma}\label{lemma: measurement error and estimation error}
Consider any moment condition $g(\cdot) \in \mathcal{G}$ in research context $Q(\cdot) \in \mathcal{Q}$. 
Then, for all $\theta \in \Theta$ and $\widehat{m}(\cdot) \in \mathcal{M}(Q, \delta)$ satisfying no training leakage,
\begin{equation}\label{equation: measurement error bound on error}
\left| \mathbb{E}_{Q}\left[ \sum_{r \in \mathcal{R}} D_r g(\widehat{m}(r), W_r; \theta) \mid T = t \right] - \mathbb{E}_{Q}\left[ \sum_{r \in \mathcal{R}} D_r g(V_r, W_r; \theta) \right] \right| \leq \overline{G} \delta.
\end{equation}
But, for all $\theta \in \Theta$, there exists $\widehat{m}(\cdot) \in \mathcal{M}(Q, \delta)$ that satisfies no training leakage such that, for $\delta(r)$ defined in the proof,
\begin{equation}\label{equation: measurement error insuff for validity}
\left| \mathbb{E}_{Q}\left[ \sum_{r \in \mathcal{R}} D_r g(\widehat{m}(r), W_r; \theta) \mid T = t \right] - \mathbb{E}_{Q}\left[ \sum_{r \in \mathcal{R}} D_r g(V_r, W_r; \theta) \right] \right| \geq \underline{G} \left( \sum_{r \in \mathcal{R}(g, Q)} |\delta(r)| q_r^{D} \right).
\end{equation}
\end{lemma}

\noindent Lemma \ref{lemma: measurement error and estimation error}(i) shows guarantee $\mathcal{M}(Q, \delta)$ bounds the LLM's error for the moment condition.  
But this is not enough for the LLM to automate the existing measurement.
Lemma \ref{lemma: measurement error and estimation error}(ii) shows text generators satisfying the guarantee $\mathcal{M}(Q, \delta)$ can still produce meaningful estimation errors.
We cannot rule out that the LLM's errors correlate with economic variables in ways that bias estimates.
This makes measurement error pernicious.

This leads to our main characterization.
Researchers can safely ignore the details of the LLM's design no matter the research context studied and economic question being asked if and only if its labels reproduce the existing measurement process everywhere. Anything less cannot guarantee valid plug-in estimation across all possible applications.
 
\begin{proposition}\label{proposition: no measurement error implies GPT for estimation community}
Suppose the LLM $\widehat{m}(\cdot; t)$ satisfies no training leakage in all research contexts $Q(\cdot) \in \mathcal{Q}$ and moment conditions $g(\cdot) \in \mathcal{G}$.
Provided there exists some $g(\cdot) \in \mathcal{G}$ that is sensitive to the economic concept for any $r \in \mathcal{R}$, then the language model is a general-purpose technology for estimation if and only if $\widehat{m}(\cdot; t)$ satisfies the guarantee $\mathcal{M}(Q, 0)$ for all research contexts $Q(\cdot)$.
\end{proposition}

\subsubsection{Some Evidence of Measurement Error}\label{section: evidence of measurement error}
Plugging in LLM outputs requires that the model reproduces the target measurement process $f^*(\cdot)$.
While their impressive performance on some tasks makes this assumption appealing, Section \ref{section: challenge of LLM evaluation} showed such intuitions about LLMs are unreliable. 
No LLM achieves perfect accuracy on benchmark evaluations, and growing evidence documents their surprising errors.
Does this matter for economics research? 
We next show that LLM errors substantially affect downstream parameter estimates in two empirical examples. 

\vspace{-1em}
\paragraph{Linear Regression with LLMs} Consider a researcher relating the economic concept $V_r$ and linked variables $W_r$ through linear regression.
The researcher may use the LLM's labels as the dependent variable:
\begin{equation}\label{equation: target regression and plug-in, LLM on LHS}
    V_r = W_r^\prime \beta^* + \epsilon_r, \mbox{ and } \widehat{m}(r; t) = W_r^\prime \beta + \widetilde{\epsilon}_r.
\end{equation}
or the independent variable:
\begin{equation}\label{equation: target regression and plug-in, LLM on RHS}
    W_r = V_r^\prime \alpha^* + \nu_r, \mbox{ and } W_r = \widehat{m}(r; t)^\prime \alpha + \widetilde{\nu}_r.
\end{equation}
The bias depends on how the LLM's error $\Delta_{r} = \widehat{m}(r; t) - V_r$ varies across text pieces. 
These are well-known results, dating back to \citet[][]{BoundEtAl(94)-ValidationData}.

\begin{proposition}\label{prop: bias characterization of LLM on LHS and RHS}
Consider the research context $Q(\cdot)$ and assume $q_{r}^{T}(t_r) = q_{r}^{T \mid D}(t_r)$ for all $r \in \mathcal{R}$. 
\begin{enumerate}
\item[(i)] Defining $\lambda_{\Delta \mid W}$ to be the coefficients in the regression of $\Delta_r = \widehat{m}(r; t) - V_r$ on $W_r$ in the research context $Q(\cdot)$, then $\beta = \beta^* + \lambda_{\Delta \mid W}$.

\item[(ii)] Defining $\lambda_{V \mid \widehat{V}}$ to be the regression coefficients of $V_r$ on $\widehat{m}(r; t)$ and $\lambda_{\eta \mid \widehat{V}}$ to be the regression coefficients of $\eta_{r}$ on $\widehat{m}(r; t)$ in the research context $Q(\cdot)$, then $\alpha = \lambda_{V \mid \widehat{V}} \alpha^* + \lambda_{\eta \mid \widehat{V}}$.
\end{enumerate}
\end{proposition}

\noindent When the economic concept is the dependent variable, the bias equals the best linear predictor of the LLM's errors given the covariates. 
When the economic concept is the independent variable, the bias has a more complex form involving attenuation and correlation between the error and the residual.
Knowing the LLM is ``accurate'' provides no guarantee about these biases. 
What matters is whether errors correlate with economic variables in the specific regression. 

\vspace{-1em}
\paragraph{Assessing Measurement Error in Financial News Headlines} 
We return to the financial news headlines dataset, focusing on headlines published in 2019 for $6,000$ publicly traded stocks. We observe each headline's text $r$, publication date and stock ticker. We merge realized returns at various horizons (1, 5, and 10 days) after publication and lagged returns before publication \citep[][]{WRDSBetaSuite}.

Financial news headlines express various economic concepts that we could measure and relate to realized stock returns. We focus on one: is the headline positive, negative or neutral news for the company? 
While we could read every single financial news headline published in 2019, this process would be painstaking.
It is natural to use LLMs to solve this text processing problem. 

We take each financial news headline $r$ and prompt LLMs to label each news headline as positive, negative or neutral news for the company it refers to, $\widehat{V}_r := \widehat{m}(r; t)$. 
This requires making several practical choices: what specific LLM to use? What prompt engineering strategy? 
If alternative choices lead to different labels and downstream estimates, this would indicate non-ignorable errors in the LLM outputs.

We prompt GPT-3.5-Turbo, GPT-4o, GPT-4o-mini, GPT-5-mini, and GPT-5-nano to label each financial news headline $r$, using nine alternative prompts to each model. 
Our two base prompts provide the LLM with the text of the headline $r$ and ask it to label whether this news is positive, negative, or neutral for the company; we also ask the LLM to provide its confidence and a magnitude in the label. 
Our two base prompts differ in how they ask the model to format its reply: filling-in-the-blanks text or a structured JavaScript Object Notation (JSON) object. 
The prompts are provided in Appendix \ref{section: prompts for empirical work}. 

We further vary these base prompts in two ways,  motivated by popular prompt engineering strategies \citep[][]{LiuEtAl(23), wei2023chainofthoughtpromptingelicitsreasoning, white2023promptpatterncatalogenhance, chen2024unleashingpotentialpromptengineering}. 
First, we request the LLM adopt one of four different ``personas," like ``knowledgeable economic agent'' or ``expert in finance.'' 
Second, we append three prompt modifiers that ask the LLM to ``think carefully'' or ``think step-by-step'' and provide an explanation for its answer.
For each financial news headline $r$, we obtain labels $\widehat{V}_{r}^{m,p}$ associated with a collection of different LLMs $m$ and prompting strategies $p$. 

For each model $m$ and prompt $p$, we regress the realized returns of each stock within 1-day, 5-days and 10-days after the headline's publication date $Y_r$ on the LLM's labels $\widehat{V}_{r}^{m,p}$, controlling for the model's reported magnitudes and lagged realized returns.
We report the coefficients $\widehat{\beta}_{m,p}$ on whether the LLM labels the headline as positive and whether the LLM labels it as negative, as well as their associated t-statistics with standard errors clustered at the date and company level. 
Figure \ref{figure: financial news headline, positive/negative news, t-stat, magnitude, 1/5/10 day realized} and Table \ref{table: financial news headline, positive/negatives, coefficient variation, magnitude} summarize the results. 
Simply changing the prompt or model yields markedly different estimates: many prompt-model combinations that produce different directions and magnitudes of the relationship between the positive/negative label and realized returns.\footnote{Appendix Figure \ref{figure: financial news headline, positive/negative news, heat maps}, shows substantial variation in the pairwise agreement in the labels produced.
There appear to be no consistent patterns in which pairs of prompting strategies tend to have the most agreement.}

\vspace{-1em}
\paragraph{Assessing Measurement Error in Congressional Legislation} 
We next return to the Congressional legislation data. 
For each bill, we observe the text of its description $r$ as well as a collection of economic variables $W_r$, such as the party affiliation of the bill's sponsor, whether the bill originated in the Senate, and an ideological score -- the DW1 roll call voting record -- of the bill's sponsor. 
A researcher might reasonably study how these variables shape the topic of each bill, $V_r$. 
Could we use an LLM to collect those labels?

We randomly select $10,000$ Congressional bills and separately prompt GPT-3.5-Turbo, GPT-4o, GPT-5-mini, and GPT-5-nano to label each Congressional bill for its policy area using alternative prompting strategies, including base prompts that modify the requested format, persona modifications, chain-of-thought modifications, and even few-shot examples (see Appendix \ref{section: prompts for empirical work} for the specific prompts we used). 

We regress the labeled economic concept $\widehat{V}_{r}^{m,p}$ --- in this case, the policy topic of the bill --- against linked covariates $W_r$, separately reporting the coefficients $\widehat{\beta}_{m,p}$ and their associated t-statistics.
Figure \ref{figure: congressional bills, LLM on LHS, full data, t stat figure} and Table \ref{table: congressional bills, coeff variation, lhs} summarize the variation in the resulting estimates across models and prompts.
For each combination of labeled policy topic and linked covariate in the Congressional bills dataset, we see substantial variability across different LLMs and prompting strategies.\footnote{Appendix Figure \ref{figure: congressional legislation, major topic heatmap} calculates the pairwise agreement in the labels produced by alternative prompting strategies $p$. We again find substantial variation in the pairwise agreement in the labels produced.}
Once again, simply changing the prompt or model yields remarkably different downstream estimates.

\subsection{Practical Guidance with Validation Data}\label{section: practical guidance with validation data}

Given the evidence on LLM brittleness in Section \ref{section: challenge of LLM evaluation}, it is difficult to defend the assumption of no measurement error in LLM outputs for estimation problems. 
The solution is to collect measurements $V_r = f^*(r)$ on a small validation sample and use them to de-bias the plug-in estimate based on the LLM labels $\widehat{V}_r$. 
The virtues of validation data have been understood for decades \citep[][]{bound1991extent, BoundEtAl(94)-ValidationData, LeeSepanski(95)-ErrorsInVariable}.
It has been recently revived in machine learning to handle the types of mismeasured covariates and outcomes produced by modern ML models; see for example \citet[][]{WangMcCormickLeek(20)}, \citet[][]{AngelopoulosEtAl(23)-PPI}, \citet[][]{EgamiEtAl(24)-ImperfectSurrogates} and \citet[][]{carlson2025unifyingframeworkrobustefficient} among others.

We illustrate the value of validation data using the familiar case of linear regression. We review the mechanics of debiasing when the LLM label appears as the dependent variable, provide intuition about when it will work well using asymptotic arguments, and demonstrate finite sample performance using Congressional legislation. 
Our discussion is pedagogical; we refer the readers to the works above for more general settings.

\vspace{-1em}
\paragraph{An Illustration with Linear Regression} 
Return to the case where the researcher wishes to regress the economic concept $V_r$ on the linked variables $W_r$, but relies on the LLM labels instead and reports the plug-in regression $\widehat{V}_r = W_r^\prime \beta + \widetilde{\epsilon}_r$. 
Appendix \ref{section: studying the asymptotic dist'n of bias corrected} considers when the economic concept $V_r$ is a covariate.

Proposition \ref{prop: bias characterization of LLM on LHS and RHS} established that the bias of the plug-in regression coefficient $\beta$ depends on how the LLM's error $\Delta_r = \widehat{m}(r; t) - V_r$ covaries with $W_r$. 
This suggests a natural solution: on a random subset of the researcher's dataset, collect $V_r$ using the existing measurement $f^*(\cdot)$. 
For example, the researcher reads and labels a subset of the text pieces themselves.
We call text pieces on which the researcher observes $(r, W_r, \widehat{m}(r; t), V_r)$ the validation sample, and we refer to the remaining text pieces on which she observes $(r, W_r, \widehat{m}(r; t))$ as the primary sample.

In the validation sample, the researcher can estimate the bias $\widehat{\lambda}_{\Delta \mid W}$ by forming $\Delta_r = \widehat{m}(r; t) - V_r$ on the validation sample and regressing it on $W_r$. 
The validation sample can therefore be used for two purposes. 
First, the validation sample provides a target to optimize when selecting a model and prompt.
For any choice of LLM $m$ and prompt $p$, the researcher can estimate the bias of the plug-in regression $\widehat{\lambda}_{\Delta \mid W}^{m, p}$ associated with the labels $\widehat{V}_{r}^{m,p}$; and thereby select the combination that results in the smallest bias.

Second, and more importantly, the validation sample enables bias correction. 
Rather than reporting the plug-in coefficient $\widehat{\beta}$, the researcher reports 
\begin{equation}
    \widehat{\beta}^{debiased} = \widehat{\beta} - \widehat{\lambda}_{\Delta \mid W}.
\end{equation}
This is exceedingly simple to implement: run two regressions -- regress $\widehat{m}(r; t)$ on $W_r$ in the primary sample and regress $\Delta_r$ on $W_r$ in the validation sample -- and subtract.
Inference is equally straightforward: for example, researchers may bootstrap the primary and validation samples. 

As discussed in Appendix \ref{section: studying the asymptotic dist'n of bias corrected}, the bias-corrected estimator has desirable theoretical properties. 
Consider a research context where the researcher randomly samples text pieces, allocating a fraction $\rho_{p}$ to the primary sample and a fraction $\rho_{v}$ to the validation sample. 
As the number of economically relevant text pieces grows large, the bias-corrected estimator $\widehat{\beta}^{debiased}$ is consistent for the target regression coefficient $\beta^*$. 
$\widehat{\beta}^{debiased}$ is asymptotically normal with limiting variance given by 
 \begin{equation}\label{equation: limiting distribution of bias correction, main text}
     \sigma^{-4}_{W} \left(\frac{1 - \rho_p}{\rho_p} \sigma^2_{\widehat{V} W} + 2 \sigma_{\widehat{V} W} \sigma_{\Delta W} + \frac{1 - \rho_v}{\rho_v} \sigma^2_{\Delta W} \right),
 \end{equation}
where $\sigma_W$ is the standard deviation of the linked variable $W_r$ across all text pieces, $\sigma_{\widehat{V} W}$ is the standard deviation of the product $\widehat{V}_r \times W_r$, and $\sigma_{\Delta W}$ is defined analogously.
The precision of the bias-corrected estimator depends on the relative size of the validation sample versus the primary sample as well as the variability of the LLM's label $\widehat{V}_{r}$ and measurement error $\Delta_r$ across text pieces. 

A natural question arises: if we collect validation data anyway, why bother with the possibly mismeasured LLM labels on the primary sample? 
The validation-only estimator $\widehat{\beta}^*$ is also consistent and asymptotically normal: its limiting variance is given by $\sigma^{-4}_{W} \frac{1 - \rho_v}{\rho_v} \sigma^2_{V W}$. 
Comparing these expressions reveals that the bias-corrected estimator is more precise when:
\begin{equation}\label{equation: variance comparison, main text}
    \frac{1 - \rho_p}{\rho_p} \sigma^2_{\widehat{V} W} + 2 \sigma_{\widehat{V} W} \sigma_{\Delta W} \leq \frac{1 - \rho_v}{\rho_v} \left( \sigma^2_{V W} - \sigma^2_{\Delta W} \right).
\end{equation}
This comparison depends on the relative standard deviation of the existing measurement $V_r$ and the LLM's error $\Delta_r$. 
Equation \eqref{equation: variance comparison, main text} implies that the bias-corrected regression coefficient can be more precisely estimated than the validation-sample-only regression coefficient if the LLM's labels are sufficiently accurate.
Correctly incorporating imperfect LLM outputs can result in tighter standard errors than ignoring the LLM altogether. This phenomenon has been documented in recent machine learning research, such as \citet[][]{AngelopoulosEtAl(23)-PPI} and associated work, that combines validation data with the outputs of machine learning models to estimate downstream parameters.

In other words, a free-lunch is possible: use the LLM to solve the text processing problem while delivering precise estimates of the target regression coefficient.
Importantly, LLM outputs are not substitutes for existing measurements; instead, they amplify a small validation sample.

Finally, validation data provides another critical benefit: it addresses concerns about specification searching that arise from LLM brittleness. 
As seen, alternative choices of prompts and models can yield dramatically different downstream estimates, creating a severe $p$-hacking risk. 
With countless possible prompts and multiple competing LLMs, researchers could search across specifications until finding desired results. 
Provided the researcher collects validation data and debiases whatever LLM output they collect, alternative choices of prompting strategy and LLM target the same empirical quantity: the parameter $\theta$ defined using the researcher's existing measurement.

\vspace{-1em}
\paragraph{Monte Carlo Simulations based on Congressional Legislation} 
The Congressional legislation data is well-suited to illustrate the value of validation data.
The Congressional Bills Project trained teams of human annotators to label the description of each Congressional bill $r$ for its major policy topic area $V_r := f^*(r)$, describing whether the bill falls into one of twenty possible policy areas.\footnote{The Congressional Bills Project states that all annotators were trained for a full academic quarter before beginning this task \citep[][]{CAPCodebook}.} 
Given its widespread use, researchers are comfortable using these measurements in downstream analyses. Can an LLM automate this measurement procedure $f^*(\cdot)$?

For a given policy topic $V_r$ (e.g., health, defense, etc.), linked covariate $W_r$ (e.g., whether the bill's sponsor was a Democrat, etc.) and pair of large language model $m$ and prompting strategy $p$, we randomly sample $5,000$ bills. 
On this random sample of $5,000$ bills, we calculate the plug-in coefficient $\widehat{\beta}$ by regressing $\widehat{V}_{r}^{m,p}$ on the linked variable $W_r$. 
We next randomly reveal the existing label $V_r$ on $250$ (i.e. 5\%) of our random sample of $5,000$ bills, which produces a validation sample. 
We then calculate the bias-corrected coefficient $\widehat{\beta}^{debiased}$. 
We repeat these steps for $1,000$ randomly sampled datasets. 
Across simulations, we calculate the average bias of the plug-in coefficient and the bias-corrected coefficient for the target regression $\beta^*$ associated with regressing the existing label $V_r$ on the linked variable $W_r$ on all $10,000$ bills. 
We repeat this exercise for each combination of bill topic $V_r$, linked covariate $W_r$, LLM $m$ (either GPT-3.5-Turbo, GPT-4o, GPT-5-mini and GPT-5-nano) and prompting strategy $p$.  
Appendix \ref{section: LLM on LHS, varying the size of the validation data} varies the size of the validation sample, finding similar results even when the validation sample contains as few as $125$ bills.

Figure \ref{figure: congressional bills, LLM on LHS, normalized bias, 5 percent validation} and the top panel of Table \ref{table: congressional bills, LLM on LHS, summary statistics, 5 percent validation} compares the average bias of the plug-in coefficient $\widehat{\beta}$ and the bias corrected coefficient $\widehat{\beta}^{debiased}$ for the target regression $\beta^*$ (normalized by their standard deviations) across possible combinations of bill topic $V_r$, linked covariate $W_r$, LLM $m$, and prompting strategy $p$. 
For most regression specifications and pairs of LLM and prompting strategy, the simple plug-in regression suffers from substantial biases. 
Using the validation sample yields estimates that are reliably unbiased --- indeed, the bias-corrected regression coefficient performs remarkably well across all regression specifications and pairs of LLM and prompting strategy.

For each regression specification and pair of LLM and prompting strategy, Table \ref{table: congressional bills, LLM on LHS, summary statistics, 5 percent validation} summarizes the fraction of simulations in which a 95\% confidence interval centered at either the plug-in coefficient or the bias-corrected coefficient includes the target regression $\beta^*$. 
The nominal $95\%$ confidence interval for the bias-corrected regression coefficient has approximately correct coverage across all regression specifications, LLMs and prompting strategies. 
By contrast, plug-in estimation often suffers from severe coverage distortions.

Finally, we can use these data to answer the question: If we already collected measurements $V_r$ in a validation sample, what is the value of the LLM labels? 
For each regression specification, LLM and prompting strategy, we compare the average mean square error of the bias-corrected coefficient and the validation-sample only coefficient for the target regression $\beta^*$. 
Figure \ref{figure: congressional bills, LLM on LHS, MSE, 5 percent validation} plots the resulting distribution across all choices of bill topic $V_r$, covariate $W_r$, and pair of model-and-prompting strategy. 
The average mean square error of the validation-sample-only coefficient is always higher (less precisely estimated) compared to the bias-corrected coefficient. 
Substantial precision improvements from using LLM outputs are possible in finite samples.  

Taken together, these results indicate that estimation is a promising use case for LLMs \textit{if} the researcher collects a validation sample to correct for LLM errors. 
LLMs can then lower the cost of data collection and improve statistical precision, while preserving the familiar econometric guarantees we desire.

\subsection{Estimation without Validation Data}

When using LLMs in estimation problems, validation data enables correct inference by allowing the researcher to estimate and correct for LLM errors. 
But what if such data cannot be collected?
Before considering possible paths forward, the researcher must make a judgment call: does there exist -- even in principle -- a measurement procedure $f^*(\cdot)$ that would produce labels $V_r$ the researcher would trust? 
This is not a statistical question, but a conceptual one about the nature of the economic concept being studied. 

\vspace{-1em} 
\paragraph{When Ground Truth Exists} Suppose the researcher believes there exists a ``true'' value of the economic concept for each text piece, even if measuring it is costly or time-consuming. 
Consider labeling Congressional bills' policy topics. 
The researcher might be confident that if they (or trained experts) carefully read each bill, they could reliably classify policy topics. 
The researcher may still not collect validation data despite believing ground truth exists -- perhaps due to budget constraints, time pressure, or confidence in LLM accuracy. In this case, researchers might pursue one of two paths that might appear defensible but ultimately are not.

First, the researcher could argue there are no errors in the LLM's outputs $\widehat{m}(r; t)$, justifying plug-in estimation.
This argument is difficult to defend at present.  
A researcher proceeding this way must somehow argue why an LLM's output exactly reproduces the economic concept, even though it is imperfect on benchmarks and why evidence on LLM brittleness does not apply. 

Second, acknowledging that LLM outputs are imperfect, the researcher might write down a statistical model of its errors $\Delta_r = \widehat{m}(r; t) - f^*(r)$, just as we would in the measurement error literature. 
Consider a stylized model: for a given LLM $m$, the errors 
errors $\Delta_{r}^{m,p} = V_r - \widehat{V}_{r}^{m,p}$ across prompting strategies $p$ are independent. 
Such an assumption would suggest particular solutions: perhaps the researcher could use one prompting strategy as an instrument for another. 
For a given LLM $m$, its labels across prompting strategies $p$ are surely correlated with one another. 
Across language models $m$, there is surely substantial overlap in training datasets. 
Consequently, the labels $\widehat{V}_r^{m,p}$ are correlated \citep[][]{kim2025correlatederrorslargelanguage}.
Taking a step back, our usual measurement error frameworks were not designed for this setting -- where the measurement comes from an algorithm whose behavior we do not fully understand, applied to a quantity we do not observe.

Given these challenges, the practical answer is clear: invest effort and collect a small validation sample.

\vspace{-1em}
\paragraph{When Ground Truth is Undefined} 
Suppose the economic concept is sufficiently abstract or subjective that the researcher is uncomfortable articulating what ``ground truth'' would even mean. 
In this case, the researcher could define the object of study as the LLM's outputs themselves. 
The concept simply is, for example, whatever GPT-4o produces with a given prompt. This makes plug-in estimation valid by definition.

But the researcher is now in uncharted territory.
This seemingly minor shift has important implications.
When a new prompt engineering strategy is produced, should we publish a new paper? 
When OpenAI inevitably releases GPT-6, should we revisit all published work that used GPT-5? 
In Section \ref{section: evidence of measurement error}, we found that variation across models and prompts can be enormous---different prompts and models can even yield estimates with different signs. If 60\% of model-prompt combinations suggest a positive relationship and 40\% suggest negative, what have we learned?

More fundamentally, defining the LLM's outputs as the object of study sidesteps the hard -- and economically interesting -- question. 
We care about the underlying economic concept, not the LLM's outputs. 
When facing such conceptual ambiguity, the scientific response is to acknowledge it and probe from multiple angles -- not to privilege one operationalization of the concept. This is especially true with LLMs, since we have found them to be brittle and unreliable on even well-defined tasks as discussed in Section \ref{section: challenge of LLM evaluation}. 

What might a more systematic investigation look like? 
As an example, suppose we compared an LLM's outputs with measurements produced by the researcher or domain experts. How do they differ? Where do they agree and disagree?
This work would help us articulate what exactly \textit{is} the economic concept we are studying.
By contrast, treating the LLM as ground truth bypasses this essential work: grappling with what the economic concept means and how best to measure it.

Altogether the researcher must ask themselves: Is it truly the case that no measurement procedure exists--even in principle--that they would trust? Or does one exist, but they are reluctant to invest in collecting even a small validation sample? 
These are fundamentally different situations. 
The former presents a genuine conceptual challenge deserving serious attention. 
The latter substitutes convenience for scientific rigor.

\section{Novel Uses of Large Language Models}

In addition to familiar prediction and estimation problems, LLMs enable exciting novel applications -- simulating human subject responses or generating research hypotheses -- that expand what we consider possible in empirical work.
We offer one interpretation by mapping these applications into our framework.
More broadly, these creative uses raise important questions about inferential goals, and we encourage further work to think rigorously about what researchers are ultimately trying to accomplish with such applications.

\subsection{Human Subject Simulation}

A growing body of research uses LLMs to simulate human subjects as ``in-silico'' subjects across economics \citep[e.g.,][]{horton2023large, manning2024automatedsocialsciencelanguage, MeiEtAl(24)}, marketing \citep[e.g.,][]{BrandEtAl(23)-LLMsForMarketResearch}, finance \citep[e.g.,][]{Bybee(24)}, political science \citep[e.g.,][]{Argyle_Busby_Fulda_Gubler_Rytting_Wingate_2023}, and computer science \citep[e.g.,][]{aher2023usinglargelanguagemodels}.

This maps into our framework by reinterpreting text pieces $r \in \mathcal{R}$ and the existing measurement $f^*(\cdot)$. 
Each text piece $r$ represents an experimental design or survey instrument, and $V_r$ might represent the average or model human response.
If the researcher collected human responses $V_r$, she would calculate downstream parameter estimates (Equation \ref{eqn: target sample parameter}).
But collecting human responses is costly, so instead the researcher substitutes LLM responses $\widehat{m}(r; t) = \widehat{V}_r$ and reports the plug-in parameter estimate (Equation \ref{eqn: plug in sample parameter}). 

\vspace{-1em}
\paragraph{Example: Testing Anomalies}
To test violations of risky choice models, behavioral economists construct ``anomalies'' -- lottery menus that highlighting flaws in the economic model, such as the Allais Paradox \citep[][]{allais1953comportement} or the Kahneman-Tversky choice experiments \citep[][]{kahneman1979prospect}. 
Could LLMs simulate human choices $\widehat{m}(r; t)$ on new anomalies? $\blacktriangle$

\vspace{-1em} 
\paragraph{Example: Large-scale Choice Experiments}
To compare risky choice models, recent work measures predictive accuracy across diverse lottery problems \citep[][]{ErevEtAl(17)-FromAnomaliesToForecasts, FudenbergEtAl(22)-Completeness}. 
\citet[][]{PetersonEtAl(21)-MLforChoiceTheory} recruited nearly 15,000 MTurk respondents to make over one million choices $V_r$ from lottery menus $r$, producing the ``Choices13K'' dataset. 
Could LLMs simulate the Choices13K dataset? $\blacktriangle$

\vspace{1em}

Viewing human subject simulation as an estimation problem implies in-silico subjects must exhibit no measurement error (Proposition \ref{proposition: no measurement error implies GPT for estimation community}) --- the LLM must reproduce human subject behavior on the researcher's experiment or survey. 
While some studies show LLMs reproduce published experiments, counterexamples abound: LLM responses to psychology experiments appear to produce more falsely significant findings than human subjects \citep[][]{cui2024aireplacehumansubjects}, cannot accurately reproduce the responses of human subjects on opinion polls \citep[][]{santurkar2023opinionslanguagemodelsreflect, boelaert_coavoux_ollion_petev_präg_2024}, and can be sensitive to prompt engineering on economic reasoning tasks \citep[][]{raman2024steerassessingeconomicrationality}. 
Moreover, estimation problems require no training leakage. Since published experiments likely enter LLM training corpora, the key question is whether LLMs can simulate behavior on entirely new experimental designs, not merely reproduce memorized results \citep[][]{manning2025generalsocialagents}.

Viewing human subject simulation as an estimation problem also implies a practical fix: collect responses from at least some real human subjects.
Consequently, in-silico subjects serve to amplify, rather than fully replace, human subjects.
We refer the reader to recent work, such as \citet[][]{BroskaEtAl(25), krsteski2025validsurveysimulationslimited, zhang2025agenticeconomicmodeling}, which provide further guidance on debiasing in-silico subjects.

\subsection{Hypothesis Generation}

Recent work uses LLMs to generate hypotheses across diverse applications: predicting user engagement from headlines \citep[][]{BatistaRoss(24), zhou2024hypothesisgenerationlargelanguage}, suggesting instrumental variables \citep[][]{Han(24)-IV}, proposing research ideas in natural language processing \citep[][]{si2024llmsgeneratenovelresearch}, and generating interpretable hypotheses that summarize estimated relationships between variables and text \citep[][]{modarressi2025causalinferenceoutcomeslearned, movva2025sparseautoencodershypothesisgeneration}.
We offer one interpretation viewing this as a type of prediction problem.

Researchers provide prompts $r$ containing one or more text pieces (e.g., a collection of headlines, a description of an empirical setting) and ask the LLM to generate a hypothesis $\widehat{m}(r; t) = \widehat{Y}_r$ that summarizes features or patterns in those texts (e.g., what drives engagement, what would be a valid instrument). 
The researcher evaluates average hypothesis quality $\frac{1}{N} \sum_{r \in \mathcal{R}} D_r \ell(\widehat{m}(r; t))$  -- though this scoring rule $\ell(\cdot)$ may be implicit or informal -- to determine whether the LLM is useful for hypothesis generation across different texts in the research context.

To assess whether the LLM is a useful tool for hypothesis generation in research context $Q(\cdot)$ -- that is, whether the quality of its hypotheses generalizes across different texts in that context -- we need no training leakage (Proposition \ref{proposition: prediction gpts iff no leakage}). 
Has the LLM seen this prompt or setting before? 
If so, it may reproduce memorized hypotheses rather than demonstrate new capability to generate insights on novel texts. 
For example, an LLM trained on text describing distance-to-college as an instrument for education returns might reproduce this strategy, leading us to overestimate its ability to identify instruments in genuinely novel settings.

Hypothesis generation is an exciting frontier.
Within economics, discussion about how machine learning and artificial intelligence are tools for hypothesis generation and scientific discovery can be found in \cite{fudenberg2019predicting}, \cite{ludwig2024machine}, \cite{agrawal2024artificial}, \cite{mullainathan2024predictive}, and \cite{mullainathan_rambachan_science}.
Further work is needed to clarify our inferential goals are in these settings -- what constitutes a ``good'' hypothesis, and how we should evaluate LLM performance in generating them.

\section{A Checklist for Empirical Research}\label{section: checklist for research}

To help use our framework in practice, we briefly summarize its key guidance for researchers incorporating LLM outputs in empirical work.

\vspace{-1em}
\paragraph{Identify Your Problem: Prediction or Estimation?} The first step is to identify your problem: are you facing a prediction problem or an estimation? 

In a prediction problem (Section \ref{section: prediction with large language models}), the researcher uses text pieces $r$ to predict some linked outcome $Y_r$ and evaluates the LLM's sample average loss $(1/N)\sum_{r\in R} D_r\ell(Y_r, \hat{m}(r;t))$. The researcher would like to understand whether this reflects the model's predictive performance. 

In an estimation problem (Section \ref{section: estimation with LLMs}), the researcher measures an economic concept expressed in text pieces $r$ to estimate downstream parameters. 
There is a measurement procedure $f^*(\cdot)$ that could be applied (e.g., the researcher reading each text piece themselves) that would produce the concept $V_r := f^*(r)$ on all text pieces and the researcher would, for example, run a regression of $V_r$ on some linked covariates $W_r$. 
Since this is costly at scale, the researcher uses LLM outputs $\hat{V}_r := \hat{m}(r;t)$ as a substitute.

\vspace{-1em}
\paragraph{For Prediction Problems, Ensure No Training Leakage} Valid conclusions in prediction problems require one condition: no training leakage between the LLM's training data and the researcher's dataset (Section \ref{section: training leakage as threat to prediction}).

Enforcing no training leakage is jointly determined by the prediction question and the model choice. Researchers must explicit about three key elements: (1) their target population -- what collection of text pieces do they ultimately want to make predictions on? (2) their sampling procedure---how do they select evaluation samples from this population?; and (3) how does this sampling procedure relate to plausible training corpora used by the LLM? 
In Section \ref{section: prediction problem guidance}, we illustrated how researchers can answer these questions in multiple common empirical settings, such as predicting on future documents, predicting on confidential documents, or constructing a random evaluation sample from a known corpus. 

When in doubt, the safest approach combines open-source models with published weights or documented training data and evaluation samples constructed to mechanically exclude training data. 
Researchers should avoid relying on closed models like the GPT from OpenAI or the Claude from Anthropic families, since their training data is undisclosed and potentially continuously updated. 
Finally, prompt engineering strategies (e.g., ``ignore information after date $\tau$'') are not reliable solutions to training leakage, as our evidence demonstrates.

\vspace{-1em}
\paragraph{For Estimation Problems, Collect Validation Data}
In estimation problems, plugging in LLM outputs requires the strong assumption that the LLM reproduces the existing measurement procedure---an assumption that is difficult to defend given evidence on LLM brittleness in computer science (see Section \ref{section: challenge of LLM evaluation} and Section \ref{section: measurement error as a threat}). 

The solution is straightforward: on a random sample of text pieces, apply the existing measurement procedure $f^*(\cdot)$ to collect some labels $V_r$. 
This validation sample allows the researcher to debias their downstream estimate for possible LLM errors. 
Section \ref{section: practical guidance with validation data} provides a pedagogical treatment for linear regression when the LLM label appears as the dependent variable. 
We found that this approach delivers approximately unbiased estimates and confidence intervals with good coverage in finite samples. 
Moreover, the debiased estimator can be more precisely estimated and yield tighter standard errors than using the validation sample alone. 
In this sense, LLM outputs amplify rather than replace existing measurements.

In our simulations, validation samples with as few as 125-250 text pieces provided substantial benefits in terms of bias and coverage. 
The optimal size of the validation sample depends on the trade-off between the cost of collecting measurements and the potential precision gains from expanding the validation sample. 
This should be approached as a design choice similar to determining sample size in surveys and experiments.
 
\section{Conclusion}
Machine learning and artificial intelligence expand the scope of empirical research in economics.
We now move beyond estimating average causal effects to learning personalized treatment effects \citep[e.g.,][]{athey2017econometrics, wager2018estimation}.
We use unstructured data, such as satellite images and digital traces, to infer outcomes at high-frequencies and granular scales \citep[e.g.,][]{DonaldsonStoreygard(16), BlumenstockEtAl(15), rambachan2024programevaluationremotelysensed}. 
We tackle prediction policy problems \citep[][]{kleinberg2015prediction, kleinberg2018human, Mullainathan(25)} and develop algorithms for hypothesis generation \citep[][]{fudenberg2019predicting,
ludwig2024machine, mullainathan2024predictive}. 
LLMs are the latest tools to enter empirical work. 

By radically reducing the cost of analyzing vast text corpora, LLMs enable economists to tackle questions previously impossible due to scale or expense. 
Using large language models, researchers can predict market reactions from earnings calls, measure sentiment across historical newspapers, track partisan polarization in social media, and simulate human responses at minimal cost. 
Yet these are complex algorithms that seemingly resist traditional econometric analysis.
Our framework shows how to harness LLMs despite their complexity. 
The straightforward practices we recommend unlock LLMs' transformative potential for empirical research.

\singlespacing 
\bibliographystyle{aea}
\bibliography{Bibliography}

@article{vafa2025has,
  title={What has a foundation model found? using inductive bias to probe for world models},
  author={Vafa, Keyon and Chang, Peter G and Rambachan, Ashesh and Mullainathan, Sendhil},
  journal={arXiv preprint arXiv:2507.06952},
  year={2025}
}

@techreport{horton2023large,
  title={Large language models as simulated economic agents: What can we learn from homo silicus?},
  author={Horton, John J},
  year={2023},
  institution={National Bureau of Economic Research}
}

@article{mancoridis2025potemkin,
  title={Potemkin Understanding in Large Language Models},
  author={Mancoridis, Marina and Weeks, Bec and Vafa, Keyon and Mullainathan, Sendhil},
  journal={arXiv preprint arXiv:2506.21521},
  year={2025}
}

@article{agrawal2024artificial,
  title={Artificial intelligence and scientific discovery: A model of prioritized search},
  author={Agrawal, Ajay and McHale, John and Oettl, Alexander},
  journal={Research Policy},
  volume={53},
  number={5},
  pages={104989},
  year={2024},
  publisher={Elsevier}
}

@inproceedings{aher2023usinglargelanguagemodels,
  author = {Aher, Gati and Arriaga, Rosa I. and Kalai, Adam Tauman},
  title = {Using large language models to simulate multiple humans and replicate human subject studies},
  year = {2023},
  publisher = {JMLR},
  booktitle = {Proceedings of the 40th International Conference on Machine Learning},
  articleno = {17},
  numpages = {35},
  location = {Honolulu, Hawaii, USA},
  series = {ICML'23}
}

@article{allais1953comportement,
  title={Le comportement de l'homme rationnel devant le risque: critique des postulats et axiomes de l'{\'e}cole am{\'e}ricaine},
  author={Allais, Maurice},
  journal={Econometrica: journal of the Econometric Society},
  pages={503--546},
  year={1953},
  publisher={JSTOR}
}

@article{Argyle_Busby_Fulda_Gubler_Rytting_Wingate_2023,
  title={Out of One, Many: Using Language Models to Simulate Human Samples}, 
  volume={31}, 
  DOI={10.1017/pan.2023.2}, 
  number={3}, 
  journal={Political Analysis}, 
  author={Argyle, Lisa P. and Busby, Ethan C. and Fulda, Nancy and Gubler, Joshua R. and Rytting, Christopher and Wingate, David}, 
  year={2023}, 
  pages={337--351}
}

@incollection{athey2017econometrics,
  title={The econometrics of randomized experiments},
  author={Athey, Susan and Imbens, Guido W.},
  booktitle={Handbook of economic field experiments},
  volume={1},
  pages={73--140},
  year={2017},
  publisher={Elsevier}
}

@inproceedings{balloccu2024leakcheatrepeatdata,
    title = "Leak, Cheat, Repeat: Data Contamination and Evaluation Malpractices in Closed-Source {LLM}s",
    author = "Balloccu, Simone  and Schmidtov{\'a}, Patr{\'\i}cia  and Lango, Mateusz  and Dusek, Ondrej",
    editor = "Graham, Yvette  and Purver, Matthew",
    booktitle = "Proceedings of the 18th Conference of the European Chapter of the Association for Computational Linguistics (Volume 1: Long Papers)",
    year = "2024",
    location = "St. Julian{'}s, Malta",
    publisher = "Association for Computational Linguistics",
    url = "https://aclanthology.org/2024.eacl-long.5",
    pages = "67--93",
}

@article{berglund2023reversal,
  title={The Reversal Curse: LLMs trained on "A is B" fail to learn "B is A"},
  author={Berglund, Lukas and Tong, Meg and Kaufmann, Max and Balesni, Mikita and Stickland, Asa Cooper and Korbak, Tomasz and Evans, Owain},
  journal={arXiv preprint arXiv:2309.12288},
  year={2023}
}

@misc{boelaert_coavoux_ollion_petev_präg_2024,
  title={How do Generative Language Models Answer Opinion Polls?},
  author={Boelaert, Julien and Coavoux, Samuel and Ollion, Etienne and Petev, Ivaylo D and Präg, Patrick},
  year={2024},
  note={\nolinkurl{https://osf.io/preprints/socarxiv/r2pnb}},
}

@article{chen2024unleashingpotentialpromptengineering,
  title={Unleashing the potential of prompt engineering in Large Language Models: a comprehensive review}, 
  author={Banghao Chen and Zhaofeng Zhang and Nicolas Langrené and Shengxin Zhu},
  year={2024},
  journal={arXiv preprint arXiv:2310.14735},
  url={https://arxiv.org/abs/2310.14735}, 
}

@article{cheng2024dateddatatracingknowledge,
  title={Dated Data: Tracing Knowledge Cutoffs in Large Language Models}, 
  author={Jeffrey Cheng and Marc Marone and Orion Weller and Dawn Lawrie and Daniel Khashabi and Benjamin Van Durme},
  year={2024},
  journal={arXiv preprint arXiv:2403.12958},
  url={https://arxiv.org/abs/2403.12958}, 
}

@article{chen2005measurement,
  title={Measurement error models with auxiliary data},
  author={Chen, Xiaohong and Hong, Han and Tamer, Elie},
  journal={The Review of Economic Studies},
  volume={72},
  number={2},
  pages={343--366},
  year={2005},
  publisher={Wiley-Blackwell}
}

@article{cui2024aireplacehumansubjects,
  title={Can AI Replace Human Subjects? A Large-Scale Replication of Psychological Experiments with LLMs}, 
  author={Ziyan Cui and Ning Li and Huaikang Zhou},
  year={2024},
  journal={arXiv preprint arXiv:2409.00128}, 
  url={https://arxiv.org/abs/2409.00128}, 
}

@article{dell2024deeplearningeconomists,
  title={Deep Learning for Economists}, 
  author={Melissa Dell},
  year={2024},
  journal={arXiv preprint arXiv:2407.15339}, 
  url={https://arxiv.org/abs/2407.15339}, 
}

@article{Donoho2024Data,
  author = {Donoho, David},
  journal = {Harvard Data Science Review},
  number = {1},
  year = {2024},
  month = {jan 29},
  url = {https://hdsr.mitpress.mit.edu/pub/g9mau4m0},
  publisher = {The MIT Press},
  title = {{Data} {Science} at the {Singularity}},
  volume = {6},
}

@article{dubey2024llama3herdmodels,
  title={The Llama 3 Herd of Models}, 
  author={Abhimanyu Dubey and Abhinav Jauhri and Abhinav Pandey and Abhishek Kadian and Ahmad Al-Dahle and Aiesha Letman and Akhil Mathur and others},
  year={2024},
  journal={arXiv preprint arXiv:2407.21783},
  url={https://arxiv.org/abs/2407.21783}, 
}

@article{fudenberg2019predicting,
  title={Predicting and understanding initial play},
  author={Fudenberg, Drew and Liang, Annie},
  journal={American Economic Review},
  volume={109},
  number={12},
  pages={4112--4141},
  year={2019},
  publisher={American Economic Association 2014 Broadway, Suite 305, Nashville, TN 37203}
}

@article{glasserman2023assessinglookaheadbiasstock,
  title={Assessing Look-Ahead Bias in Stock Return Predictions Generated By GPT Sentiment Analysis}, 
  author={Paul Glasserman and Caden Lin},
  year={2023},
  journal={arXiv preprint arXiv:2309.17322}, 
  url={https://arxiv.org/abs/2309.17322}, 
}

@article{golchin2024timetravelllmstracing,
  title={Time Travel in LLMs: Tracing Data Contamination in Large Language Models}, 
  author={Shahriar Golchin and Mihai Surdeanu},
  year={2024},
  journal={arXiv preprint arXiv:2308.08493}, 
  url={https://arxiv.org/abs/2308.08493}, 
}

@article{hendrycks2020measuring,
  title={Measuring massive multitask language understanding},
  author={Hendrycks, Dan and Burns, Collin and Basart, Steven and Zou, Andy and Mazeika, Mantas and Song, Dawn and Steinhardt, Jacob},
  journal={arXiv preprint arXiv:2009.03300},
  year={2020}
}

@inproceedings{jacovi-etal-2023-stop,
  title = "Stop Uploading Test Data in Plain Text: Practical Strategies for Mitigating Data Contamination by Evaluation Benchmarks",
  author = "Jacovi, Alon  and
    Caciularu, Avi  and
    Goldman, Omer  and
    Goldberg, Yoav",
  editor = "Bouamor, Houda  and
    Pino, Juan  and
    Bali, Kalika",
  booktitle = "Proceedings of the 2023 Conference on Empirical Methods in Natural Language Processing",
  month = dec,
  year = "2023",
  location = "Singapore",
  publisher = "Association for Computational Linguistics",
  url = "https://aclanthology.org/2023.emnlp-main.308",
  doi = "10.18653/v1/2023.emnlp-main.308",
  pages = "5075--5084"
}

@article{kahneman1979prospect,
  title={Prospect theory: An analysis of decision under risk},
  author={Kahneman, Daniel and Tversky, Amos},
  journal={Econometrica},
  volume={47},
  number={2},
  pages={363--391},
  year={1979}
}

@article{kleinberg2018human,
  title={Human decisions and machine predictions},
  author={Kleinberg, Jon and Lakkaraju, Himabindu and Leskovec, Jure and Ludwig, Jens and Mullainathan, Sendhil},
  journal={The quarterly journal of economics},
  volume={133},
  number={1},
  pages={237--293},
  year={2018},
  publisher={Oxford University Press}
}

@article{kleinberg2015prediction,
  title={Prediction policy problems},
  author={Kleinberg, Jon and Ludwig, Jens and Mullainathan, Sendhil and Obermeyer, Ziad},
  journal={American Economic Review},
  volume={105},
  number={5},
  pages={491--495},
  year={2015},
  publisher={American Economic Association 2014 Broadway, Suite 305, Nashville, TN 37203}
}

@article{li_general_2017,
  title = {General {{Forms}} of {{Finite Population Central Limit Theorems}} with {{Applications}} to {{Causal Inference}}},
  author = {Li, Xinran and Ding, Peng},
  year = {2017},
  journal = {Journal of the American Statistical Association},
  volume = {112},
  number = {520},
  pages = {1759--1769},
  issn = {0162-1459},
  doi = {10.1080/01621459.2017.1295865},
  url = {https://doi.org/10.1080/01621459.2017.1295865}
}

@article{lopezlira2024chatgptforecaststockprice,
  title={Can ChatGPT Forecast Stock Price Movements? Return Predictability and Large Language Models}, 
  author={Alejandro Lopez-Lira and Yuehua Tang},
  year={2024},
  journal={arXiv preprint arXiv:2304.07619},
  url={https://arxiv.org/abs/2304.07619}, 
}

@article{ludwig2024machine,
  title={Machine learning as a tool for hypothesis generation},
  author={Ludwig, Jens and Mullainathan, Sendhil},
  journal={The Quarterly Journal of Economics},
  volume={139},
  number={2},
  pages={751--827},
  year={2024},
  publisher={Oxford University Press}
}

@article{manning2024automatedsocialsciencelanguage,
  title={Automated Social Science: Language Models as Scientist and Subjects}, 
  author={Benjamin S. Manning and Kehang Zhu and John J. Horton},
  year={2024},
  journal={arXiv preprint arXiv:2404.11794},
  url={https://arxiv.org/abs/2404.11794}, 
}

@article{mccoy2023embersautoregressionunderstandinglarge,
  author = {R. Thomas McCoy  and Shunyu Yao  and Dan Friedman  and Mathew D. Hardy  and Thomas L. Griffiths },
  title = {Embers of Autoregression: Understanding Large Language Models Through the Problem They are Trained to Solve},
  journal = {Proceedings of the National Academy of Sciences},
  volume = {121},
  number = {41},
  pages = {e2322420121},
  year = {2024},
  doi = {10.1073/pnas.2322420121},
  URL = {https://www.pnas.org/doi/abs/10.1073/pnas.2322420121},
  eprint = {https://www.pnas.org/doi/pdf/10.1073/pnas.2322420121},
}

@techreport{mullainathan2024predictive,
  title = "From predictive algorithms to automatic generation of anomalies",
  author = "Mullainathan, Sendhil and Rambachan, Ashesh",
  year = "2024",
  institution = "National Bureau of Economic Research",
  type = "Working Paper",
  series = "Working Paper Series",
  number = "32422",
  doi = {10.3386/w32422},
  URL = "http://www.nber.org/papers/w32422"
}

@article{nezhurina2024alicewonderlandsimpletasks,
  title={Alice in Wonderland: Simple Tasks Showing Complete Reasoning Breakdown in State-Of-the-Art Large Language Models}, 
  author={Marianna Nezhurina and Lucia Cipolina-Kun and Mehdi Cherti and Jenia Jitsev},
  year={2024},
  journal={arXiv preprint arXiv:2406.02061},
  url={https://arxiv.org/abs/2406.02061}, 
}

@article{raman2024steerassessingeconomicrationality,
  title={STEER: Assessing the Economic Rationality of Large Language Models}, 
  author={Narun Raman and Taylor Lundy and Samuel Amouyal and Yoav Levine and Kevin Leyton-Brown and Moshe Tennenholtz},
  year={2024},
  journal={arXiv preprint arXiv:2402.09552},
  url={https://arxiv.org/abs/2402.09552}, 
}

@article{rambachan2024programevaluationremotelysensed,
  title={Program Evaluation with Remotely Sensed Outcomes}, 
  author={Ashesh Rambachan and Rahul Singh and Davide Viviano},
  year={2024},
  journal={arXiv preprint arXiv:2411.10959},
  url={https://arxiv.org/abs/2411.10959}, 
}

@article{ravaut2024largelanguagemodelscontaminated,
  title={How Much are Large Language Models Contaminated? A Comprehensive Survey and the LLMSanitize Library}, 
  author={Mathieu Ravaut and Bosheng Ding and Fangkai Jiao and Hailin Chen and Xingxuan Li and Ruochen Zhao and Chengwei Qin and Caiming Xiong and Shafiq Joty},
  year={2024},
  journal={arXiv preprint arXiv:2404.00699},
  url={https://arxiv.org/abs/2404.00699}, 
}

@inproceedings{sainz2023nlpevaluationtroubleneed,
  title = "{NLP} Evaluation in trouble: On the Need to Measure {LLM} Data Contamination for each Benchmark",
  author = "Sainz, Oscar  and
    Campos, Jon  and
    Garc{\'\i}a-Ferrero, Iker  and
    Etxaniz, Julen  and
    de Lacalle, Oier Lopez  and
    Agirre, Eneko",
  editor = "Bouamor, Houda  and
    Pino, Juan  and
    Bali, Kalika",
  booktitle = "Findings of the Association for Computational Linguistics: EMNLP 2023",
  month = dec,
  year = "2023",
  location = "Singapore",
  publisher = "Association for Computational Linguistics",
  url = "https://aclanthology.org/2023.findings-emnlp.722",
  doi = "10.18653/v1/2023.findings-emnlp.722",
  pages = "10776--10787",
}

@inproceedings{santurkar2023opinionslanguagemodelsreflect,
author = {Santurkar, Shibani and Durmus, Esin and Ladhak, Faisal and Lee, Cinoo and Liang, Percy and Hashimoto, Tatsunori},
title = {Whose opinions do language models reflect?},
year = {2023},
publisher = {JMLR},
booktitle = {Proceedings of the 40th International Conference on Machine Learning},
articleno = {1244},
numpages = {34},
location = {Honolulu, Hawaii, USA},
series = {ICML'23}
}

@article{sarkar2024lookahead,
  title={Lookahead bias in pretrained language models},
  author={Sarkar, Suproteem K and Vafa, Keyon},
  journal={Available at SSRN},
  year={2024}
}

@article{schennach2016recent,
  title={Recent advances in the measurement error literature},
  author={Schennach, Susanne M},
  journal={Annual Review of Economics},
  volume={8},
  number={1},
  pages={341--377},
  year={2016},
  publisher={Annual Reviews}
}

@article{si2024llmsgeneratenovelresearch,
  title={Can LLMs Generate Novel Research Ideas? A Large-Scale Human Study with 100+ NLP Researchers}, 
  author={Chenglei Si and Diyi Yang and Tatsunori Hashimoto},
  year={2024},
  journal={arXiv preprint arXiv:2409.04109},
  url={https://arxiv.org/abs/2409.04109}, 
}

@article{srivastava2022beyond,
  title={Beyond the imitation game: Quantifying and extrapolating the capabilities of language models},
  author={Srivastava, Aarohi and Rastogi, Abhinav and Rao, Abhishek and Shoeb, Abu Awal Md and Abid, Abubakar and Fisch, Adam and Brown, Adam R and others},
  journal={arXiv preprint arXiv:2206.04615},
  year={2022}
}

@article{touvron2023llama2openfoundation,
  title={Llama 2: Open Foundation and Fine-Tuned Chat Models}, 
  author={Hugo Touvron and Louis Martin and Kevin Stone and Peter Albert and Amjad Almahairi and Yasmine Babaei and Nikolay Bashlykov and others},
  year={2023},
  journal={arXiv preprint arXiv:2307.09288},
  url={https://arxiv.org/abs/2307.09288}, 
}

@article{vafa2024humangeneralization,
  title={Do Large Language Models Generalize the Way People Expect? A Benchmark for Evaluation},
  author={Vafa, Keyon and Rambachan, Ashesh and Mullainathan, Sendhil},
  journal={arXiv preprint arXiv:2406.01382},
  year={2024}
}

@article{wager2018estimation,
  title={Estimation and inference of heterogeneous treatment effects using random forests},
  author={Wager, Stefan and Athey, Susan},
  journal={Journal of the American Statistical Association},
  volume={113},
  number={523},
  pages={1228--1242},
  year={2018},
  publisher={Taylor \& Francis}
}

@inproceedings{wei2023chainofthoughtpromptingelicitsreasoning,
author = {Wei, Jason and Wang, Xuezhi and Schuurmans, Dale and Bosma, Maarten and Ichter, Brian and Xia, Fei and Chi, Ed H. and Le, Quoc V. and Zhou, Denny},
title = {Chain-of-Thought Prompting Elicits Reasoning in Large Language Models},
year = {2024},
isbn = {9781713871088},
publisher = {Curran Associates Inc.},
booktitle = {Proceedings of the 36th International Conference on Neural Information Processing Systems},
articleno = {1800},
numpages = {14},
series = {NIPS '22}
}

@article{white2023promptpatterncatalogenhance,
  title={A Prompt Pattern Catalog to Enhance Prompt Engineering with ChatGPT}, 
  author={Jules White and Quchen Fu and Sam Hays and Michael Sandborn and Carlos Olea and Henry Gilbert and Ashraf Elnashar and Jesse Spencer-Smith and Douglas C. Schmidt},
  year={2023},
  journal={arXiv preprint arXiv:2302.11382},
  url={https://arxiv.org/abs/2302.11382}, 
}

@inproceedings{wu2024reasoningrecitingexploringcapabilities,
  title = "Reasoning or Reciting? Exploring the Capabilities and Limitations of Language Models Through Counterfactual Tasks",
  author = {Wu, Zhaofeng  and Qiu, Linlu  and Ross, Alexis  and Aky{\"u}rek, Ekin  and Chen, Boyuan  and Wang, Bailin  and Kim, Najoung  and Andreas, Jacob  and Kim, Yoon},
  editor = "Duh, Kevin  and Gomez, Helena  and Bethard, Steven",
  booktitle = "Proceedings of the 2024 Conference of the North American Chapter of the Association for Computational Linguistics: Human Language Technologies (Volume 1: Long Papers)",
  year = "2024",
  location = "Mexico City, Mexico",
  publisher = "Association for Computational Linguistics",
  url = "https://aclanthology.org/2024.naacl-long.102",
  doi = "10.18653/v1/2024.naacl-long.102",
  pages = "1819--1862",
}

@article{zhou2024hypothesisgenerationlargelanguage,
  title={Hypothesis Generation with Large Language Models}, 
  author={Yangqiaoyu Zhou and Haokun Liu and Tejes Srivastava and Hongyuan Mei and Chenhao Tan},
  year={2024},
  journal={arXiv preprint arXiv:2404.04326},
  url={https://arxiv.org/abs/2404.04326}, 
}

@InProceedings{zong2024foolvisionandlanguage,
  title = 	 {Fool Your ({V}ision and) Language Model with Embarrassingly Simple Permutations},
  author =       {Zong, Yongshuo and Yu, Tingyang and Chavhan, Ruchika and Zhao, Bingchen and Hospedales, Timothy},
  booktitle = 	 {Proceedings of the 41st International Conference on Machine Learning},
  pages = 	 {62892--62913},
  year = 	 {2024},
  editor = 	 {Salakhutdinov, Ruslan and Kolter, Zico and Heller, Katherine and Weller, Adrian and Oliver, Nuria and Scarlett, Jonathan and Berkenkamp, Felix},
  volume = 	 {235},
  series = 	 {Proceedings of Machine Learning Research},
  publisher =    {PMLR},
  url = 	 {https://proceedings.mlr.press/v235/zong24b.html},
}

@misc{lm_index,
  author={{LM Contamination Index}},
  year={2024},
  note={\url{https://hitz-zentroa.github.io/lm-contamination} (accessed December 5, 2024)}
}

@misc{CAPCodebook,
  author = {Jones, Bryan D. and Baumgartner, Frank R. and Theriault, Sean M. and Epp, Derek A. and Lee, Cheyenne and Sullivan, Miranda E.},
  year = {2023},
  title = {Policy Agendas Project: Codebook}
}

@book{CBP,
  author = {Adler, E Scott and Wilkerson, John},
  year = {2020},
  title = {Congressional Bills Project, NSF 00880066 and 00880061},
  publisher = {\url{http://congressionalbills.org/download.html} (accessed July 5, 2024)}
}

@book{Headlines,
  author       = {Aenlle, Miguel},
  year         = {2020},
  title        = {Daily Financial News for 6000+ Stocks},
  publisher    = {\url{https://www.kaggle.com/datasets/miguelaenlle/massive-stock-news-analysis-db-for-nlpbacktests} (accessed August 1, 2024)}
}

@book{WRDSBetaSuite,
  author       = {{Beta Suite by WRDS}},
  year         = {2024},
  publisher    = {Provided by Wharton Research Data Services. \url{https://wrds-www.wharton.upenn.edu/pages/grid-items/beta-suite-wrds} (accessed August 1, 2024)}
}

@book{CAP,
  author = {Wilkerson, John and Adler, E. Scott and Jones, Bryan D. and Baumgartner, Frank R. and Freedman, Guy and Theriault, Sean M. and Craig, Alison and Epp, Derek A. and Lee, Cheyenne and Sullivan, Miranda E.},
  year = {2023},
  title = {Policy Agendas Project: Congressional Bills},
  publisher = {\url{https://minio.la.utexas.edu/compagendas/datasetfiles/US-Legislative-congressional_bills_19.3_3_3.csv} (accessed July 5, 2024)}
}

@article{bound1991extent,
  title={The extent of measurement error in longitudinal earnings data: Do two wrongs make a right?},
  author={Bound, John and Krueger, Alan B},
  journal={Journal of labor economics},
  volume={9},
  number={1},
  pages={1--24},
  year={1991},
  publisher={University of Chicago Press}
}

@article{vafa2024worldmodel,
  title={Evaluating the World Model Implicit in a Generative Model},
  author={Vafa, Keyon and Chen, Justin Y. and Kleinberg, Jon and Mullainathan, Sendhil and Rambachan, Ashesh},
  journal={Harvard Working Paper},
  year={2024}
}

@misc{kim2025correlatederrorslargelanguage,
      title={Correlated Errors in Large Language Models}, 
      author={Elliot Kim and Avi Garg and Kenny Peng and Nikhil Garg},
      year={2025},
      eprint={2506.07962},
      archivePrefix={arXiv},
      primaryClass={cs.CL},
      url={https://arxiv.org/abs/2506.07962}, 
}

@misc{park2023generativeagentsinteractivesimulacra,
      title={Generative Agents: Interactive Simulacra of Human Behavior}, 
      author={Joon Sung Park and Joseph C. O'Brien and Carrie J. Cai and Meredith Ringel Morris and Percy Liang and Michael S. Bernstein},
      year={2023},
      eprint={2304.03442},
      archivePrefix={arXiv},
      primaryClass={cs.HC},
      url={https://arxiv.org/abs/2304.03442}, 
}

@misc{park2024generativeagentsimulations1000,
      title={Generative Agent Simulations of 1,000 People}, 
      author={Joon Sung Park and Carolyn Q. Zou and Aaron Shaw and Benjamin Mako Hill and Carrie Cai and Meredith Ringel Morris and Robb Willer and Percy Liang and Michael S. Bernstein},
      year={2024},
      eprint={2411.10109},
      archivePrefix={arXiv},
      primaryClass={cs.AI},
      url={https://arxiv.org/abs/2411.10109}, 
}

@misc{lopezlira2025memorizationproblemtrustllms,
      title={The Memorization Problem: Can We Trust LLMs' Economic Forecasts?}, 
      author={Alejandro Lopez-Lira and Yuehua Tang and Mingyin Zhu},
      year={2025},
      eprint={2504.14765},
      archivePrefix={arXiv},
      primaryClass={q-fin.GN},
      url={https://arxiv.org/abs/2504.14765}, 
}

@misc{krsteski2025validsurveysimulationslimited,
      title={Valid Survey Simulations with Limited Human Data: The Roles of Prompting, Fine-Tuning, and Rectification}, 
      author={Stefan Krsteski and Giuseppe Russo and Serina Chang and Robert West and Kristina Gligorić},
      year={2025},
      eprint={2510.11408},
      archivePrefix={arXiv},
      primaryClass={cs.CL},
      url={https://arxiv.org/abs/2510.11408}, 
}

@misc{zhang2025agenticeconomicmodeling,
      title={Agentic Economic Modeling}, 
      author={Bohan Zhang and Jiaxuan Li and Ali Hortaçsu and Xiaoyang Ye and Victor Chernozhukov and Angelo Ni and Edward Huang},
      year={2025},
      eprint={2510.25743},
      archivePrefix={arXiv},
      primaryClass={econ.EM},
      url={https://arxiv.org/abs/2510.25743}, 
}

@misc{modarressi2025causalinferenceoutcomeslearned,
      title={Causal Inference on Outcomes Learned from Text}, 
      author={Iman Modarressi and Jann Spiess and Amar Venugopal},
      year={2025},
      eprint={2503.00725},
      archivePrefix={arXiv},
      primaryClass={econ.EM},
      url={https://arxiv.org/abs/2503.00725}, 
}

@misc{movva2025sparseautoencodershypothesisgeneration,
      title={Sparse Autoencoders for Hypothesis Generation}, 
      author={Rajiv Movva and Kenny Peng and Nikhil Garg and Jon Kleinberg and Emma Pierson},
      year={2025},
      eprint={2502.04382},
      archivePrefix={arXiv},
      primaryClass={cs.CL},
      url={https://arxiv.org/abs/2502.04382}, 
}

@misc{jimenez2024swebenchlanguagemodelsresolve,
      title={SWE-bench: Can Language Models Resolve Real-World GitHub Issues?}, 
      author={Carlos E. Jimenez and John Yang and Alexander Wettig and Shunyu Yao and Kexin Pei and Ofir Press and Karthik Narasimhan},
      year={2024},
      eprint={2310.06770},
      archivePrefix={arXiv},
      primaryClass={cs.CL},
      url={https://arxiv.org/abs/2310.06770}, 
}

@misc{cobbe2021trainingverifierssolvemath,
      title={Training Verifiers to Solve Math Word Problems}, 
      author={Karl Cobbe and Vineet Kosaraju and Mohammad Bavarian and Mark Chen and Heewoo Jun and Lukasz Kaiser and Matthias Plappert and Jerry Tworek and Jacob Hilton and Reiichiro Nakano and Christopher Hesse and John Schulman},
      year={2021},
      eprint={2110.14168},
      archivePrefix={arXiv},
      primaryClass={cs.LG},
      url={https://arxiv.org/abs/2110.14168}, 
}

@misc{carlson2025unifyingframeworkrobustefficient,
      title={A Unifying Framework for Robust and Efficient Inference with Unstructured Data}, 
      author={Jacob Carlson and Melissa Dell},
      year={2025},
      eprint={2505.00282},
      archivePrefix={arXiv},
      primaryClass={econ.EM},
      url={https://arxiv.org/abs/2505.00282}, 
}

@misc{he2025chronologicallyconsistentlargelanguage,
      title={Chronologically Consistent Large Language Models}, 
      author={Songrun He and Linying Lv and Asaf Manela and Jimmy Wu},
      year={2025},
      eprint={2502.21206},
      archivePrefix={arXiv},
      primaryClass={q-fin.GN},
      url={https://arxiv.org/abs/2502.21206}, 
}

@techreport{ash2024llm,
  title        = {Large Language Models in Economics},
  author       = {Ash, Elliott and Hansen, Stephen and Muvdi, Yony},
  year         = {2024},
  month        = {September},
  institution  = {Centre for Economic Policy Research},
  type         = {CEPR Discussion Paper},
  number       = {19479},
  address      = {London},
  url          = {https://cepr.org/publications/dp19479}
}

@misc{minaee2025largelanguagemodelssurvey,
      title={Large Language Models: A Survey}, 
      author={Shervin Minaee and Tomas Mikolov and Narjes Nikzad and Meysam Chenaghlu and Richard Socher and Xavier Amatriain and Jianfeng Gao},
      year={2025},
      eprint={2402.06196},
      archivePrefix={arXiv},
      primaryClass={cs.CL},
      url={https://arxiv.org/abs/2402.06196}, 
}

@misc{zhao2025surveylargelanguagemodels,
      title={A Survey of Large Language Models}, 
      author={Wayne Xin Zhao and Kun Zhou and Junyi Li and Tianyi Tang and Xiaolei Wang and Yupeng Hou and Yingqian Min and Beichen Zhang and Junjie Zhang and Zican Dong and Yifan Du and Chen Yang and Yushuo Chen and Zhipeng Chen and Jinhao Jiang and Ruiyang Ren and Yifan Li and Xinyu Tang and Zikang Liu and Peiyu Liu and Jian-Yun Nie and Ji-Rong Wen},
      year={2025},
      eprint={2303.18223},
      archivePrefix={arXiv},
      primaryClass={cs.CL},
      url={https://arxiv.org/abs/2303.18223}, 
}

@article{chang2024_llmevals,
author = {Chang, Yupeng and Wang, Xu and Wang, Jindong and Wu, Yuan and Yang, Linyi and Zhu, Kaijie and Chen, Hao and Yi, Xiaoyuan and Wang, Cunxiang and Wang, Yidong and Ye, Wei and Zhang, Yue and Chang, Yi and Yu, Philip S. and Yang, Qiang and Xie, Xing},
title = {A Survey on Evaluation of Large Language Models},
year = {2024},
issue_date = {June 2024},
publisher = {Association for Computing Machinery},
address = {New York, NY, USA},
volume = {15},
number = {3},
issn = {2157-6904},
url = {https://doi.org/10.1145/3641289},
doi = {10.1145/3641289},
journal = {ACM Trans. Intell. Syst. Technol.},
month = mar,
articleno = {39},
numpages = {45}
}

@misc{manning2025generalsocialagents,
      title={General Social Agents}, 
      author={Benjamin S. Manning and John J. Horton},
      year={2025},
      eprint={2508.17407},
      archivePrefix={arXiv},
      primaryClass={econ.GN},
      url={https://arxiv.org/abs/2508.17407}, 
}

@incollection{mullainathan_rambachan_science,
  author    = {Mullainathan, Sendhil and Rambachan, Ashesh},
  title     = {Science in the Age of Algorithms},
  booktitle = {The Economics of Transformative AI},
  publisher = {University of Chicago Press},
  year      = {2025},
  chapter   = {13},
  url       = {https://www.nber.org/books-and-chapters/economics-transformative-ai/science-age-algorithms}
}

@misc{wongchamcharoen2025largelanguagemodelsllms,
      title={Do Large Language Models (LLMs) Understand Chronology?}, 
      author={Pattaraphon Kenny Wongchamcharoen and Paul Glasserman},
      year={2025},
      eprint={2511.14214},
      archivePrefix={arXiv},
      primaryClass={cs.AI},
      url={https://arxiv.org/abs/2511.14214}, 
}

@misc{li2024emergentworldrepresentationsexploring,
      title={Emergent World Representations: Exploring a Sequence Model Trained on a Synthetic Task}, 
      author={Kenneth Li and Aspen K. Hopkins and David Bau and Fernanda Viégas and Hanspeter Pfister and Martin Wattenberg},
      year={2024},
      eprint={2210.13382},
      archivePrefix={arXiv},
      primaryClass={cs.LG},
      url={https://arxiv.org/abs/2210.13382}, 
}

@misc{nanda2023emergentlinearrepresentationsworld,
      title={Emergent Linear Representations in World Models of Self-Supervised Sequence Models}, 
      author={Neel Nanda and Andrew Lee and Martin Wattenberg},
      year={2023},
      eprint={2309.00941},
      archivePrefix={arXiv},
      primaryClass={cs.LG},
      url={https://arxiv.org/abs/2309.00941}, 
}

@misc{nikankin2025arithmeticalgorithmslanguagemodels,
      title={Arithmetic Without Algorithms: Language Models Solve Math With a Bag of Heuristics}, 
      author={Yaniv Nikankin and Anja Reusch and Aaron Mueller and Yonatan Belinkov},
      year={2025},
      eprint={2410.21272},
      archivePrefix={arXiv},
      primaryClass={cs.CL},
      url={https://arxiv.org/abs/2410.21272}, 
}

@misc{jylin04_2024_othellogpt,
  author = {jylin04 and JackS and Adam Karvonen and Can},
  title = {{OthelloGPT} Learned a Bag of Heuristics},
  year = {2024},
  month = {July},
  howpublished = {LessWrong},
  url = {https://www.lesswrong.com/posts/gcpNuEZnxAPayaKBY/othellogpt-learned-a-bag-of-heuristics-1},
  urldate = {2024-12-01}
}

\clearpage \newpage
\section*{Main Figures and Tables}
\begin{figure}[htb!]
\includegraphics[width=0.77\textwidth]{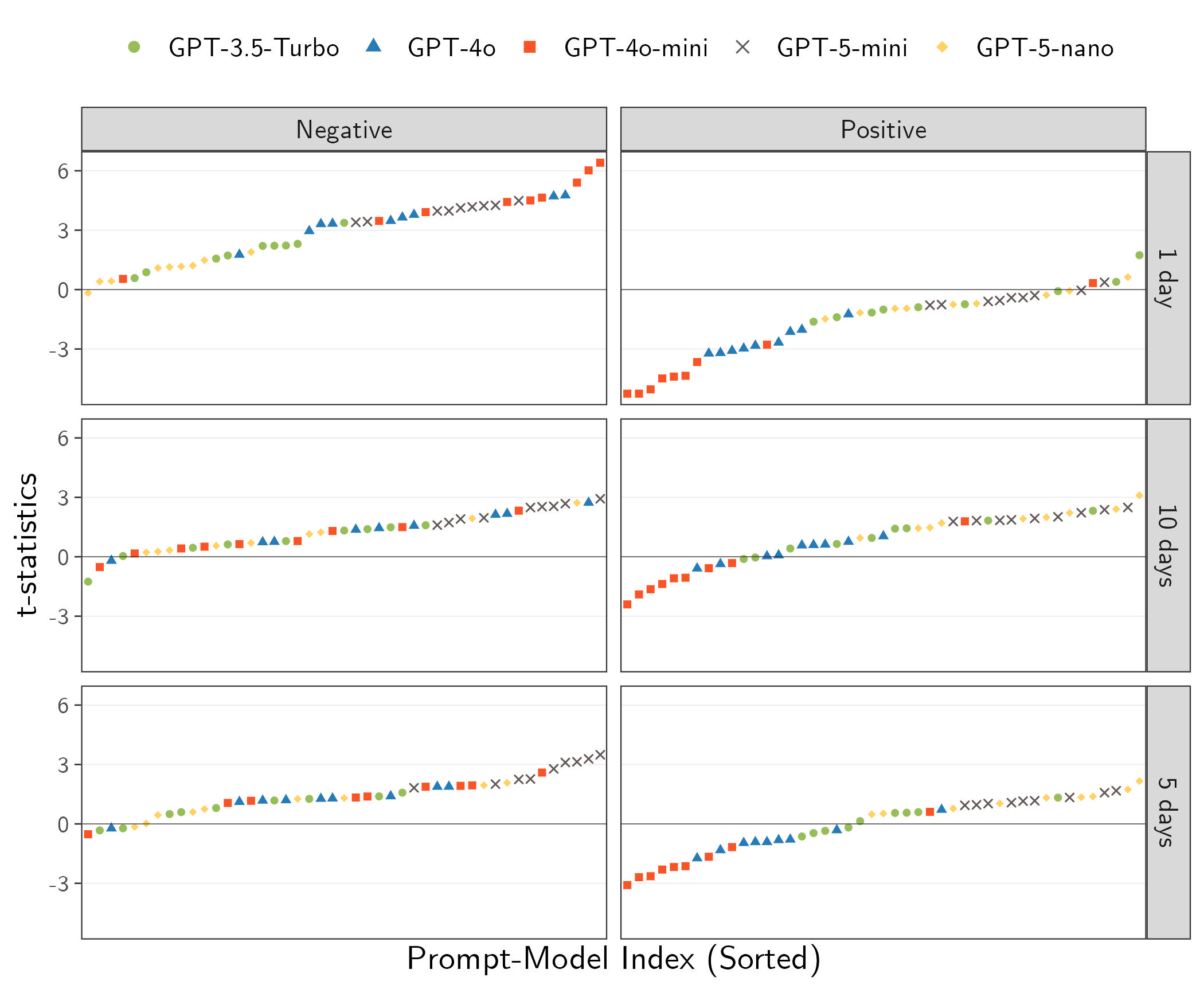}
\caption{Variation in t-statistics for realized returns across large language models and prompting strategies on financial news headlines.}
\floatfoot{\textit{Notes}: On financial news headlines from 2019, we prompt GPT-3.5-Turbo, GPT-4o-mini, GPT-4o, GPT-5-mini, and GPT-5-nano to label each headline for whether it expressed positive, negative or uncertain news about the respective company using alternative prompting strategies. 
For each model $m$ and prompt $p$, we regress the realized returns of each stock within 1 day, 5 days or 10 days of the headline's publication date on each large language model's labels $\widehat{V}_{r}^{m,p}$, the large language model's assessed magnitude denoted $S_{r}^{m,p}$ and their interaction, controlling for lagged realized returns. 
We separately report the t-statistics associated with the regression coefficients on whether the headline is labeled as positive or negative news (standard errors are two-way clustered at the date and company level). 
In each subplot, the t-statistics are sorted in ascending order for clarity.
See Section \ref{section: evidence of measurement error} for discussion.
}
\label{figure: financial news headline, positive/negative news, t-stat, magnitude, 1/5/10 day realized}
\end{figure}

\begin{figure}[htbp!]
\includegraphics[width=\textwidth]{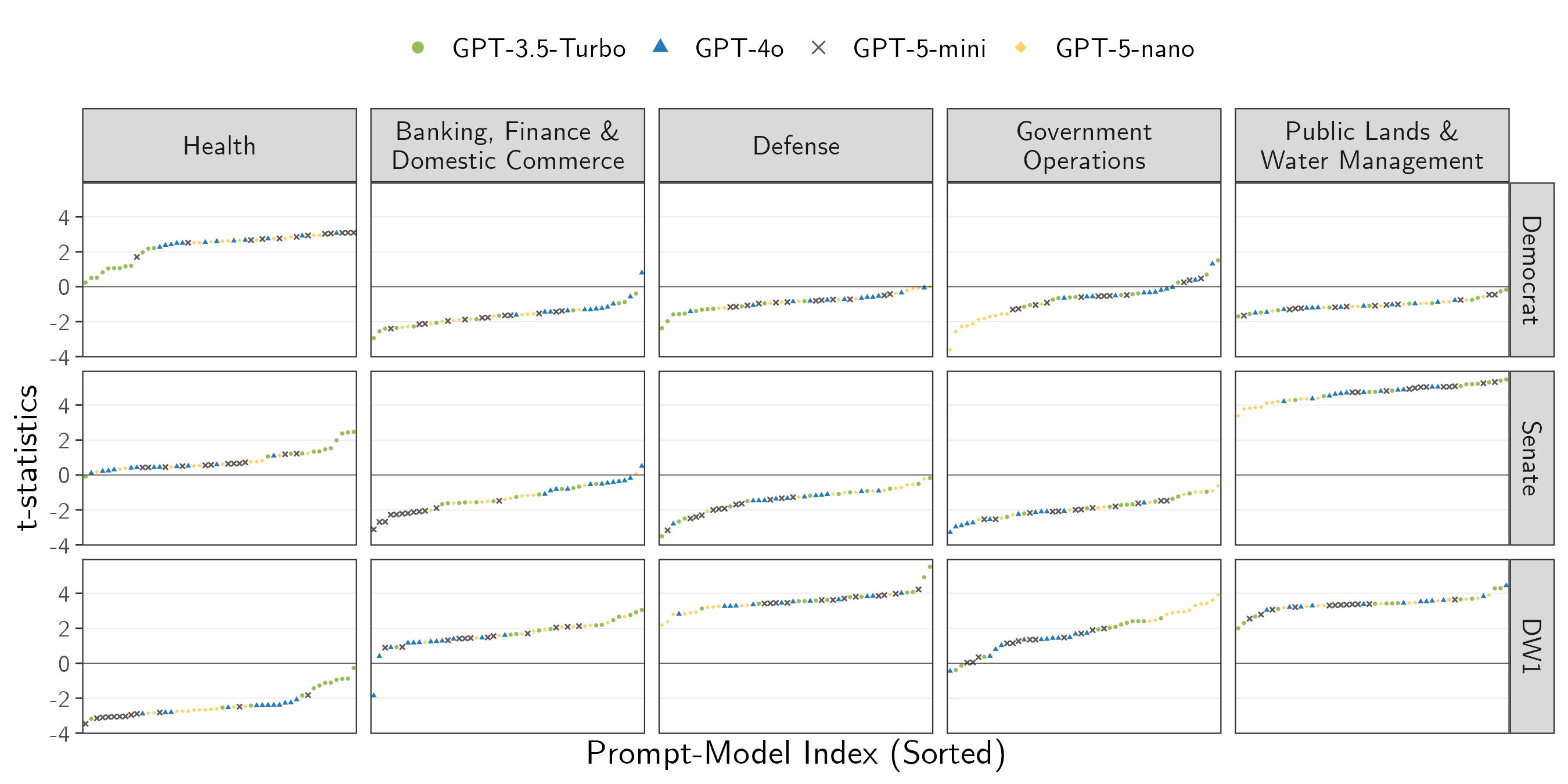}
\caption{Variation in t-statistics across large language models and prompting strategies on congressional legislation.}
\floatfoot{\textit{Notes}: 
On 10,000 Congressional bills, we prompt GPT-3.5-Turbo, GPT-4o, GPT-5-mini, and GPT-5-nano to label each description for its policy topic area using alternative prompting strategies. 
For each model $m$ and prompt $p$, we regress $\widehat{V}_r^{m,p}$ on the linked covariate $W_r$, where $\widehat{V}_{r}^{m,p}$ are indicators for the policy topic of the bill and the covariates $W_r$ are whether the bill's sponsor was a Democrat, whether the bill originated in the Senate, and the DW1 score of the bill's sponsor.
In each subplot, the t-statistics were sorted in ascending order for clarity.
See Section \ref{section: evidence of measurement error} for discussion.
}
\label{figure: congressional bills, LLM on LHS, full data, t stat figure}
\end{figure}

\begin{figure}[htbp!]
\includegraphics[width=\textwidth]{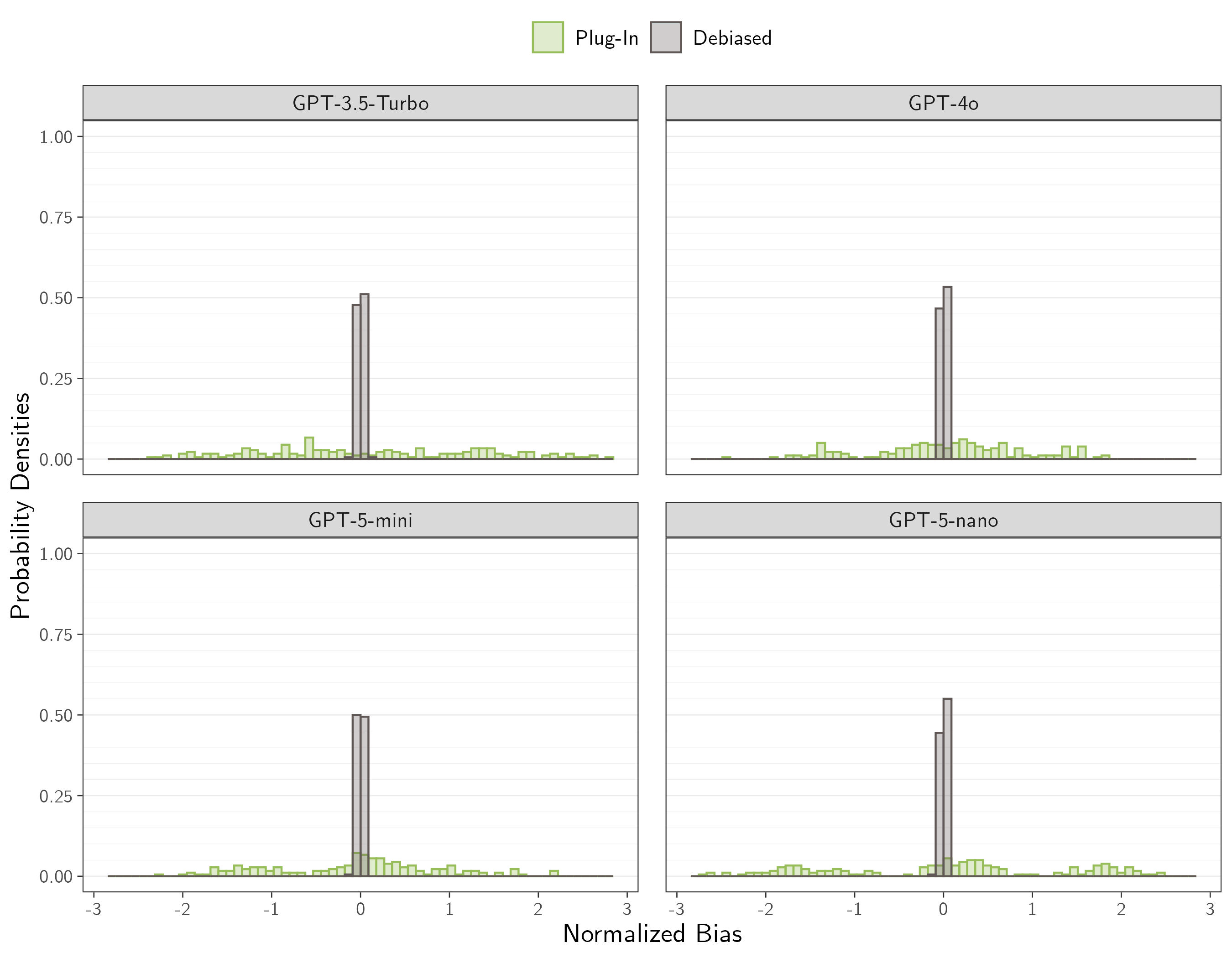}
\caption{Normalized bias of the plug-in regression and bias-corrected regression across Monte Carlo simulations based on congressional legislation.}
\floatfoot{\textit{Notes}: The normalized bias reports the average bias of the plug-in regression coefficient $\widehat{\beta}$ and the debiased coefficient $\widehat{\beta}^{debiased}$ for the target regression coefficient divided by their respective standard deviations across simulations. 
We summarize the distribution of normalized bias and coverage across regression specifications, choice of large language model and prompting strategies.
For each combination of model topic $V_r$, linked covariate $W_r$, large language model $m$ and prompting strategy $p$, we randomly sample $5,000$ Congressional bills and calculate the plug-in regression coefficient $\widehat{\beta}$ and the bias-corrected regression coefficient $\widehat{\beta}^{debiased}$ based on a 5\% validation sample.
See Section \ref{section: practical guidance with validation data} for discussion.
}
\label{figure: congressional bills, LLM on LHS, normalized bias, 5 percent validation}
\end{figure}

\begin{figure}[htbp!]
\includegraphics[width=\textwidth]{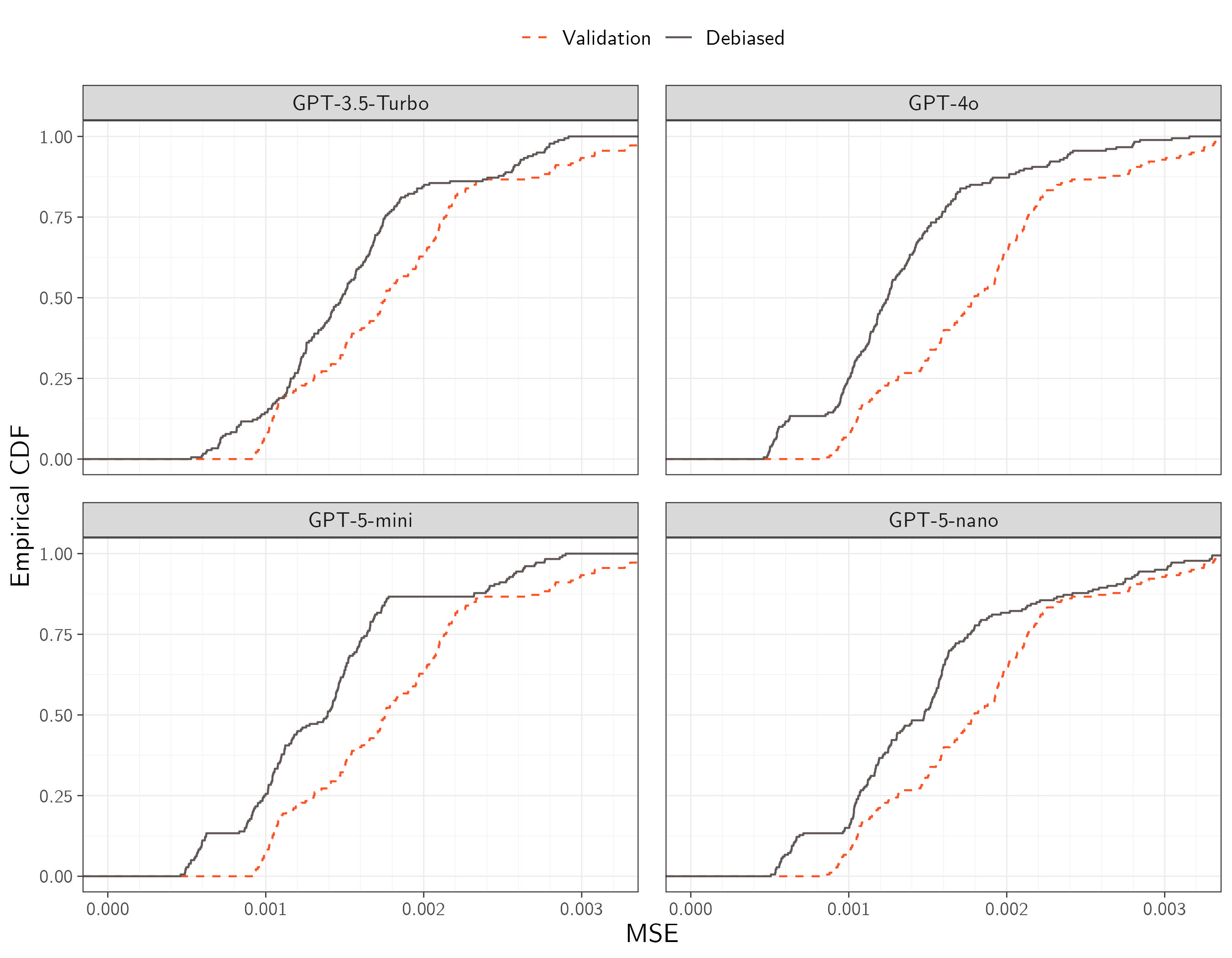}
\caption{Cumulative distribution function of mean square error for the bias-corrected estimator against validation-sample only estimator.}
\floatfoot{\textit{Notes}:
For each combination of model topic $V_r$, covariate $W_r$, large language model $m$ and prompting strategy $p$, we randomly sample $5,000$ Congressional bills and calculate the bias-corrected regression coefficient using a 5\% validation sample and the validation-sample only regression coefficient. We calculate the mean square error of $\widehat{\beta}^{debiased}$ and $\widehat{\beta}^*$ for the target regression, and we average the results over $1,000$ simulations.
We summarize the distribution of average mean square error across regression specifications, choice of large language model and prompting strategies.
See Section \ref{section: practical guidance with validation data} for discussion.
}
\label{figure: congressional bills, LLM on LHS, MSE, 5 percent validation}
\end{figure}
\clearpage \newpage
\begin{table}[htb!]
\begin{subtable}[t]{0.49\textwidth}

\begin{tabular}{lrrr}
\toprule
\multicolumn{1}{c}{\em{ }} & \multicolumn{3}{c}{\em{Return Horizon}} \\
\cmidrule(l{3pt}r{3pt}){2-4}
Point Estimates & 1 day & 5 days & 10 days\\
\midrule
Mean & -0.677 & -0.290 & 0.289\\
Median & -0.221 & 0.141 & 0.637\\
$5^\text{th}$ Percentile & -2.475 & -1.787 & -1.584\\
$95^\text{th}$ Percentile & 0.079 & 0.465 & 1.186\\
\midrule
Sample Average & 0.045 & 0.316 & 0.588\\
\bottomrule
\end{tabular}

\caption{Positive Labels}
\end{subtable} %
\begin{subtable}[t]{0.49\textwidth}

\begin{tabular}{lrrr}
\toprule
\multicolumn{1}{c}{\em{ }} & \multicolumn{3}{c}{\em{Return Horizon}} \\
\cmidrule(l{3pt}r{3pt}){2-4}
Point Estimates & 1 day & 5 days & 10 days\\
\midrule
Mean & 1.219 & 1.027 & 1.088\\
Median & 0.862 & 1.082 & 1.156\\
$5^\text{th}$ Percentile & 0.092 & -0.201 & -0.126\\
$95^\text{th}$ Percentile & 3.474 & 2.325 & 2.505\\
\midrule
Sample Average & 0.045 & 0.316 & 0.588\\
\bottomrule
\end{tabular}

\caption{Negative Labels}
\end{subtable}
\caption{Variation in point estimates across large language models and prompting strategies on financial news headlines.}
\floatfoot{\textit{Notes}:
On financial news headlines from 2019, we prompt GPT-3.5-Turbo, GPT-4o-mini, GPT-4o, GPT-5-mini, and GPT-5-nano to label each headline for whether it expressed positive, negative or uncertain news about the respective company using alternative prompting strategies.
For each model $m$ and prompt $p$, we regress the realized returns of each stock within 1-day of the headline's publication date on each large language model's labels $\widehat{V}_{r}^{m,p}$, the large language model's assessed magnitude denoted $S_{r}^{m,p}$ and their interaction, controlling for lagged realized returns. 
See Section \ref{section: evidence of measurement error} for discussion.
}
\label{table: financial news headline, positive/negatives, coefficient variation, magnitude}
\end{table}

\begin{table}[htb!]

\begin{tabular}{llrrrrr}
\toprule
\multirow{2}{*}{Policy Topic} & \multirow{2}{*}{Covariate} & \multicolumn{4}{c}{\textit{Point Estimates}} & Sample\\
\cmidrule(l{3pt}r{3pt}){3-6}
& & Mean & Median & 5\% & 95\% & Average\\
\midrule
Health & DW1 & -0.018 & -0.020 & -0.023 & -0.006 & 0.150\\
Health & Democrat & 0.012 & 0.014 & 0.003 & 0.016 & 0.150\\
Health & Senate & 0.005 & 0.004 & 0.001 & 0.012 & 0.150\\
\midrule
Banking, Finance \& Domestic Com. & DW1 & 0.013 & 0.013 & 0.007 & 0.022 & 0.127\\
Banking, Finance \& Domestic Com. & Democrat & -0.010 & -0.010 & -0.015 & -0.004 & 0.127\\
Banking, Finance \& Domestic Com. & Senate & -0.008 & -0.009 & -0.016 & -0.001 & 0.127\\
\midrule
Defense & DW1 & 0.022 & 0.022 & 0.015 & 0.034 & 0.204\\
Defense & Democrat & -0.005 & -0.004 & -0.009 & 0.000 & 0.204\\
Defense & Senate & -0.008 & -0.006 & -0.019 & -0.003 & 0.204\\
\midrule
Government Operations & DW1 & 0.015 & 0.015 & 0.000 & 0.028 & 0.281\\
Government Operations & Democrat & -0.005 & -0.004 & -0.014 & 0.003 & 0.281\\
Government Operations & Senate & -0.013 & -0.013 & -0.020 & -0.006 & 0.281\\
\midrule
Public Lands \& Water Management & DW1 & 0.027 & 0.027 & 0.020 & 0.030 & 0.238\\
Public Lands \& Water Management & Democrat & -0.006 & -0.007 & -0.010 & -0.003 & 0.238\\
Public Lands \& Water Management & Senate & 0.032 & 0.032 & 0.025 & 0.036 & 0.238\\
\bottomrule
\end{tabular}

\caption{Variation in point estimates across large language models and prompting strategies on Congressional bills.}
\floatfoot{\textit{Notes}:
On 10,000 Congressional bills, we prompt GPT-3.5-Turbo, GPT-4o, GPT-5-mini, and GPT-5-nano to label each Congressional bill for its policy topic using alternative prompting strategies. For each model $m$ and prompt $p$, we regress an indicator for whether the large language model labeled a particular policy topic $1\{\widehat{V}_r^{m,p} = v\}$ on alternative covariates $W_r$. 
For comparison, the final column (``Sample Average'') reports the fraction of all Congressional bills assigned to the policy topic $1\{V_r = v\}$.
See Section \ref{section: evidence of measurement error} for discussion.
}
\label{table: congressional bills, coeff variation, lhs}
\end{table}

\begin{table}[htbp!]
\begin{subtable}[t]{0.475\textwidth}

\begin{tabular}{r|rrr}
\toprule
  & Median & 5\% & 95\%\\
\midrule
\multicolumn{1}{l|}{\textit{Normalized Bias}}\\
Plug-In & -0.023 & -1.899 & 2.211\\
Debiased & 0.001 & -0.055 & 0.066\\
\midrule
\multicolumn{1}{l|}{\textit{Coverage}}\\
Plug-In & 0.820 & 0.381 & 0.945\\
Debiased & 0.930 & 0.910 & 0.945\\
\bottomrule
\end{tabular}

\caption{GPT-3.5-turbo}
\end{subtable}
\hfill
\begin{subtable}[t]{0.475\textwidth}

\begin{tabular}{r|rrr}
\toprule
  & Median & 5\% & 95\%\\
\midrule
\multicolumn{1}{l|}{\textit{Normalized Bias}}\\
Plug-In & 0.084 & -1.411 & 1.514\\
Debiased & 0.001 & -0.055 & 0.054\\
\midrule
\multicolumn{1}{l|}{\textit{Coverage}}\\
Plug-In & 0.920 & 0.637 & 0.954\\
Debiased & 0.927 & 0.902 & 0.945\\
\bottomrule
\end{tabular}

\caption{GPT-4o}
\end{subtable} \\
\begin{subtable}[t]{0.475\textwidth}

\begin{tabular}{r|rrr}
\toprule
  & Median & 5\% & 95\%\\
\midrule
\multicolumn{1}{l|}{\textit{Normalized Bias}}\\
Plug-In & 0.030 & -1.646 & 1.567\\
Debiased & -0.002 & -0.052 & 0.054\\
\midrule
\multicolumn{1}{l|}{\textit{Coverage}}\\
Plug-In & 0.906 & 0.589 & 0.953\\
Debiased & 0.927 & 0.903 & 0.945\\
\bottomrule
\end{tabular}

\caption{GPT-5-mini}
\end{subtable}
\hfill
\begin{subtable}[t]{0.475\textwidth}

\begin{tabular}{r|rrr}
\toprule
  & Median & 5\% & 95\%\\
\midrule
\multicolumn{1}{l|}{\textit{Normalized Bias}}\\
Plug-In & 0.168 & -2.079 & 2.092\\
Debiased & 0.005 & -0.052 & 0.060\\
\midrule
\multicolumn{1}{l|}{\textit{Coverage}}\\
Plug-In & 0.779 & 0.387 & 0.953\\
Debiased & 0.930 & 0.906 & 0.946\\
\bottomrule
\end{tabular}

\caption{GPT-5-nano}
\end{subtable}
\caption{Summary statistics for normalized bias and coverage across Monte Carlo simulations based on Congressional legislation.}
\floatfoot{\textit{Notes}: 
The normalized bias reports the average bias of the plug-in regression coefficient $\widehat{\beta}$ and the debiased coefficient $\widehat{\beta}^{debiased}$ for the target regression coefficient divided by their respective standard deviations across simulations. 
The coverage reports the fraction of simulations in which a 95\% nominal confidence interval centered around the plug-in regression coefficient $\widehat{\beta}$ and the bias-corrected coefficient $\widehat{\beta}^{debiased}$ cover the target regression coefficient.
We summarize the distribution of normalized bias and coverage across regression specifications, choice of large language model and prompting strategies.
For each combination of model topic $V_r$, covariate $W_r$, large language model $m$ and prompting strategy $p$, we randomly sample $5,000$ Congressional bills and calculate the plug-in regression coefficient $\widehat{\beta}$ and the bias-corrected regression coefficient $\widehat{\beta}^{debiased}$ using a 5\% validation sample. Results are averaged over $1,000$ simulations. 
See Section \ref{section: practical guidance with validation data} for discussion.
}
\label{table: congressional bills, LLM on LHS, summary statistics, 5 percent validation}
\end{table}

\clearpage
\newpage

\appendix
\singlespacing

\begin{center}
{\Large \textbf{Large Language Models:}} \\ {\Large \textbf{An Applied Econometric Perspective}} \medskip \\
{\Large \textit{Online Appendix}} \medskip \\
\large Jens Ludwig \& Sendhil Mullainathan \& Ashesh Rambachan \medskip \\
\end{center}

\section{Appendix Figures and Tables}
\renewcommand{\thefigure}{A\arabic{figure}}
\setcounter{figure}{0}
\renewcommand{\thetable}{A\arabic{table}}
\setcounter{table}{0}

\begin{table}[htb!]
\centering
\caption{Accuracy, true positive rate (TPR), and false positive rate (FPR) of GPT-4o's predictions on Congressional legislation.}
\begin{subtable}[t]{0.475\textwidth}
\centering

\begin{tabular}{lrrr}
\toprule
  & Accuracy & TPR & FPR\\
\midrule
House & 0.912 & 0.198 & 0.031\\
Senate & 0.925 & 0.225 & 0.031\\
\bottomrule
\end{tabular}

\caption{Base prompt}
\end{subtable}
\hfill
\begin{subtable}[t]{0.475\textwidth}
\centering

\begin{tabular}{lrrr}
\toprule
  & Accuracy & TPR & FPR\\
\midrule
House & 0.644 & 0.691 & 0.359\\
Senate & 0.695 & 0.711 & 0.306\\
\bottomrule
\end{tabular}

\caption{Prompt with date restriction}
\end{subtable}
\floatfoot{\textit{Notes}:
We prompt GPT-4o to predict whether 10,000 randomly selected Congressional bills would pass the Senate or the House based on its text description. 
This table reports the accuracy, true positive rate (TPR), and false positive rate (FPR) of GPT-4o's predictions.
Table (a) provides results for the base prompt, and Table (b) provides results for the base prompt with the additional date restriction.
See Section \ref{section: evidence of training leakage} for discussion.}
\label{tab: congressional bills prediction, performance of GPT-4o}
\end{table}

\begin{table}[htbp!]
\centering
\caption{Embedding distance between GPT-4o's completed bill descriptions and original bill descriptions.}
\begin{subtable}[t]{0.475\textwidth}
\centering

\begin{tabular}{lrr}
\toprule
Metric & Average & Benchmark\\
\midrule
Cosine similarity & 0.830 & 0.379\\
Euclidean distance & 0.536 & 1.110\\
\bottomrule
\end{tabular}

\caption{Base prompt}
\end{subtable}
\hfill
\begin{subtable}[t]{0.475\textwidth}
\centering

\begin{tabular}{lrr}
\toprule
Metric & Average & Benchmark\\
\midrule
Cosine similarity & 0.830 & 0.379\\
Euclidean distance & 0.536 & 1.110\\
\bottomrule
\end{tabular}

\caption{Prompt with date restriction}
\end{subtable}
\floatfoot{\textit{Notes}: This table calculates the cosine similarity and Euclidean distance between embeddings of GPT-4o's completed bill descriptions and embeddings of the original bill descriptions.
We construct embeddings using OpenAI's text-embedding-3-small model.
As a benchmark, we calculate the average cosine similarity and Euclidean distance between 10,000 randomly selected pairs of original bill descriptions. 
Table (a) provides results for the base prompt and Table (b) provides results for the base prompt with the additional date restriction.
The results in Table (a) and Table (b) are the same up to 3 decimal places.
See Section \ref{section: evidence of training leakage} for discussion.}
\label{table: congressional bills prediction, embedding closeness}
\end{table}

\begin{table}[htbp!]
\centering
\caption{Embedding distance between GPT-4o's completed financial news headlines and original financial news headlines.}
\begin{subtable}[t]{0.475\textwidth}
\centering

\begin{tabular}{lrr}
\toprule
Metric & Average & Benchmark\\
\midrule
Cosine similarity & 0.880 & 0.309\\
Euclidean distance & 0.455 & 1.172\\
\bottomrule
\end{tabular}

\caption{Base prompt}
\end{subtable}
\hfill
\begin{subtable}[t]{0.475\textwidth}
\centering

\begin{tabular}{lrr}
\toprule
Metric & Average & Benchmark\\
\midrule
Cosine similarity & 0.880 & 0.309\\
Euclidean distance & 0.455 & 1.172\\
\bottomrule
\end{tabular}

\caption{Prompt with date restriction}
\end{subtable}
\floatfoot{\textit{Notes}: This table calculates the cosine similarity and Euclidean distance between embeddings of GPT-4o's completed financial news headlines and embeddings of the original financial news headlines.
We construct embeddings using OpenAI's text-embedding-3-small model.
As a benchmark, we calculate the average cosine similarity and Euclidean distance between 10,000 randomly selected pairs of original financial news headlines. 
Table (a) provides results for the base prompt and Table (b) provides results for the base prompt with the additional date restriction.
The results in Table (a) and Table (b) are the same up to 3 decimal places.
See Section \ref{section: evidence of training leakage} for discussion.}
\label{table: financial news headlines prediction, embedding closeness}
\end{table}
\begin{figure}[htb!]
\input{Figures_AnnualReview/CongressionalBills_Prediction/fig_completion_exact_gpt4o_cb_base_prompt}
\caption{Two examples of GPT-4o completions that exactly match original descriptions of congressional legislation.}
\floatfoot{\textit{Notes}: On 10,000 randomly sampled congressional bills, we prompted GPT-4o to complete the description of the congressional bill based on 50\% of its text. See Section \ref{section: evidence of training leakage}.}
\label{figure: congressional bills prediction, no date restriction, memorization examples}
\end{figure}

\begin{figure}[htbp!]
\input{Figures_AnnualReview/FinancialNewsHeadlines_Prediction/fig_completion_exact_gpt4o_headlines_base_prompt}
\caption{Two examples of GPT-4o completions that exactly match original financial news headlines.}
\floatfoot{\textit{Notes}: On 10,000 randomly sampled financial news headlines from 2019, we prompted GPT-4o to complete the financial news headline based on 50\% of its text. See Section \ref{section: evidence of training leakage}.
}
\label{figure: financial news headlines prediction, no date restriction, memorization examples}
\end{figure}

\begin{figure}[htb!]
\includegraphics[width=\textwidth]{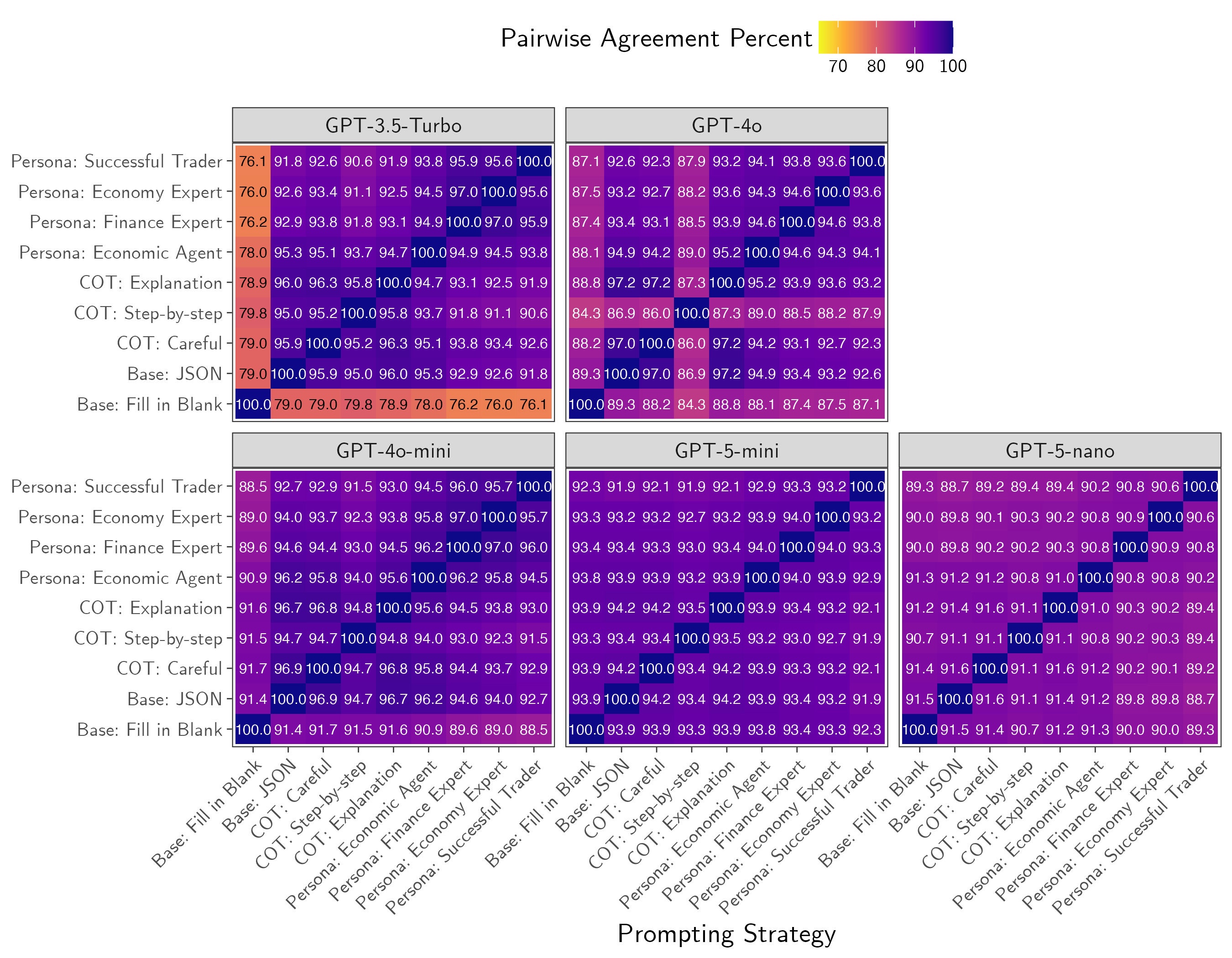}
\caption{Variation in pairwise agreement between large language model labels across prompting strategies on financial news headlines.}
\floatfoot{\textit{Notes}:
On financial news headlines from 2019, we prompt GPT-3.5-Turbo, GPT-4o, GPT-4o-mini, GPT-5-mini, and GPT-5-nano to label each headline for whether it expressed positive, negative or uncertain news about the respective company using alternative prompting strategies.
For each pair of prompting strategies, we calculate the fraction of financial news headlines that receive the same label by the two prompting strategies, separately by large language model. 
See Section \ref{section: evidence of measurement error} for discussion.
}
\label{figure: financial news headline, positive/negative news, heat maps}
\end{figure}

\begin{figure}[htb!]
\includegraphics[width=0.9\textwidth]{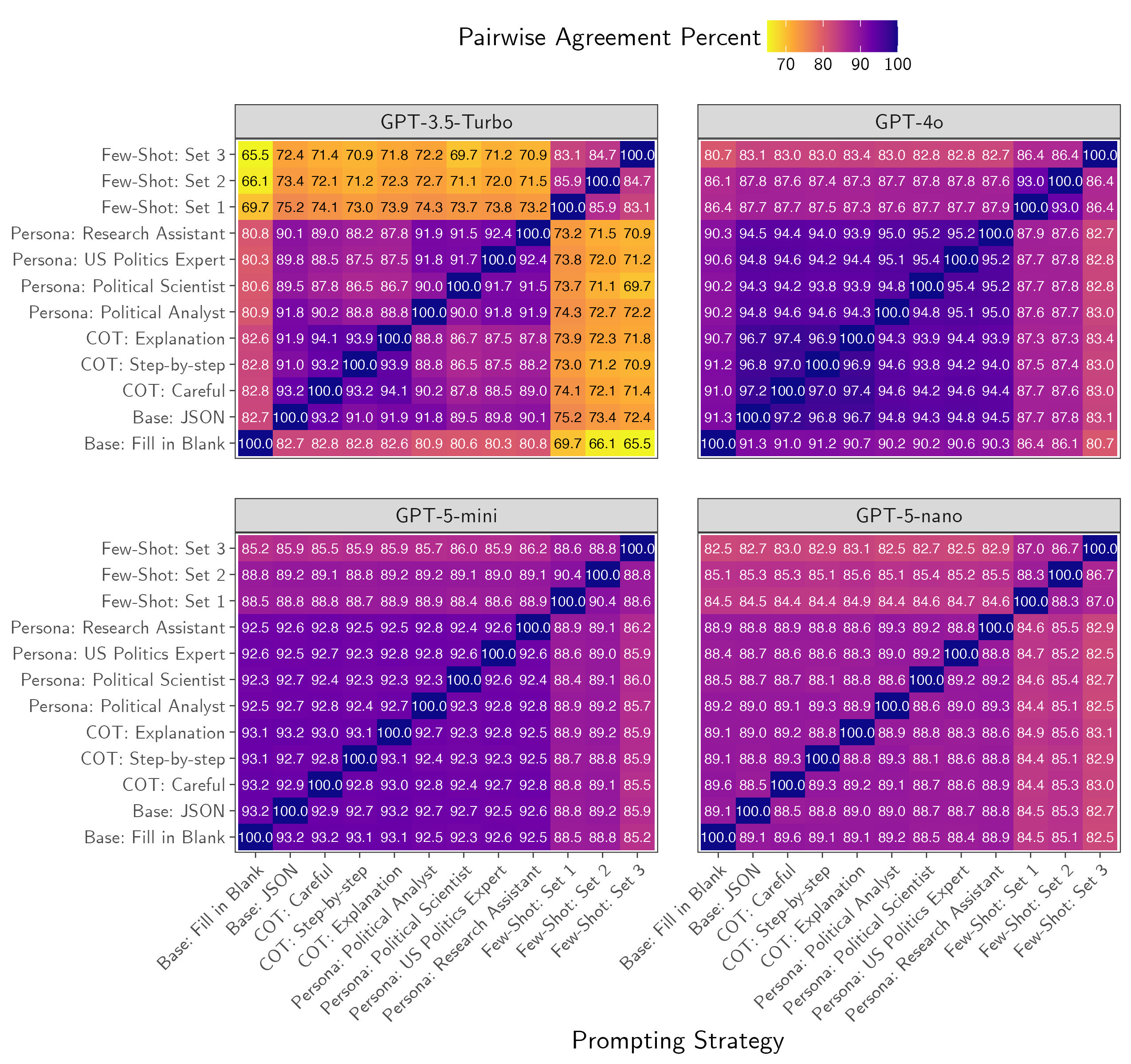}
\caption{Variation in pairwise agreement between large language model labels across prompting strategies on congressional legislation.}
\floatfoot{\textit{Notes}: 
On 10,000 randomly sampled Congressional bills, we prompt GPT-3.5-turbo, GPT-4o, GPT-5-mini, and GPT-5-nano to label the policy topic of each Congressional bill. 
For each pair of prompting strategies, we calculate the fraction of congressional bills that receive the same label, separately by large language model.
See Section \ref{section: evidence of measurement error} for discussion.}
\label{figure: congressional legislation, major topic heatmap}
\end{figure}

\begin{figure}[htbp!]
\includegraphics[width=\textwidth]{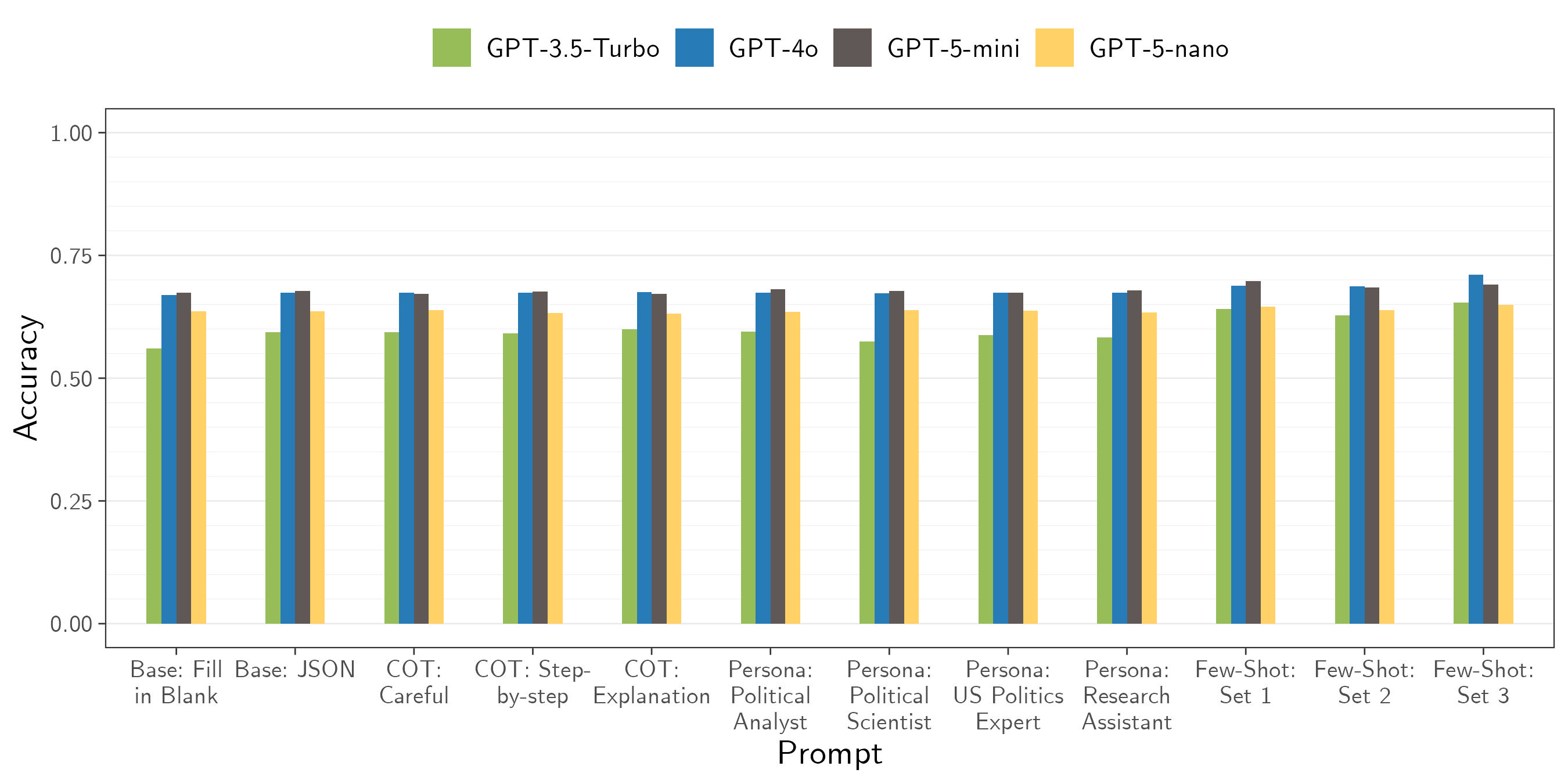}
\caption{Accuracy of large language model labels of bill topic across model and prompt variation.}
\floatfoot{\textit{Notes}: On 10,000 Congressional bills, we prompt GPT-3.5-Turbo, GPT-4o, GPT-5-mini, and GPT-5-nano to label each description for its policy topic area using alternative prompting strategies. 
For each combination of model $m$ and prompt $p$, we calculate the accuracy of the labels $\widehat{V}_{r}^{m,p}$ for the ground-truth label $V_r$.
See Section \ref{section: evidence of measurement error} for discussion.}
\label{figure: congressional bills, accuracy across models and prompts}
\end{figure}

\newpage \clearpage
\section{Proofs of Main Results}

\subsection{Proof of Lemma \ref{lemma: leakage decomposition} and Proposition \ref{proposition: prediction gpts iff no leakage}}

Proposition \ref{proposition: prediction gpts iff no leakage} is an immediate consequence of Lemma \ref{lemma: leakage decomposition}. 
To prove Lemma \ref{lemma: leakage decomposition}, observe that, for any $Q(\cdot) \in \mathcal{Q}$,
$$
\mathbb{E}_{Q}[\sum_{r \in \mathcal{R}} D_{r} \ell(Y_r, \widehat{m}(r))] - \mathbb{E}_{Q}[\sum_{r \in \mathcal{R}} D_r \ell(Y_r, \widehat{m}(r)) \mid T = t] = 
$$
$$
\sum_{r \in \mathcal{R}} q^{D}_r \ell(Y_r, \widehat{m}(r)) - \sum_{r \in \mathcal{R}} q_{r}^{D \mid T}(t_r) \ell(Y_r, \widehat{m}(r)) =
\sum_{r \in \mathcal{R}} \left( q_{r}^{D} - q_{r}^{D \mid T}(t_{r}) \right) \ell(Y_r, \widehat{m}(r)).
$$
Under Assumption \ref{asm: research contexts}, for any text piece $r \in \mathcal{R}$, we can rewrite $q_{r}^{D} - q_{r}^{D \mid T}(t_{r})$ as $q_{r}^{D} \left(1 - \frac{q_{r}^{T \mid D}(t_r)}{q_{r}^T(t_r)} \right)$ by Bayes' rule.
We therefore have 
$$
\mathbb{E}_{Q}[\sum_{r \in \mathcal{R}} D_{r} \ell(Y_r, \widehat{m}(r))] - \mathbb{E}_{Q}[\sum_{r \in \mathcal{R}} D_r \ell(Y_r, \widehat{m}(r)) \mid T = t] = \sum_{r \in \mathcal{R}} q_{r}^{D} \left( 1 - \frac{q_{r}^{T \mid D}(t_r)}{q_{r}^{T}(t_r)} \right) \ell(Y_r, \widehat{m}(r)).
$$
Lemma \ref{lemma: leakage decomposition} then follows immediately. 
$\Box$

\subsection{Proof of Lemma \ref{lemma: training leakage decomp for estimation}}
To show this result, rewrite 
\begin{equation*}
    \mathbb{E}_{Q}[\sum_{r \in \mathcal{R}} D_r g(\widehat{m}(r), W_r; \theta) \mid T = t] - \mathbb{E}_{Q}[\sum_{r \in \mathcal{R}} D_r g(V_r, W_r; \theta)]
\end{equation*}
as
\begin{align*}
   & \left( \mathbb{E}_{Q}[\sum_{r \in \mathcal{R}} D_r g(\widehat{m}(r), W_r; \theta) \mid T = t] - \mathbb{E}_{Q}[\sum_{r \in \mathcal{R}} D_r g(\widehat{m}(r), W_r; \theta)] \right) + \\
   & \left( \mathbb{E}_{Q}[\sum_{r \in \mathcal{R}} D_r g(\widehat{m}(r), W_r; \theta)] - \mathbb{E}_{Q}[\sum_{r \in \mathcal{R}} D_r g(V_r, W_r; \theta)] \right).
\end{align*}
The result then follows by applying the same argument as the proof of Lemma \ref{lemma: leakage decomposition} to rewrite the first term as $\mathbb{E}_{Q}[ \sum_{r \in \mathcal{R}} D_r \left( \frac{q^{T \mid D}_r(t_r)}{q_{r}^{T}(t_r)} - 1  \right) g(\widehat{m}(r), W_r; \theta) ]$. $\Box$

\subsection{Proof of Lemma \ref{lemma: measurement error and estimation error} and Proposition \ref{proposition: no measurement error implies GPT for estimation community}}
Proposition \ref{proposition: no measurement error implies GPT for estimation community} is an immediate consequence of Lemma \ref{lemma: measurement error and estimation error}. 
As a result, we focus on proving Lemma \ref{lemma: measurement error and estimation error}.

We first prove the claim in Equation \eqref{equation: measurement error bound on error}.
Consider any $Q(\cdot) \in \mathcal{Q}$ and $g(\cdot) \in \mathcal{G}$.
Since no training leakage is satisfied, by Lemma \ref{lemma: leakage decomposition}, we may write 
$$
\mathbb{E}_{Q}[\sum_{r \in \mathcal{R}} D_r g(\widehat{m}(r), W_r; \theta) \mid T = t] - \mathbb{E}[\sum_{r \in \mathcal{R}} D_r g(V_r, W_r; \theta)] =
$$
$$
\mathbb{E}_{Q}[\sum_{r \in \mathcal{R}} D_r \left( g(\widehat{m}(r), W_r; \theta) - g(V_r, W_r; \theta) \right)].
$$
Defining $\Delta(r) = \widehat{m}(r) - f^*(r)$, the previous display can be further written as 
$$
\sum_{r \in \mathcal{R}} q_{r}^{D} \left( g(f^*(r) + \Delta(r), W_r; \theta) - g(f^*(r), W_r; \theta)\right) = 
$$
$$
\sum_{r \in \mathcal{R}} q_{r}^{D} \frac{\partial g(\xi(t, W_r, \theta), W_r; \theta)}{\partial v} \Delta(r)
$$
where the equality applies the mean value theorem for some $\xi(t, x; \theta)$ in between $f^*(r) + \Delta(r)$ and $f^*(r)$. It therefore follows that 
$$
\left| \mathbb{E}_{Q}[\sum_{r \in \mathcal{R}} D_r g(\widehat{m}(r), W_r; \theta) \mid T = t] - \mathbb{E}[\sum_{r \in \mathcal{R}} D_r g(V_t, W_r; \theta)] \right| \leq
$$
$$
\sum_{r \in \mathcal{R}} q_{r}^{D} \left| \frac{\partial g(\xi(t, W_r, \theta), W_r; \theta)}{\partial v} \right| \left| \Delta(r) \right| \leq
\overline{G} \sum_{r \in \mathcal{R}} q_{r}^{D} \left| \Delta(r) \right|,
$$
where the last inequality follows by Assumption \ref{asm: moment conditions}. 
The result in Equation \eqref{equation: measurement error bound on error} is immediate following the definition of $\mathcal{M}(Q, \delta)$.

To prove Equation \eqref{equation: measurement error insuff for validity}, consider any $Q(\cdot) \in \mathcal{Q}$ and $g(\cdot) \in \mathcal{G}$.
Since no training leakage is satisfied, we can again write, for any $\widehat{m}(\cdot) \in \mathcal{M}(Q, \delta)$,
$$
\left| \mathbb{E}_{Q}[\sum_{r \in \mathcal{R}} D_r g(\widehat{m}(r), W_r; \theta) \mid T = t] - \mathbb{E}_{Q}[\sum_{r \in \mathcal{R}} D_r g(V_r, W_r; \theta) ] \right| = 
$$
$$
\left| \sum_{r \in \mathcal{R}} q_{t}^{D} \left( g(\widehat{m}(r), W_r; \theta) - g(V_r, W_r; \theta) \right) \right|.
$$
Again, defining $\Delta(r) = \widehat{m}(r) - f^*(r)$ and $\Delta(Q, \delta) = \{ \Delta(r) \colon -\delta \leq \Delta(r) \leq \delta \mbox{ for } r \mbox{ with } q_{r}^{D} > 0 \}$, we have that 
$$
\sup_{\widehat{m}(\cdot) \in \mathcal{M}(Q,\delta)} \left| \mathbb{E}_{Q}[\sum_{r \in \mathcal{R}} D_r g(\widehat{m}(r), W_r; \theta) \mid T = t] - \mathbb{E}_{Q}[\sum_{r \in \mathcal{R}} D_r g(f^*(r), W_r; \theta) ] \right| = 
$$
$$
\sup_{\Delta(\cdot) \in \Delta(Q,\delta)} \left| \sum_{r \in \mathcal{R}} q_{r}^{D} \left( g(f^*(r) + \Delta(r), W_r; \theta) - g(f^*(r), W_r; \theta) \right) \right|
$$
Consider the following choice of $\delta(r)$. 
Define $\tilde{\Delta}(r) = \arg \max_{-\delta \leq \tilde{\delta} \leq \delta} g(f^*(r) + \tilde{\delta}, W_r; \theta) - g(f^*(r), W_r; \theta)$, and let $\delta(r) = \tilde{\Delta}(r) 1\{ g(f^*(r) + \tilde{\Delta}(r)\}, W_r; \theta) - g(f^*(r), W_r; \theta)  \geq 0\}$.
This choice is feasible, and so it follows that 
$$
\sup_{\Delta(\cdot) \in \Delta(\delta, Q)} \left| \sum_{r \in \mathcal{R}} q_{r}^{D} \left( g(f^*(r) + \Delta(r), W_r; \theta) - g(f^*(r), W_r; \theta) \right) \right| \geq 
$$
$$
\sum_{r \in \mathcal{R}} q_r^D \left| g(f^*(r) + \delta(r), W_r; \theta) - g(f^*(r), W_r; \theta) \right|,
$$
where we further used that the triangle inequality holds with equality when all terms in a summation are non-negative. By a similar argument as given in the proof of Equation \eqref{equation: measurement error bound on error}, we can apply the mean value theorem and the definition of sensitive text pieces to obtain the lower bound 
$$
\sup_{\widehat{m}(\cdot) \in \mathcal{M}(\delta, Q)} \left| \mathbb{E}_{Q}[\sum_{r \in \mathcal{R}} D_r g(\widehat{m}(r), W_r; \theta) \mid T = t] - \mathbb{E}_{Q}[\sum_{r \in \mathcal{R}} D_r g(f^*(r), W_r; \theta) ] \right| \geq \underline{G} \sum_{r \in \mathcal{R}_{g,Q}} \delta(r) q_{r}^{D}. 
$$ $\Box$

\subsection{Proof of Proposition \ref{prop: bias characterization of LLM on LHS and RHS}}

To show (i), given that $q_{r}^{T}(t_r) = q_{r}^{T \mid D}(t_r)$ for all $r \in \mathcal{R}$, it follows that 
$$
\beta = \left( \sum_{r \in \mathcal{R}} q_r^{D} W_r W_r^\prime \right)^{-1} \left( \sum_{r \in \mathcal{R}} q_{r}^{D} W_r \widehat{V}_r \right) \mbox{ and } \beta^* = \left( \sum_{r \in \mathcal{R}} q_r^{D} W_r W_r^\prime \right)^{-1} \left( \sum_{r \in \mathcal{R}} q_{r}^{D} W_r V_r \right).
$$
But, of course, since $\widehat{V}_r = V_r + \Delta_r$, it then follows that
$$
\beta = \left( \sum_{r \in \mathcal{R}} q_r^{D} W_r W_r^\prime \right)^{-1} \left( \sum_{r \in \mathcal{R}} q_{r}^{D} W_r V_r \right) + \left( \sum_{r \in \mathcal{R}} q_r^{D} W_r W_r^\prime \right)^{-1} \left( \sum_{r \in \mathcal{R}} q_{r}^{D} W_r \Delta_r \right).
$$
The result is then immediate from the definition of $\beta^*$ and the best linear projection of $\Delta_r$ onto $W_r$.

To show (ii), since there is again no training leakage by assumption, it follows that 
$$
\beta = \left( \sum_{r \in \mathcal{R}} q_{r}^{D} \widehat{V}_r \widehat{V}_r^\prime \right)^{-1} \left( \sum_{r \in \mathcal{R}} q_{r}^{D} \widehat{V}_r W_r \right) \mbox{ and } \beta^* = \left( \sum_{r \in \mathcal{R}} q_{r}^{D} V_r V_r^\prime \right)^{-1} \left( \sum_{r \in \mathcal{R}} q_{r}^{D} V_r W_r \right).
$$
But, of course, since $W_r = V_r^\prime \beta^* + \epsilon_r$ for $\epsilon_r$ the residual from the best-linear projection, it then follows that 
\begin{align*}
    & \beta = \left( \sum_{r \in \mathcal{R}} q_r^D \widehat{V}_r \widehat{V}_r^\prime \right)^{-1} \left( \sum_{r \in \mathcal{R}} q_r^D \widehat{V}_r V_r^\prime \right) \beta^* + \left( \sum_{r \in \mathcal{R}} q_r^D \widehat{V}_r \widehat{V}_r^\prime \right)^{-1} \left( \sum_{r \in \mathcal{R}} q_r^D \widehat{V}_r \epsilon_r \right).
\end{align*}
The result follows by the definition of the best linear projections of $V_r$ onto $\widehat{V}_r$ and $\epsilon_r$ onto $\widehat{V}_r$ in the research context $Q(\cdot)$. $\Box$

\section{Additional Theoretical Results}\label{section: additional theoretical results}

In this section, we collect together additional theoretical results that are referenced in the main text. 

\subsection{Analyzing the Researcher's Sample Average Loss and Sample Moment Condition}\label{section: prediction and estimation, large sample}

\subsubsection{The Researcher's Sample Average Loss}
To tackle the prediction problem, the researcher calculates the sample average loss of the large language model's predictions:
\begin{equation*}
    \frac{1}{N} \sum_{r \in \mathcal{R}} D_r \ell(Y_r, \widehat{m}(r; t))
\end{equation*}
where $N = \sum_{r \in \mathcal{R}} D_r$ is the number of text pieces collected by the researcher. 
Under Assumption \ref{asm: research contexts}(i), for all values $d$, $Q(D = d, T = t) = \Pi_{\sigma \in \Sigma^*} Q(D_{\sigma} = d_{\sigma}, T_{\sigma} = t_{\sigma})$, and therefore $Q(T = t) = \Pi_{\sigma \in \Sigma^*} Q(T_{\sigma} = t_{\sigma})$. 
We can then write $Q(D = d \mid T = t) = \Pi_{\sigma \in \Sigma^*} Q(D_{\sigma} = d_{\sigma} \mid T_{\sigma} = t_{\sigma})$, and the researcher's sampling distribution over text pieces is also independent but not identically distributed over text pieces, conditional on the large language model's realized training dataset. 

Consequently, we can re-interpret the researcher's sampling distribution over text pieces conditional on the large language model's realized training dataset as i.n.i.d sampling from the finite population of text pieces; and the researcher's sample average loss calculates the sample mean of the finite population characteristics $\ell(Y_r, \widehat{m}(r; t))$. 
Existing results on finite-population inference, such as those given in \citet{AAIW(20)}, \citet{Xu(20)} and \citet{RambachanRoth(24)}, provide regularity conditions under which Equation \eqref{equation: convergence of sample avg loss} holds and
\begin{equation*}
    \frac{1}{N} \sum_{r \in \mathcal{R}} D_{r} \ell(Y_r; \widehat{m}(r; t)) - \frac{1}{\mathbb{E}_{Q}\left[\sum_{r \in \mathcal{R}} D_r \mid T = t\right]} \mathbb{E}_{Q}\left[\sum_{r \in \mathcal{R}} D_{r} \ell(Y_r; \widehat{m}(r; t)) \mid T = t\right] \xrightarrow{p} 0,
\end{equation*}
as the number of text pieces grows large.

\subsubsection{The Researcher's Sample Moment Condition}
To tackle the estimation problem, recall that the researcher would like to calculate the sample moment function using the true economic concept:
$$
    \frac{1}{N} \sum_{r \in \mathcal{R}} D_r g(V_r, W_r; \theta),
$$
where $N = \sum_{r \in \mathcal{R}} D_r$ is the number of text pieces collected by the researcher. 
Under Assumption \ref{asm: research contexts}(i), for all values $d$, $Q(D = d, T = t) = \Pi_{\sigma \in \Sigma^*} Q(D_\sigma = d_\sigma, T_\sigma = t_{\sigma})$ and therefore $Q(D = d) = \Pi_{\sigma \in \Sigma^*} Q(D_{\sigma} = d_{\sigma})$. 
We can therefore interpret the researcher's sampling distribution over text pieces as independent but not identically distributed sampling from the finite population; and the researcher's sample moment function calculates the sample mean of the finite population characteristic $g(V_r, W_r; \theta)$. 
As for the researcher's sample average loss, existing results in the finite-population literature imply that 
\begin{equation*}
    \frac{1}{N} \sum_{r \in \mathcal{R}} D_{r} g(W_r, V_r; \theta) - \frac{1}{\mathbb{E}_{Q}\left[\sum_{r \in \mathcal{R}} D_r\right]} \mathbb{E}_{Q}\left[\sum_{r \in \mathcal{R}} D_{r} g(W_r, V_r; \theta)\right] \xrightarrow{p} 0,
\end{equation*}
as the number of text pieces grow large.

Due to the text processing problem, the researcher instead constructs the large large language model's labels of the economic concept and calculates the plug-in, sample moment function:
$$
    \frac{1}{N} \sum_{r \in \mathcal{R}} D_r g(\widehat{m}(r; t), W_r; \theta).
$$
By the same argument, we can interpret the researcher sampling distribution over text pieces conditional on the large language model's realized training dataset as i.n.i.d sampling from the finite population of text pieces; and the researcher's plug-in moment function then calculates the sample mean of the finite population characteristics $g(\widehat{m}(r; t), W_r; \theta)$. 
Existing results then provide regularity conditions under which 
$$
\frac{1}{N} \sum_{r \in \mathcal{R}} D_r g(\widehat{m}(r; t), W_r; \theta) - \frac{1}{\mathbb{E}\left[\sum_{r \in \mathcal{R}} D_r \mid T = t\right]} \mathbb{E}\left[\sum_{r \in \mathcal{R}} D_r g(\widehat{m}(r; t), W_r; \theta) \mid T = t\right] \xrightarrow{p} 0
$$
as the number of text pieces grow large.

\subsection{Analyzing the Asymptotic Distribution of Bias-Corrected Coefficient}\label{section: studying the asymptotic dist'n of bias corrected}

In this section, we separately analyze the asymptotic distribution of the bias-corrected regression coefficient introduced in Section \ref{section: practical guidance with validation data} in two separate cases: first, when the economic concept $V_r$ is the dependent variable; and second, when the economic concept $V_r$ is the independent variable.

\subsubsection{Linear Regression with Large Language Model Labels as the Dependent Variable}\label{section: asymptotics for LLM on LHS}

As discussed in Section \ref{section: practical guidance with validation data}, we study the limiting distribution of the bias-corrected linear regression in which the researcher uses the economic concept as the dependent variable.
It is convenient to now define the researcher's sampling indicator as taking three possible $D_r \in \{0, 1, 2\}$, where $D_r = 0$ denotes the researcher does not sample the text piece $r$, $D_r = 1$ denotes that the researcher samples the text piece in the primary sample and observes $(\widehat{m}(r; t), W_r)$, and $D_r = 2$ denotes that the researcher samples the text piece in the validation sample and observes $(\widehat{m}(r; t), V_r, W_r)$. 
Altogether the researcher observes $(\widehat{m}(r; t), W_r)$ for all $r \in \mathcal{R}$ with $D_r = 1$ and $(\widehat{m}(r; t), V_r, W_r)$ for all $r \in \mathcal{R}$ with $D_r = 2$. 

On the primary sample, the researcher calculates the plug-in regression coefficient
$$
    \widehat{\beta} = \left( \frac{1}{N_p} \sum_{r \in \mathcal{R}} 1\{D_r = 1\} W_r W_r^\prime \right)^{-1} \left( \frac{1}{N_p} \sum_{r \in \mathcal{R}} 1\{D_r = 1\} W_r \widehat{m}(r; t) \right)
$$
for $N_p = \sum_{r} 1\{D_r = 1\}$ the size of the primary sample.
On the validation sample, the researcher estimates the measurement error regression coefficient 
$$
    \widehat{\lambda}_{\Delta \mid W} = \left( \frac{1}{N_v} \sum_{r \in \mathcal{R}} 1\{D_r = 2\} W_r W_r^\prime \right)^{-1} \left( \frac{1}{N_v} \sum_{r \in \mathcal{R}} 1\{D_r = 2\} W_r \Delta_r \right)
$$
for $N_v = \sum_{r} 1\{D_r = 2\}$ the size of the validation sample. 
The bias-corrected regression coefficient is then given by $\widehat{\beta}^{debiased} = \widehat{\beta} - \widehat{\lambda}$.
The researcher's validation-sample only regression coefficient is 
$$
    \widehat{\beta}^{validation} = \left( \frac{1}{N_v} \sum_{r \in \mathcal{R}} 1\{D_r = 2\} W_r W_r^\prime \right)^{-1} \left( \frac{1}{N_v} \sum_{r \in \mathcal{R}} 1\{D_r = 2\} W_r V_r \right).
$$
As further notation, let $N = N_p + N_v$ denote the size of the researcher's dataset, $N_{R}$ be the total number of text pieces, and let $N_{o} = N_{R} - N$ denote the number of text pieces that are not sampled by the researcher.

To derive the limiting distribution as the number of economically relevant text pieces $N_R$ grows large, we make three simplifying assumptions. 
First, we assume that $W_r$ is a scalar, which is not technically necessary but will simplify the resulting expressions.
Second, we assume the large language model satisfies no training leakage as mentioned in the the main text. 
Third, we analyze a research context $Q(\cdot)$ in which the researcher randomly samples text pieces into their dataset and further randomly partitions the collected text pieces into the primary and validation sample. 
More formally, the text pieces are randomly sampled into three groups of size $N_o, N_p, N_v$ respectively and the probability that the vector $D$ takes a particular value $d$ is given by $N_o! N_p! N_v!/N_R!$, where $d$ satisfies $\sum_{r \in \mathcal{R}} 1\{D_r = 0\} = N_o$, $\sum_{r \in \mathcal{R}} 1\{ D_r = 1 \} = N_p$, $\sum_{r \in \mathcal{R}} 1\{ D_r = 2 \} = N_v$. 
Finally, we will assume there exists some finite constant $M > 0$ such that $-M \leq W_r, V_r, \widehat{m}(r; t) \leq M$ for all $r \in \mathcal{R}$.
The last two assumptions enable us to apply existing finite-population central limit theorem in deriving limiting distributions

We study the properties of the bias-corrected regression and the validation-sample only regression along a sequence of finite populations satisfying $N_{R} \rightarrow \infty$, $N_{v} / N_{R} =\rho_{v} > 0$, $N_{p} / N_{R} = \rho_{p} > 0$. 
Under these stated conditions, results in \cite{li_general_2017} imply that $\frac{1}{N_p} \sum_{r \in \mathcal{R}} 1\{D_r = 1\} W_r^2 - \frac{1}{N_R} \sum_{r \in \mathcal{R}} W_r^2 \xrightarrow{p} 0$ and $\frac{1}{N_v} \sum_{r \in \mathcal{R}} 1\{D_r = 2\} W_r^2 - \frac{1}{N_R} \sum_{r \in \mathcal{R}} W_r^2 \xrightarrow{p} 0$. 
We therefore focus on analyzing the properties of $\frac{1}{N_p} \sum_{r \in \mathcal{R}} 1\{D_r = 1\} W_r \widehat{m}(r; t)$ and $\frac{1}{N_v} \sum_{r \in \mathcal{R}} 1\{D_r = 1\} W_r \Delta_r$.

Towards this, let us define $X_{r} = W_r \widehat{m}(r; t)$ and $Z_{r} = W_r \Delta_r$ as convenient shorthand. 
We then write $\bar{X}_{p} = \frac{1}{N_p} \sum_{r \in \mathcal{R}} 1\{D_r = 1\} X_{r}$ and $\bar{Z}_{v} = \frac{1}{N_p} \sum_{r \in \mathcal{R}} 1\{D_r = 2\} Z_{r}$.
Define the finite population quantities $\bar{X}_{N} = \frac{1}{N_R} \sum_{r \in \mathcal{R}} X_{r}$ and $\bar{Z}_{N} = \frac{1}{N_R} \sum_{r \in \mathcal{R}} Z_{r}$, and  
$\sigma^2_{X, N} = \frac{1}{N-1} \sum_{r \in \mathcal{R}} (X_r -  \bar{X}_{N})^2$, $\sigma^2_{Z, N} = \frac{1}{N-1} \sum_{r \in \mathcal{R}} (Z_r -  \bar{Z}_{N})^2$. 

Proposition 2 in \cite{li_general_2017} implies that 
$$
Var_{Q}\left( (\bar{X}_p, \bar{Z}_v)^\prime \right) = N^{-1}_{R} \begin{pmatrix} 
    \frac{1 - \rho_p}{\rho_p} \sigma^2_{X,N} & -\sigma_{X,N} \sigma_{Z,N} \\
    -\sigma_{X,N} \sigma_{Z,N} & \frac{1 - \rho_v}{\rho_v} \sigma^2_{Z,N}
\end{pmatrix}.
$$
Consequently, provided $\sigma^2_{X,N} \rightarrow \sigma^2_{X}$ and $\sigma^2_{Z, N} \rightarrow \sigma^2_{Z}$ as $N_R \rightarrow \infty$, Theorem 5 in \cite{li_general_2017} implies that 
$$
\sqrt{N_{R}} \left( (\bar{X}_p, \bar{Z}_v)^\prime - (\bar{X}_N, \bar{Z}_N)^\prime \right) \xrightarrow{d} N\left(0, \begin{pmatrix} 
    \frac{1 - \rho_p}{\rho_p} \sigma^2_{X} & -\sigma_{X} \sigma_{Z} \\
    -\sigma_{X} \sigma_{Z} & \frac{1 - \rho_v}{\rho_v} \sigma^2_{Z}
\end{pmatrix} \right).
$$
We can therefore characterize the limiting distribution of the bias-corrected regression coefficient by an application of Slutsky's theorem and the Delta method. In particular, the previous display implies that
\begin{equation*}
\sqrt{N}_R \left( \widehat{\beta}^{debiased} - \beta^* \right) \xrightarrow{d} N(0, \Omega^{debiased})   
\end{equation*}
for 
\begin{equation*}
    \Omega^{debiased} = \sigma^{-4}_{W} \left(\frac{1 - \rho_p}{\rho_p} \sigma^2_{X} + 2 \sigma_X \sigma_{Z} + \frac{1 - \rho_v}{\rho_v} \sigma^2_Z \right).
\end{equation*}
and $\sigma^2_{W}$ the limit of $\frac{1}{N} \sum_{r \in \mathcal{R}} W_r^2$.
This delivers Equation \eqref{equation: limiting distribution of bias correction, main text} given in Section \ref{section: practical guidance with validation data} of the main text.
By a similar argument, we can show that the validation-sample only regression coefficient has a limiting distribution given by 
\begin{equation*}
    \sqrt{N}_R \left( \widehat{\beta}^{validation} - \beta^* \right) \xrightarrow{d} N(0, \Omega^{validation})   
\end{equation*}
for $\Omega^{validation} = \sigma^{-4}_{W} \frac{1 - \rho_{V}}{\rho_{V}} \sigma^2_{WV}$, as stated in Section \ref{section: practical guidance with validation data} of the main text.

\subsubsection{Linear Regression with Large Language Model Labels as Covariates}\label{section: linear regression with LLMs on RHS, theory}

We next discuss how the researcher using the economic concept as a covariate in a linear regression could bias correct their estimates using a small validation sample.
Towards this, recall that the target regression and plug-in regression are given by
$$
W_r = V_r^\prime \alpha^* + \nu_r, \mbox{ and } W_r = \widehat{m}(r; t)^\prime \alpha + \widetilde{\nu}_r.
$$
The researcher again observes $(\widehat{m}(r; t), W_r)$ for all $r \in \mathcal{R}$ with $D_r = 1$ and $(\widehat{m}(r; t), V_r, W_r)$ for all $r \in \mathcal{R}$ with $D_r = 2$. 

We will estimate the target regression using the validation sample and the primary sample in the following manner.
On the primary sample, the researcher separately calculates
$$
\widehat{\Sigma}_{\hat{V} \hat{V}}^{primary} = \frac{1}{N_p} \sum_{r \in \mathcal{R}} 1\{D_r = 1\} \widehat{m}(r; t) \widehat{m}(r; t)^\prime \mbox{ and } \widehat{\Sigma}_{\widehat{V} W}^{primary} = \frac{1}{N_p} \sum_{r \in \mathcal{R}} 1\{D_r = 1\} \widehat{m}(r; t) W_r.
$$
On the validation sample, the researcher separately calculates
$$
    \widehat{\Lambda}_{\hat{V} V}^{validation} \frac{1}{N_v} \sum_{r \in \mathcal{R}} 1\{D_r = 2\} \left( \widehat{m}(r; t) \widehat{m}(r; t)^\prime - V_r V_r^\prime \right) \mbox{ and } \widehat{\Lambda}_{\Delta W}^{validation} = \frac{1}{N_v} \sum_{r \in \mathcal{R}} 1\{D_r = 2\} \Delta_r W_r.
$$
The bias-corrected regression coefficient is then given by 
$$
\widehat{\alpha}^{debiased} = \left( \widehat{\Sigma}_{\hat{V} \hat{V}}^{primary} - \widehat{\Lambda}_{\hat{V} V}^{validation} \right)^{-1} \left( \widehat{\Sigma}_{\widehat{V} W}^{primary} - \widehat{\Lambda}_{\Delta W}^{validation} \right).
$$
The researcher's validation-sample only regression coefficient is 
$$
    \widehat{\alpha}^{validation} = \left( \frac{1}{N_v} \sum_{r \in \mathcal{R}} 1\{D_r = 2\} V_r V_r^\prime \right)^{-1} \left( \frac{1}{N_v} \sum_{r \in \mathcal{R}} 1\{D_r = 2\} V_r W_r \right).
$$
To analyze the limiting distribution as the number of economically relevant text pieces $N_R$ grows large, we make the same simplifying assumptions as in Appendix Section \ref{section: asymptotics for LLM on LHS} with the modification that $V_r$ is a scalar.
By the same arguments, it can be shown that 
$$
    \sqrt{N_R} \left( \widehat{\alpha}^{debiased} - \alpha^* \right) \xrightarrow{d} N(0, \Omega^{debiased})
$$
for 
$$
    \Omega^{debiased} = \sigma_V^{-4} \left( \frac{1 - \rho_p}{\rho_p} \sigma^2_{\widehat{V} W} + 2 \sigma_{\widehat{V} W} \sigma_{\Delta W} + \frac{1 - \rho_v}{\rho_v} \sigma_{\Delta W}^2 \right),
$$
where $\sigma_{\widehat{V} W}^2$, for example, is the finite-population limit of the variance of $\widehat{V}_r W_r$ across text pieces and the remaining terms are defined analogously.
It can be analogously shown that 
$$
    \sqrt{N_R} \left( \hat{\alpha}^{validation} - \alpha^* \right) \xrightarrow{d} N(0, \Omega^{validation})
$$
for $\Omega^{validation} = \sigma^4_{-V} \frac{1 - \rho_v}{\rho_v} \sigma_{V W}^2$. 
Consequently, we can compare the limiting variances, and again observe that the bias-corrected regression coefficient has a smaller limiting variance if $\frac{1 - \rho_p}{\rho_p} \sigma^2_{\widehat{V} W} + 2 \sigma_{\widehat{V} W} \leq \frac{1 - \rho_v}{\rho_v} \left( \sigma_{V W}^2 -  \sigma_{\Delta W}^2 \right)$. 
This can be satisfied provided the LLM's errors in reproducing the existing measurement are sufficiently small.

\section{Additional Monte Carlo Simulations based on Congressional Legislation}\label{section: additional empirical results on congressional legislation}

In this section, we report additional Monte Carlo simulations based on the data from the Congressional Bills Project \citep[][]{CBP, CAP}. 
We first illustrate how the performance of the bias-corrected regression coefficient varies with the size of the validation data.
We further illustrate that the performance of the bias-corrected regression when the economic concept is used as a covariate in the linear regression, as described in Appendix \ref{section: linear regression with LLMs on RHS, theory}.

\subsection{Varying the Size of the Validation Data}\label{section: LLM on LHS, varying the size of the validation data}

In Section \ref{section: practical guidance with validation data} of the main text, we evaluated the performance of the plug-in regression coefficient against the bias-corrected estimator using a 5\% validation sample. 
We explore how performance varies as we vary the size of the validation sample.

For a given bill topic $V_r$, covariate $W_r$, and pair of large language model and prompting strategy, we randomly draw a sample of $5,000$ bills. 
On this random sample, we first calculate the plug-in regression coefficient $\widehat{\beta}$. 
We next randomly reveal the ground-truth label $V_r$ on 2.5\% (125 bills), 5\% (250 bills), 10\% (500 bills), 25\% (1250 bills), and 50\% (2500 bills) of the random sample of $5,000$ bills. 
We then calculate the bias-corrected coefficient $\widehat{\beta}^{debiased}$ on each validation sample. 
We repeat these steps for $1,000$ randomly sampled datasets, and we calculate the average bias of these alternative estimates for the target regression $\beta^*$ of the ground-truth concept $V_r$ on the chosen covariate $W_r$ on all 10,000 bills as well as the coverage of conventional confidence intervals. 
We repeat this exercise for each possible combination of bill topic $V_r$, linked covariate $W_r$, large language model $m$, and prompting strategy $p$. 
This allows us to summarize how the plug-in regression performs against the bias-corrected regression across a wide variety of possible regression specifications, choices of large language model and prompting strategies. 

Appendix Figure \ref{figure: congressional bills simulation, LLM on LHS, normalized bias varying validation size} illustrates the distribution of normalized bias across possible combinations of bill topic $V_r$, linked covariate $W_r$, large language model $m$, and prompting strategy $p$, as the size of the validation sample changes. 
The top panels of Appendix Tables \ref{table: congressional bills, LLM on LHS, summary statistics, GPT-3.5-turbo, varying validation sample}-\ref{table: congressional bills, LLM on LHS, summary statistics, GPT-5-nano, varying validation sample} report summary statistics for labels produced by each model respectively.
While we often see severe biases for the plug-in regression, by contrast the bias-corrected regression coefficient is on average equal to the target regression coefficient for all sizes of the validation sample. 

The bottom panels of Appendix Tables \ref{table: congressional bills, LLM on LHS, summary statistics, GPT-3.5-turbo, varying validation sample}-\ref{table: congressional bills, LLM on LHS, summary statistics, GPT-5-nano, varying validation sample} provide summary statistics of the coverage of conventional confidence intervals for the target regression. 
We see substantial coverage distortions for the plug-in regression, whereas the bias-corrected regression delivers approximately correct coverage for all sizes of the validation sample. 

Finally, Appendix Figure \ref{figure: congressional bills, LLM on LHS, MSE, varying validation sample} compares the mean square error of the bias-corrected coefficient versus the validation-sample only estimate of the target regression as we vary the size of the validation sample.
The bias-corrected coefficient obtains noticeable improvements in mean square error for the validation proportions equal to 2.5\%, 5\% and 10\%. 
The bias-corrected coefficient performs similarly to the validation-sample only estimator for validation proportions equal to 25\%, although it is likely unrealistic that the researcher would collect such large validation samples in an empirical application.

\begin{figure}[htbp!]
\includegraphics[width=\textwidth]{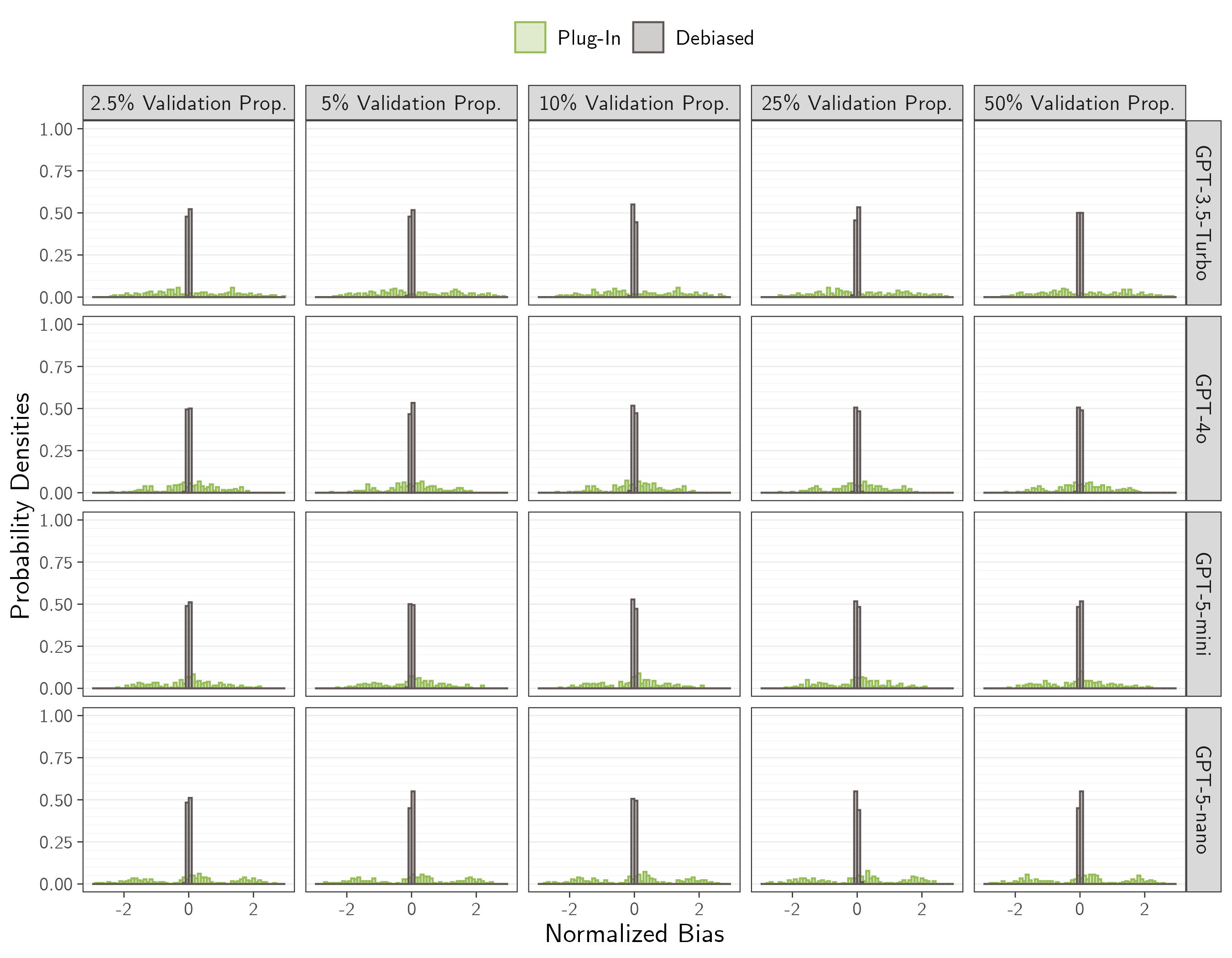}
\caption{Normalized bias of the plug-in regression and bias-corrected regression across Monte Carlo simulations based on congressional legislation as the validation sample size varies.}
\floatfoot{\textit{Notes}: The normalized bias reports the average bias of the plug-in regression coefficient $\widehat{\beta}$ and the bias-corrected coefficient $\widehat{\beta}^{debiased}$ for the target regression coefficient divided by their respective standard deviations across simulations. 
For each combination of model topic $V_r$, covariate $W_r$, large language model $m$ and prompting strategy $p$, we randomly sample $5,000$ Congressional bills and calculate the plug-in regression coefficient $\widehat{\beta}$ and the bias-corrected regression coefficient $\widehat{\beta}^{debiased}$. We vary the size of the validation sample over 2.5\%, 5\%, 10\%, 25\% and 50\%.
Results are averaged over $1,000$ simulations.
We summarize the distribution of normalized bias and coverage across regression specifications, choice of large language model and prompting strategies.
See Appendix \ref{section: LLM on LHS, varying the size of the validation data}.
}
\label{figure: congressional bills simulation, LLM on LHS, normalized bias varying validation size}
\end{figure}

\begin{table}[htbp!]
\begin{subtable}[t]{0.475\textwidth}

\begin{tabular}{r|rrr}
\toprule
Validation Prop. & Median & 5\% & 95\%\\
\midrule
\multicolumn{1}{l|}{\textit{Normalized Bias}}\\
2.5\% & -0.015 & -1.907 & 2.159\\
5\% & -0.023 & -1.899 & 2.211\\
10\% & -0.005 & -1.863 & 2.179\\
25\% & 0.007 & -1.837 & 2.264\\
50\% & -0.004 & -1.864 & 2.172\\
\midrule
\multicolumn{1}{l|}{\textit{Coverage}}\\
2.5\% & 0.806 & 0.381 & 0.946\\
5\% & 0.820 & 0.381 & 0.945\\
10\% & 0.812 & 0.383 & 0.949\\
25\% & 0.815 & 0.362 & 0.945\\
50\% & 0.816 & 0.369 & 0.950\\
\bottomrule
\end{tabular}

\caption{Plug-in regression}
\end{subtable}
\hfill
\begin{subtable}[t]{0.475\textwidth}

\begin{tabular}{r|rrr}
\toprule
Validation Prop. & Median & 5\% & 95\%\\
\midrule
\multicolumn{1}{l|}{\textit{Normalized Bias}}\\
2.5\% & 0.003 & -0.039 & 0.050\\
5\% & 0.001 & -0.055 & 0.066\\
10\% & -0.005 & -0.053 & 0.049\\
25\% & 0.002 & -0.044 & 0.053\\
50\% & 0.000 & -0.050 & 0.049\\
\midrule
\multicolumn{1}{l|}{\textit{Coverage}}\\
2.5\% & 0.901 & 0.862 & 0.927\\
5\% & 0.930 & 0.910 & 0.945\\
10\% & 0.941 & 0.927 & 0.952\\
25\% & 0.946 & 0.934 & 0.957\\
50\% & 0.948 & 0.934 & 0.959\\
\bottomrule
\end{tabular}

\caption{Debiased regression}
\end{subtable}
\caption{Summary statistics for normalized bias and coverage for Monte Carlo simulations on congressional legislation for GPT-3.5-Turbo, varying the size of the validation sample}
\floatfoot{\textit{Notes}: The normalized bias reports the average bias of the plug-in regression coefficient $\widehat{\beta}$ and the bias-corrected coefficient $\widehat{\beta}^{debiased}$ for the target regression coefficient divided by their respective standard deviations across simulations. 
The coverage reports the fraction of simulations in which a 95\% nominal confidence interval centered around the plug-in regression coefficient $\widehat{\beta}$ and the bias-corrected coefficient $\widehat{\beta}^{debiased}$ cover the target regression coefficient $\beta^*$.
For each combination of model topic $V_r$, covariate $W_r$, large language model $m$ and prompting strategy $p$, we randomly sample $5,000$ Congressional bills and calculate the plug-in regression coefficient $\widehat{\beta}$ and the bias-corrected regression coefficient $\widehat{\beta}^{debiased}$. We vary the size of the validation sample over 2.5\%, 5\%, 10\%, 25\% and 50\%.
Results are averaged over $1,000$ simulations.
We summarize the distribution of normalized bias and coverage across regression specifications, choice of large language model and prompting strategies. See Appendix \ref{section: LLM on LHS, varying the size of the validation data}.
}
\label{table: congressional bills, LLM on LHS, summary statistics, GPT-3.5-turbo, varying validation sample}
\end{table}

\begin{table}[htbp!]
\begin{subtable}[t]{0.475\textwidth}

\begin{tabular}{r|rrr}
\toprule
Validation Prop. & Median & 5\% & 95\%\\
\midrule
\multicolumn{1}{l|}{\textit{Normalized Bias}}\\
2.5\% & 0.051 & -1.456 & 1.540\\
5\% & 0.084 & -1.411 & 1.514\\
10\% & 0.059 & -1.447 & 1.507\\
25\% & 0.056 & -1.422 & 1.463\\
50\% & 0.070 & -1.441 & 1.510\\
\midrule
\multicolumn{1}{l|}{\textit{Coverage}}\\
2.5\% & 0.920 & 0.630 & 0.954\\
5\% & 0.920 & 0.637 & 0.954\\
10\% & 0.919 & 0.642 & 0.952\\
25\% & 0.920 & 0.625 & 0.954\\
50\% & 0.919 & 0.635 & 0.950\\
\bottomrule
\end{tabular}

\caption{Plug-in regression}
\end{subtable}
\hfill
\begin{subtable}[t]{0.475\textwidth}

\begin{tabular}{r|rrr}
\toprule
Validation Prop. & Median & 5\% & 95\%\\
\midrule
\multicolumn{1}{l|}{\textit{Normalized Bias}}\\
2.5\% & 0.000 & -0.058 & 0.046\\
5\% & 0.001 & -0.055 & 0.054\\
10\% & 0.000 & -0.066 & 0.050\\
25\% & -0.001 & -0.053 & 0.060\\
50\% & -0.001 & -0.045 & 0.059\\
\midrule
\multicolumn{1}{l|}{\textit{Coverage}}\\
2.5\% & 0.893 & 0.846 & 0.926\\
5\% & 0.927 & 0.902 & 0.945\\
10\% & 0.941 & 0.926 & 0.953\\
25\% & 0.946 & 0.934 & 0.958\\
50\% & 0.948 & 0.935 & 0.959\\
\bottomrule
\end{tabular}

\caption{Debiased regression}
\end{subtable}
\caption{Summary statistics for normalized bias and coverage for Monte Carlo simulations on congressional legislation for GPT-4o, varying the size of the validation sample.}
\floatfoot{\textit{Notes}: The normalized bias reports the average bias of the plug-in regression coefficient $\widehat{\beta}$ and the debiased coefficient $\widehat{\beta}^{debiased}$ for the target regression coefficient divided by their respective standard deviations across simulations. 
The coverage reports the fraction of simulations in which a 95\% nominal confidence interval centered around the plug-in regression coefficient $\widehat{\beta}$ and the debiased coefficient $\widehat{\beta}^{debiased}$ cover the target regression coefficient.
For each combination of model topic $V_r$, covariate $W_r$, large language model $m$ and prompting strategy $p$, we randomly sample $5,000$ Congressional bills and calculate the plug-in regression $\widehat{V}_{r}^{m,p} = \alpha + \beta W_{r}$ and the debiased regression coefficient. 
We vary the size of the validation sample over 2.5\%, 5\%, 10\%, 25\% and 50\%.
Results are averaged over $1,000$ simulations.
See Appendix \ref{section: LLM on LHS, varying the size of the validation data}.
}
\label{table: congressional bills, LLM on LHS, summary statistics, GPT-4o, varying validation sample}
\end{table}

\begin{table}[htbp!]
\begin{subtable}[t]{0.475\textwidth}

\begin{tabular}{r|rrr}
\toprule
Validation Prop. & Median & 5\% & 95\%\\
\midrule
\multicolumn{1}{l|}{\textit{Normalized Bias}}\\
2.5\% & 0.062 & -1.612 & 1.566\\
5\% & 0.030 & -1.646 & 1.567\\
10\% & 0.058 & -1.624 & 1.545\\
25\% & 0.040 & -1.569 & 1.533\\
50\% & 0.040 & -1.619 & 1.572\\
\midrule
\multicolumn{1}{l|}{\textit{Coverage}}\\
2.5\% & 0.907 & 0.570 & 0.954\\
5\% & 0.906 & 0.589 & 0.953\\
10\% & 0.909 & 0.558 & 0.954\\
25\% & 0.900 & 0.585 & 0.955\\
50\% & 0.905 & 0.583 & 0.954\\
\bottomrule
\end{tabular}

\caption{Plug-in regression}
\end{subtable}
\hfill
\begin{subtable}[t]{0.475\textwidth}

\begin{tabular}{r|rrr}
\toprule
Validation Prop. & Median & 5\% & 95\%\\
\midrule
\multicolumn{1}{l|}{\textit{Normalized Bias}}\\
2.5\% & 0.000 & -0.041 & 0.051\\
5\% & -0.002 & -0.052 & 0.054\\
10\% & -0.002 & -0.066 & 0.049\\
25\% & -0.003 & -0.050 & 0.055\\
50\% & 0.003 & -0.048 & 0.047\\
\midrule
\multicolumn{1}{l|}{\textit{Coverage}}\\
2.5\% & 0.895 & 0.843 & 0.925\\
5\% & 0.927 & 0.903 & 0.945\\
10\% & 0.939 & 0.925 & 0.952\\
25\% & 0.947 & 0.936 & 0.958\\
50\% & 0.948 & 0.934 & 0.957\\
\bottomrule
\end{tabular}

\caption{Debiased regression}
\end{subtable}
\caption{Summary statistics for normalized bias and coverage for Monte Carlo simulations on congressional legislation for GPT-5-mini, varying the size of the validation sample.}
\floatfoot{\textit{Notes}: The normalized bias reports the average bias of the plug-in regression coefficient $\widehat{\beta}$ and the debiased coefficient $\widehat{\beta}^{debiased}$ for the target regression coefficient divided by their respective standard deviations across simulations. 
The coverage reports the fraction of simulations in which a 95\% nominal confidence interval centered around the plug-in regression coefficient $\widehat{\beta}$ and the debiased coefficient $\widehat{\beta}^{debiased}$ cover the target regression coefficient.
For each combination of model topic $V_r$, covariate $W_r$, large language model $m$ and prompting strategy $p$, we randomly sample $5,000$ Congressional bills and calculate the plug-in regression $\widehat{V}_{r}^{m,p} = \alpha + \beta W_{r}$ and the debiased regression coefficient. 
We vary the size of the validation sample over 2.5\%, 5\%, 10\%, 25\% and 50\%.
Results are averaged over $1,000$ simulations.
See Appendix \ref{section: LLM on LHS, varying the size of the validation data}.
}
\label{table: congressional bills, LLM on LHS, summary statistics, GPT-5-mini, varying validation sample}
\end{table}

\begin{table}[htbp!]
\begin{subtable}[t]{0.475\textwidth}

\begin{tabular}{r|rrr}
\toprule
Validation Prop. & Median & 5\% & 95\%\\
\midrule
\multicolumn{1}{l|}{\textit{Normalized Bias}}\\
2.5\% & 0.151 & -2.072 & 2.113\\
5\% & 0.168 & -2.079 & 2.092\\
10\% & 0.141 & -2.017 & 2.116\\
25\% & 0.160 & -2.058 & 2.085\\
50\% & 0.171 & -2.021 & 2.123\\
\midrule
\multicolumn{1}{l|}{\textit{Coverage}}\\
2.5\% & 0.771 & 0.379 & 0.952\\
5\% & 0.779 & 0.387 & 0.953\\
10\% & 0.781 & 0.361 & 0.951\\
25\% & 0.780 & 0.390 & 0.951\\
50\% & 0.785 & 0.377 & 0.956\\
\bottomrule
\end{tabular}

\caption{Plug-in regression}
\end{subtable}
\hfill
\begin{subtable}[t]{0.475\textwidth}

\begin{tabular}{r|rrr}
\toprule
Validation Prop. & Median & 5\% & 95\%\\
\midrule
\multicolumn{1}{l|}{\textit{Normalized Bias}}\\
2.5\% & 0.000 & -0.050 & 0.042\\
5\% & 0.005 & -0.052 & 0.060\\
10\% & -0.001 & -0.052 & 0.054\\
25\% & -0.002 & -0.058 & 0.062\\
50\% & 0.002 & -0.052 & 0.054\\
\midrule
\multicolumn{1}{l|}{\textit{Coverage}}\\
2.5\% & 0.902 & 0.850 & 0.929\\
5\% & 0.930 & 0.906 & 0.946\\
10\% & 0.942 & 0.927 & 0.955\\
25\% & 0.947 & 0.935 & 0.959\\
50\% & 0.946 & 0.935 & 0.958\\
\bottomrule
\end{tabular}

\caption{Debiased regression}
\end{subtable}
\caption{Summary statistics for normalized bias and coverage for Monte Carlo simulations on congressional legislation for GPT-5-nano, varying the size of the validation sample.}
\floatfoot{\textit{Notes}: The normalized bias reports the average bias of the plug-in regression coefficient $\widehat{\beta}$ and the debiased coefficient $\widehat{\beta}^{debiased}$ for the target regression coefficient divided by their respective standard deviations across simulations. 
The coverage reports the fraction of simulations in which a 95\% nominal confidence interval centered around the plug-in regression coefficient $\widehat{\beta}$ and the debiased coefficient $\widehat{\beta}^{debiased}$ cover the target regression coefficient.
For each combination of model topic $V_r$, covariate $W_r$, large language model $m$ and prompting strategy $p$, we randomly sample $5,000$ Congressional bills and calculate the plug-in regression $\widehat{V}_{r}^{m,p} = \alpha + \beta W_{r}$ and the debiased regression coefficient. 
We vary the size of the validation sample over 2.5\%, 5\%, 10\%, 25\% and 50\%.
Results are averaged over $1,000$ simulations.
See Appendix \ref{section: LLM on LHS, varying the size of the validation data}.
}
\label{table: congressional bills, LLM on LHS, summary statistics, GPT-5-nano, varying validation sample}
\end{table}

\begin{figure}[htbp!]
\includegraphics[width=\textwidth]{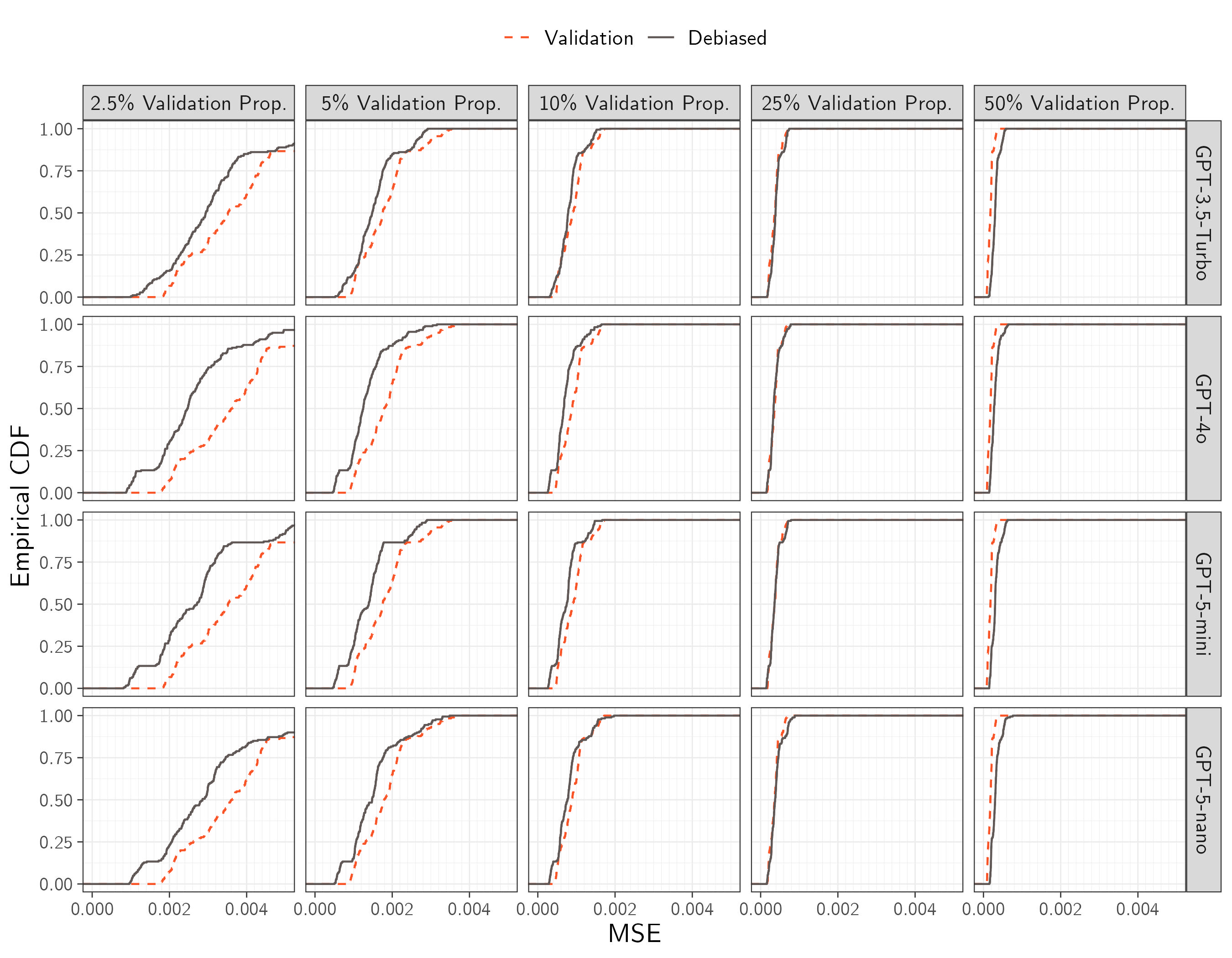}
\caption{Cumulative distribution function of mean square error for the bias-corrected estimator against validation-sample only estimator, varying the size of the validation sample.}
\floatfoot{\textit{Notes}: 
For each combination of model topic $V_r$, covariate $W_r$, large language model $m$ and prompting strategy $p$, we randomly sample $5,000$ Congressional bills and calculate the bias-corrected regression coefficient $\widehat{\beta}^{debiased}$ and the validation-sample only regression coefficient $\widehat{\beta}^*$.
We calculate the mean square error of $\widehat{\beta}^{debiased}$ and $\widehat{\beta}^*$ for the target regression. 
We vary the size of the validation sample over 2.5\%, 5\%, 10\%, 25\% and 50\%, and we average the results over $1,000$ simulations.
We summarize the distribution of average mean square error across regression specifications, choice of large language model and prompting strategies.
See Appendix \ref{section: LLM on LHS, varying the size of the validation data}.
}
\label{figure: congressional bills, LLM on LHS, MSE, varying validation sample}
\end{figure}

\clearpage
\subsection{Large Language Model Labels as Covariates}\label{section: LLM on RHS simulations}

In this section, we extend our analysis using data from the Congressional Bills Project to explore the biases that can arise from using large language model labels as covariates in a linear regression and whether the resulting biases can be corrected using a small collection of validation data. 

We use the same random sample of $10,000$ Congressional bills from the main text, and we now regress alternative linked economic variables on dummy indicators for the large language model's labeled economic concept -- in this case, the policy topic of the bill. For alternative dependent variables such as whether the bill's sponsor was a Democrat, whether the bill originated in the Senate, and the DW1 score of the bill's sponsor, we run the regression $W_r = \widehat{V}_{r}^{m,p} \beta + \epsilon$ for each possible pair of large language model $m$ and prompting strategy $p$. 
In Appendix Figure \ref{figure: congressional bills, LLM on RHS, t stat variation}, each row considers a different regression for a linked covariate $W_r$ as the dependent variable, and each column plots the t-statistic for different large language model labels $\widehat{V}_{r}^{m,p}$ associated with alternative bill topics.
For every combination of the linked variable $W_r$ and policy topic area, we see substantial variation in the t-statistics across alternative large language models and prompting strategies.
Appendix Table \ref{table: congressional bills, coeff variation, rhs} summarizes the coefficient estimates across models and prompts for each choice of labeled policy topic and the covariate.

\begin{figure}[htb]
\includegraphics[width=\textwidth]{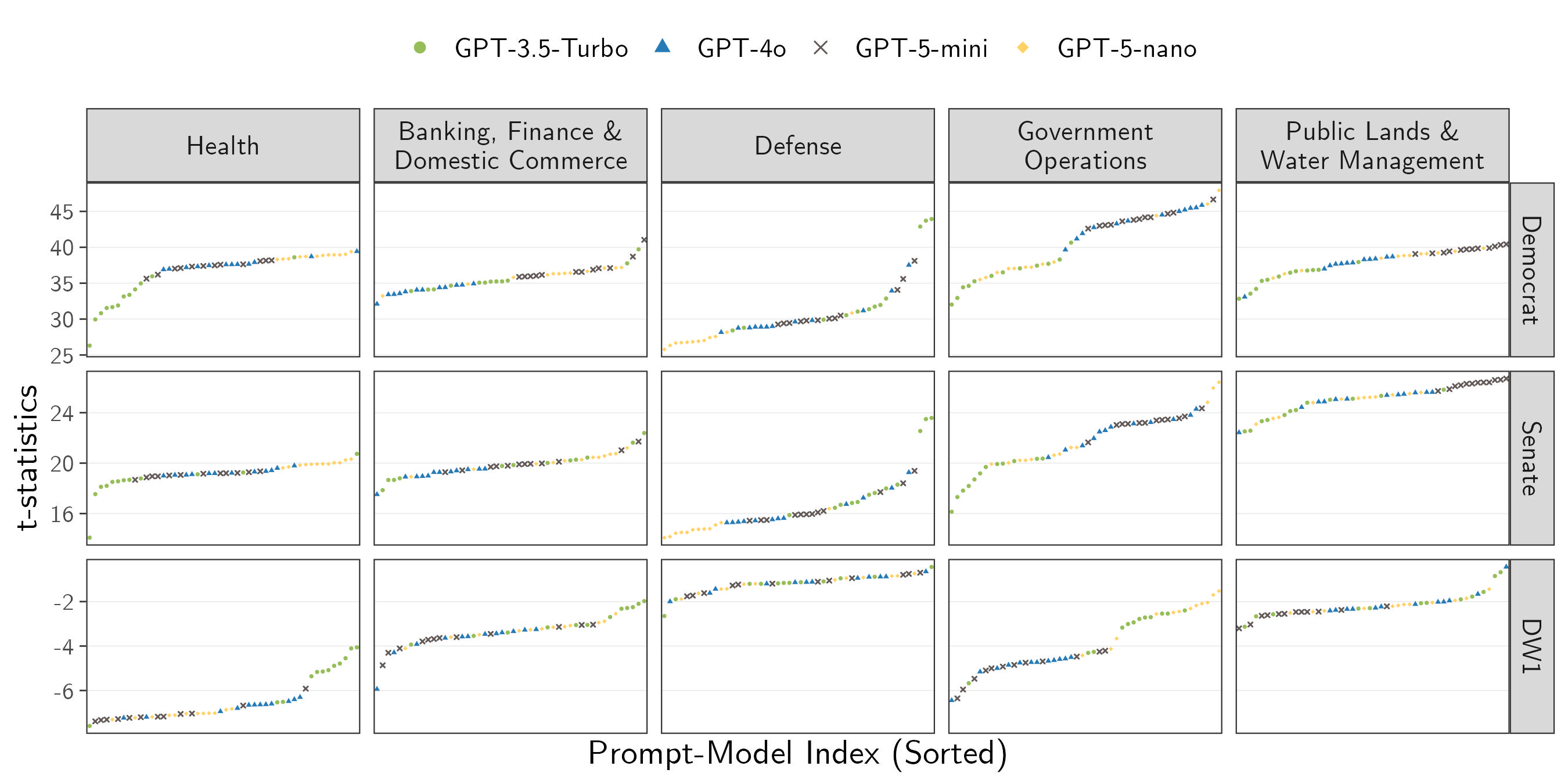}
\caption{Variation in t-statistics across large language models and prompting strategies on congressional legislation, using the economic concept as a covariate.}
\floatfoot{\textit{Notes}: 
On 10,000 Congressional bills, we prompt GPT-3.5-Turbo, GPT-4o, GPT-5-mini, and GPT-5-nano to label each description for its policy topic area using alternative prompting strategies. 
For each model $m$ and prompt $p$, we regress a linked variable $W_r$ on indicators $\widehat{V}_r^{m,p}$ for the large language model's labeled policy topic. 
In each subplot, the t-statistic estimates are sorted in ascending order for clarity.
See Appendix \ref{section: LLM on RHS simulations}.
}
\label{figure: congressional bills, LLM on RHS, t stat variation}
\end{figure}

\begin{table}[htb!]

\begin{tabular}{llrrrrr}
\toprule
\multirow{2}{*}{Covariate} & \multirow{2}{*}{Policy Topic} & \multicolumn{4}{c}{\textit{Point Estimates}} & Sample\\
\cmidrule(l{3pt}r{3pt}){3-6}
& & Mean & Median & 5\% & 95\% & Average\\
\midrule
DW1 & Health & -0.096 & -0.097 & -0.105 & -0.078 & -0.062\\
DW1 & Banking, Finance \& Domestic Com. & -0.043 & -0.043 & -0.053 & -0.029 & -0.062\\
DW1 & Defense & -0.016 & -0.016 & -0.025 & -0.010 & -0.062\\
DW1 & Government Operations & -0.043 & -0.046 & -0.064 & -0.024 & -0.062\\
DW1 & Public Lands \& Water Management & -0.025 & -0.026 & -0.033 & -0.013 & -0.062\\
\midrule
Democrat & Health & 0.643 & 0.646 & 0.617 & 0.655 & 0.604\\
Democrat & Banking, Finance \& Domestic Com. & 0.579 & 0.579 & 0.568 & 0.593 & 0.604\\
Democrat & Defense & 0.588 & 0.588 & 0.577 & 0.602 & 0.604\\
Democrat & Government Operations & 0.594 & 0.596 & 0.571 & 0.614 & 0.604\\
Democrat & Public Lands \& Water Management & 0.589 & 0.588 & 0.581 & 0.598 & 0.604\\
\midrule
Senate & Health & 0.331 & 0.327 & 0.320 & 0.359 & 0.317\\
Senate & Banking, Finance \& Domestic Com. & 0.299 & 0.298 & 0.282 & 0.313 & 0.317\\
Senate & Defense & 0.293 & 0.294 & 0.276 & 0.308 & 0.317\\
Senate & Government Operations & 0.292 & 0.292 & 0.282 & 0.305 & 0.317\\
Senate & Public Lands \& Water Management & 0.386 & 0.386 & 0.372 & 0.398 & 0.317\\
\bottomrule
\end{tabular}

\caption{Variation in point estimates across large language models and prompting strategies on Congressional bills, using the economic concept as a covariate.}
\floatfoot{\textit{Notes}:
On 10,000 Congressional bills, we prompt GPT-3.5-Turbo, GPT-4o, GPT-5-mini, and GPT-5-nano to label each description for its policy topic area using alternative prompting strategies. 
For each model $m$ and prompt $p$, we regress a linked variable $W_r$ on indicators for whether the large language model labeled a particular policy topic $1\{\widehat{V}_r^{m,p} = v\}$. 
The final column (``Sample Average'') reports the average of the linked variable $W_r$ across all Congressional bills.
See Appendix \ref{section: LLM on RHS simulations}.
}
\label{table: congressional bills, coeff variation, rhs}
\end{table}

We next explore whether these biases can be addressed by collecting a small validation sample and implementing the bias-corrected procedure described in Appendix \ref{section: linear regression with LLMs on RHS, theory} can address these issues. 
We leverage the same Monte Carlo simulation design as described in Section \ref{section: practical guidance with validation data} of the main text.

For each linked variable $W_r$ and pair of large language model $m$ and prompting strategy $p$, we randomly sample $5,000$ bills from our dataset of $10,000$ bills. 
On this random sample of $5,000$ bills, we calculate the plug-in coefficients $\widehat{\beta}$ byregressing $W_r$ on $\widehat{V}_{r}^{m,p}$ (for $\widehat{V}_{r}^{m,p}$ a vector of indicators for the labeled policy topic). 
We next randomly reveal the ground-truth label $V_r$ on 5\% of our random sample of $5,000$ bills, which produces a validation sample. 
We calculate the bias-corrected coefficients $\widehat{\beta}^{debiased}$ as described in Appendix \ref{section: linear regression with LLMs on RHS, theory}.
We repeat these steps for $1,000$ randomly sampled datasets. 
We repeat this exercise for each possible combination of linked variable $W_r$, large language model $m$ (either GPT-3.5-turbo or GPT-4o) and prompting strategy $p$. 
This allows to summarize how the plug-in regression performs against the bias-corrected regression across a wide variety of possible regression specifications, choices of large language model and prompting strategies. 

Appendix Figure \ref{figure: normalized bias, LLM on RHS, validation prop 5} and Appendix Table \ref{table: congressional bills, LLM on RHS, summary statistics, 5 percent validation} summarizes our results. 
The plug-in regression suffers from substantial biases for almost all combinations of linked variable $W_r$, large language model $m$, and prompting strategy $p$. 
By contrast, using the validation sample for bias correction effectively eliminates these biases. 
Furthermore, the bottom panel of Appendix Table \ref{table: congressional bills, LLM on RHS, summary statistics, 5 percent validation} further illustrates the coverage comparison between the plug-in regression and the bias-corrected estimator --- while confidence intervals centered at the plug-in regression are significantly distorted, bias-correction restores nominal coverage.

Finally, Appendix Figure \ref{figure: congressional bills, LLM on RHS, MSE, 5 percent validation} compares the mean square error of the bias-corrected regression against directly estimating the target regression on the validation sample. 
For many regression specifications, choices of language model and prompting strategies, we again find that the MSE of the bias-corrected regression is smaller than that of the validation-sample only regression. 

\begin{figure}[htbp!]
\includegraphics[width=\textwidth]{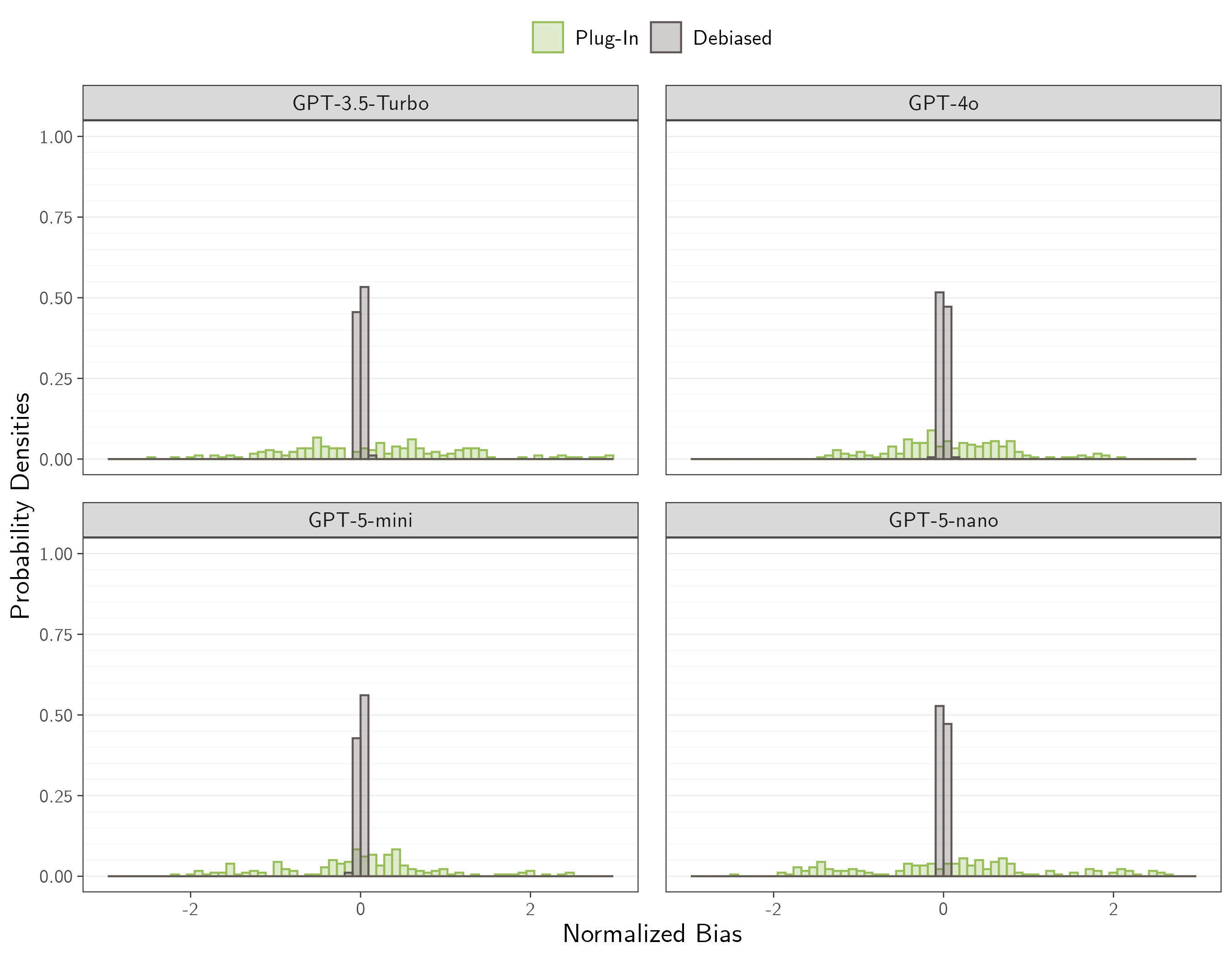}
\caption{Normalized bias of the plug-in regression and bias-corrected regression using policy topic as a covariate across Monte Carlo simulations based on congressional legislation}
\floatfoot{\textit{Notes}: 
The normalized bias reports the average bias of the plug-in regression coefficient $\widehat{\beta}$ and the bias-corrected coefficient $\widehat{\beta}^{debiased}$ for the target regression coefficient divided by their respective standard deviations across simulations.  
For each combination of left hand side variable $W_r$, large language model $m$ and prompting strategy $p$, we randomly sample $5,000$ Congressional bills and calculate the plug-in regression $\widehat{\beta}$ and bias-corrected coefficient $\widehat{\beta}^*$ with the policy topic as a covariate using a 5\% validation sample.
Results are averaged over $1,000$ simulations.
See Appendix \ref{section: LLM on RHS simulations}.
}
\label{figure: normalized bias, LLM on RHS, validation prop 5}
\end{figure}

\begin{table}[htbp!]
\begin{subtable}[t]{0.475\textwidth}

\begin{tabular}{r|rrr}
\toprule
  & Median & 5\% & 95\%\\
\midrule
\multicolumn{1}{l|}{\textit{Normalized Bias}}\\
Plug-In & 0.144 & -1.514 & 2.083\\
Debiased & 0.003 & -0.051 & 0.060\\
\midrule
\multicolumn{1}{l|}{\textit{Coverage}}\\
Plug-In & 0.901 & 0.363 & 0.950\\
Debiased & 0.933 & 0.907 & 0.955\\
\bottomrule
\end{tabular}

\caption{GPT-3.5-Turbo}
\end{subtable}
\hfill
\begin{subtable}[t]{0.475\textwidth}

\begin{tabular}{r|rrr}
\toprule
  & Median & 5\% & 95\%\\
\midrule
\multicolumn{1}{l|}{\textit{Normalized Bias}}\\
Plug-In & 0.042 & -1.146 & 1.558\\
Debiased & -0.003 & -0.063 & 0.053\\
\midrule
\multicolumn{1}{l|}{\textit{Coverage}}\\
Plug-In & 0.928 & 0.640 & 0.957\\
Debiased & 0.930 & 0.900 & 0.952\\
\bottomrule
\end{tabular}

\caption{GPT-4o}
\end{subtable} \\
\begin{subtable}[t]{0.475\textwidth}

\begin{tabular}{r|rrr}
\toprule
  & Median & 5\% & 95\%\\
\midrule
\multicolumn{1}{l|}{\textit{Normalized Bias}}\\
Plug-In & 0.034 & -1.615 & 1.828\\
Debiased & 0.003 & -0.058 & 0.058\\
\midrule
\multicolumn{1}{l|}{\textit{Coverage}}\\
Plug-In & 0.927 & 0.521 & 0.953\\
Debiased & 0.932 & 0.906 & 0.955\\
\bottomrule
\end{tabular}

\caption{GPT-5-mini}
\end{subtable}
\hfill
\begin{subtable}[t]{0.475\textwidth}

\begin{tabular}{r|rrr}
\toprule
  & Median & 5\% & 95\%\\
\midrule
\multicolumn{1}{l|}{\textit{Normalized Bias}}\\
Plug-In & 0.126 & -1.664 & 2.172\\
Debiased & -0.002 & -0.064 & 0.058\\
\midrule
\multicolumn{1}{l|}{\textit{Coverage}}\\
Plug-In & 0.900 & 0.405 & 0.950\\
Debiased & 0.934 & 0.908 & 0.958\\
\bottomrule
\end{tabular}

\caption{GPT-5-nano}
\end{subtable}
\caption{Summary statistics for normalized bias and coverage for Monte Carlo simulations on congressional legislation using policy topic as a covariate.}
\floatfoot{\textit{Notes}: The normalized bias reports the average bias of the plug-in regression coefficient $\widehat{\beta}$ and the debiased coefficient $\widehat{\beta}^{debiased}$ for the target regression coefficient divided by their respective standard deviations across simulations. 
The coverage reports the fraction of simulations in which a 95\% nominal confidence interval centered around the plug-in regression coefficient $\widehat{\beta}$ and the debiased coefficient $\widehat{\beta}^{debiased}$ cover the target regression coefficient.
For each combination of left hand side variable $W_r$, large language model $m$ and prompting strategy $p$, we randomly sample $5,000$ Congressional bills and calculate the plug-in regression $\widehat{\beta}$ and bias-corrected coefficient $\widehat{\beta}^*$ with the policy topic as a covariate using a 5\% validation sample.
Results are averaged over $1,000$ simulations.
See Appendix \ref{section: LLM on RHS simulations}.}
\label{table: congressional bills, LLM on RHS, summary statistics, 5 percent validation}
\end{table}

\begin{figure}[htbp!]
\includegraphics[width=\textwidth]{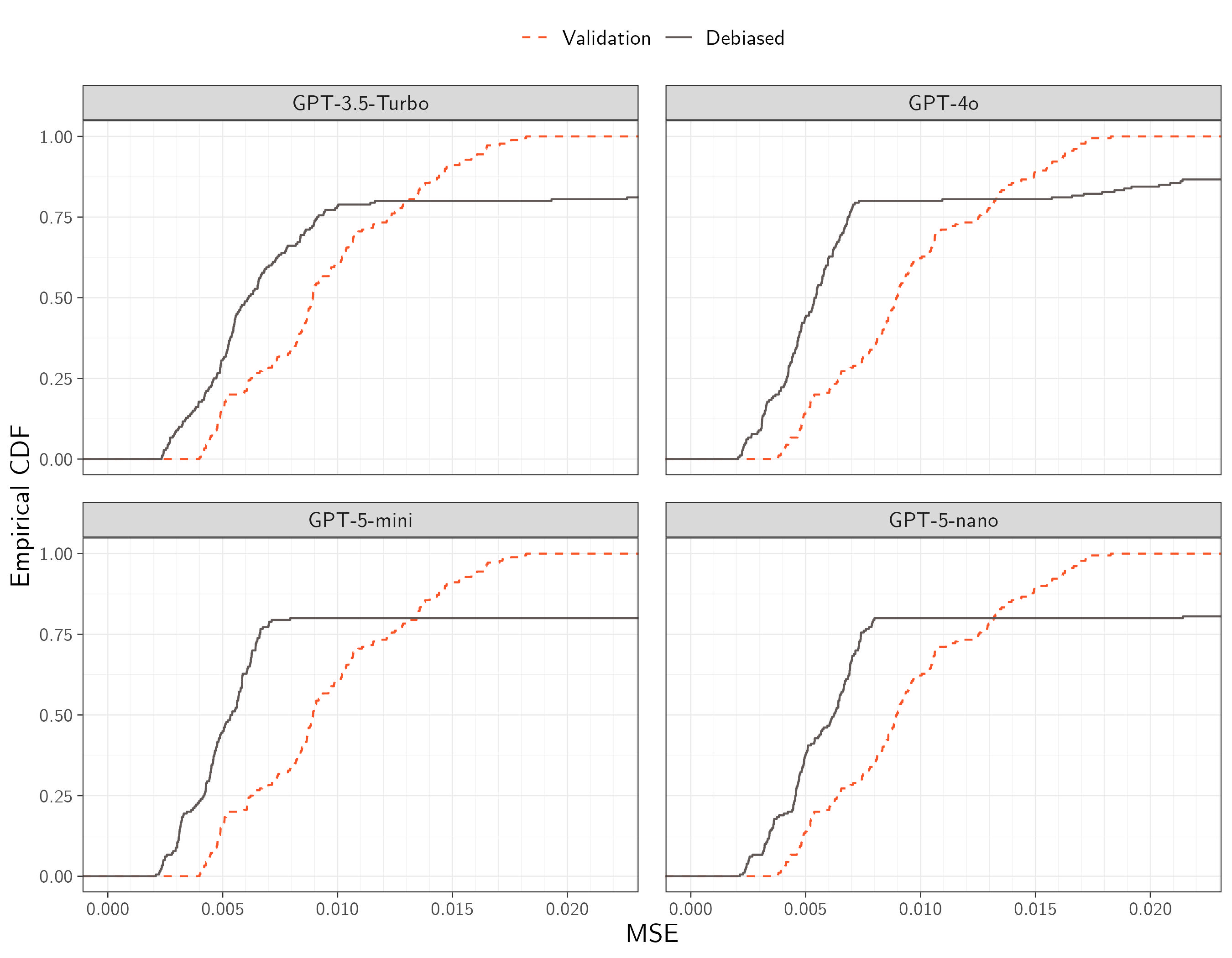}
\caption{Cumulative distribution function of mean square error for the bias-corrected estimator against validation-sample only estimator using policy topic as a covariate.}
\floatfoot{\textit{Notes}: For each combination of left hand side variable $W_r$, large language model $m$ and prompting strategy $p$, we randomly sample $5,000$ Congressional bills and calculate the plug-in regression $\widehat{\beta}$ and bias-corrected coefficient $\widehat{\beta}^*$ with the policy topic as a covariate using a 5\% validation sample. 
We calculate the mean square error of $\widehat{\beta}^{debiased}$ and $\widehat{\beta}^*$ for the target regression $\beta^*$.
Results are averaged over $1,000$ simulations.
We summarize the distribution of average mean square error across regression specifications, choice of large language model and prompting strategies.
See Appendix \ref{section: LLM on RHS simulations}.}
\label{figure: congressional bills, LLM on RHS, MSE, 5 percent validation}
\end{figure}

\subsubsection{Varying the Size of the Validation Data:}
We evaluated the performance of bias-correcting linear regression that use large language model labels as covariates using a 5\% validation sample. 
We finally explore how the performance of the bias-corrected regression coefficient varies as we vary the size of the validation sample. 
We repeat our Monte Carlo simulations now varying the size of the validation sample by randomly revealing the measurements $V_r$ on 2.5\% (125 bills), 5\% (250 bills), 10\% (500 bills), 25\% (1250 bills), and 50\% (2500 bills) of the random sample of $5,000$ bills.
The results are summarized in Appendix Figure \ref{figure: normalized bias, LLM on RHS, validation all prop}, Appendix Tables \ref{table: congressional bills, LLM on RHS, summary statistics, GPT-3.5-turbo, varying validation sample}-\ref{table: congressional bills, LLM on RHS, summary statistics, GPT-5-nano, varying validation sample} and Appendix Figure \ref{figure: congressional bills, LLM on RHS, MSE, all prop}. 
We continue to find that the bias-corrected regression performs well in finite samples, even when the validation sample only contains $125$ bills.

\begin{figure}[htbp!]
\includegraphics[width=\textwidth]{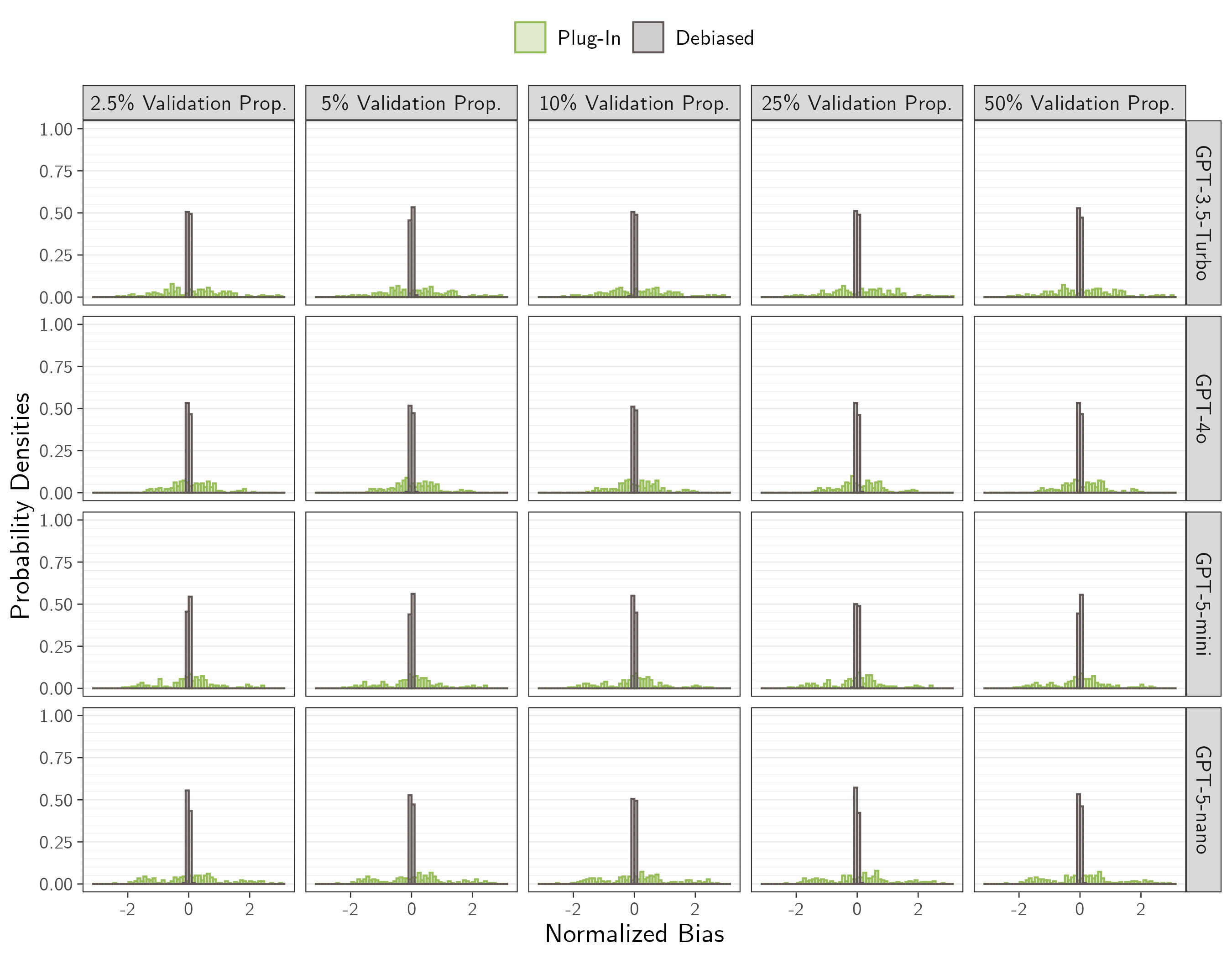}
\caption{Normalized bias of the plug-in regression and bias-corrected regression using policy topic as a covariate as the validation sample size varies.}
\floatfoot{\textit{Notes}: 
The normalized bias reports the average bias of the plug-in regression coefficient $\widehat{\beta}$ and the bias-corrected coefficient $\widehat{\beta}^{debiased}$ for the target regression coefficient divided by their respective standard deviations across simulations.  
For each combination of left hand side variable $W_r$, large language model $m$ and prompting strategy $p$, we randomly sample $5,000$ Congressional bills and calculate the plug-in regression $\widehat{\beta}$ and bias-corrected coefficient $\widehat{\beta}^*$ with the policy topic as a covariate.
We vary the size of the validation sample over 2.5\%, 5\%, 10\%, 25\% and 50\%. 
We average the results over $1,000$ simulations.
Results are averaged over $1,000$ simulations.
See Appendix \ref{section: LLM on RHS simulations}.
}
\label{figure: normalized bias, LLM on RHS, validation all prop}
\end{figure}

\begin{table}[htbp!]
\begin{subtable}[t]{0.475\textwidth}

\begin{tabular}{r|rrr}
\toprule
Validation Prop. & Median & 5\% & 95\%\\
\midrule
\multicolumn{1}{l|}{\textit{Normalized Bias}}\\
2.5\% & 0.120 & -1.376 & 2.030\\
5\% & 0.144 & -1.514 & 2.083\\
10\% & 0.147 & -1.437 & 2.099\\
25\% & 0.136 & -1.469 & 2.024\\
50\% & 0.127 & -1.486 & 2.043\\
\midrule
\multicolumn{1}{l|}{\textit{Coverage}}\\
2.5\% & 0.900 & 0.373 & 0.951\\
5\% & 0.901 & 0.363 & 0.950\\
10\% & 0.897 & 0.353 & 0.949\\
25\% & 0.893 & 0.391 & 0.948\\
50\% & 0.893 & 0.368 & 0.951\\
\bottomrule
\end{tabular}

\caption{Plug-in regression}
\end{subtable}
\hfill
\begin{subtable}[t]{0.475\textwidth}

\begin{tabular}{r|rrr}
\toprule
Validation Prop. & Median & 5\% & 95\%\\
\midrule
\multicolumn{1}{l|}{\textit{Normalized Bias}}\\
2.5\% & 0.000 & -0.057 & 0.050\\
5\% & 0.003 & -0.051 & 0.060\\
10\% & -0.002 & -0.055 & 0.057\\
25\% & -0.001 & -0.053 & 0.052\\
50\% & -0.002 & -0.055 & 0.056\\
\midrule
\multicolumn{1}{l|}{\textit{Coverage}}\\
2.5\% & 0.909 & 0.868 & 0.959\\
5\% & 0.933 & 0.907 & 0.955\\
10\% & 0.941 & 0.927 & 0.954\\
25\% & 0.946 & 0.935 & 0.958\\
50\% & 0.947 & 0.936 & 0.959\\
\bottomrule
\end{tabular}

\caption{Debiased regression}
\end{subtable}
\caption{Summary statistics for normalized bias and coverage using policy topic as a covariate for GPT-3.5-Turbo, varying the size of the validation sample.}
\floatfoot{\textit{Notes}: The normalized bias reports the average bias of the plug-in regression coefficient $\widehat{\beta}$ and the debiased coefficient $\widehat{\beta}^{debiased}$ for the target regression coefficient divided by their respective standard deviations across simulations. 
The coverage reports the fraction of simulations in which a 95\% nominal confidence interval centered around the plug-in regression coefficient $\widehat{\beta}$ and the debiased coefficient $\widehat{\beta}^{debiased}$ cover the target regression coefficient.
For each combination of left hand side variable $W_r$, large language model $m$ and prompting strategy $p$, we randomly sample $5,000$ Congressional bills and calculate the plug-in regression $\widehat{\beta}$ and bias-corrected coefficient $\widehat{\beta}^*$ with the policy topic as a covariate.
We vary the size of the validation sample over 2.5\%, 5\%, 10\%, 25\% and 50\%.
Results are averaged over $1,000$ simulations.
See Appendix \ref{section: LLM on RHS simulations}.
}
\label{table: congressional bills, LLM on RHS, summary statistics, GPT-3.5-turbo, varying validation sample}
\end{table}

\begin{table}[htbp!]
\begin{subtable}[t]{0.475\textwidth}

\begin{tabular}{r|rrr}
\toprule
Validation Prop. & Median & 5\% & 95\%\\
\midrule
\multicolumn{1}{l|}{\textit{Normalized Bias}}\\
2.5\% & 0.038 & -1.087 & 1.509\\
5\% & 0.042 & -1.146 & 1.558\\
10\% & 0.024 & -1.118 & 1.624\\
25\% & 0.039 & -1.145 & 1.541\\
50\% & 0.029 & -1.111 & 1.490\\
\midrule
\multicolumn{1}{l|}{\textit{Coverage}}\\
2.5\% & 0.928 & 0.675 & 0.954\\
5\% & 0.928 & 0.640 & 0.957\\
10\% & 0.927 & 0.642 & 0.952\\
25\% & 0.926 & 0.639 & 0.953\\
50\% & 0.922 & 0.648 & 0.953\\
\bottomrule
\end{tabular}

\caption{Plug-in regression}
\end{subtable}
\hfill
\begin{subtable}[t]{0.475\textwidth}

\begin{tabular}{r|rrr}
\toprule
Validation Prop. & Median & 5\% & 95\%\\
\midrule
\multicolumn{1}{l|}{\textit{Normalized Bias}}\\
2.5\% & -0.003 & -0.065 & 0.052\\
5\% & -0.003 & -0.063 & 0.053\\
10\% & -0.001 & -0.059 & 0.055\\
25\% & -0.002 & -0.062 & 0.053\\
50\% & -0.003 & -0.057 & 0.052\\
\midrule
\multicolumn{1}{l|}{\textit{Coverage}}\\
2.5\% & 0.904 & 0.860 & 0.948\\
5\% & 0.930 & 0.900 & 0.952\\
10\% & 0.943 & 0.928 & 0.952\\
25\% & 0.947 & 0.933 & 0.957\\
50\% & 0.949 & 0.935 & 0.959\\
\bottomrule
\end{tabular}

\caption{Debiased regression}
\end{subtable}
\caption{Summary statistics for normalized bias and coverage using policy topic as a covariate for GPT-4o, varying the size of the validation sample.}
\floatfoot{\textit{Notes}: The normalized bias reports the average bias of the plug-in regression coefficient $\widehat{\beta}$ and the debiased coefficient $\widehat{\beta}^{debiased}$ for the target regression coefficient divided by their respective standard deviations across simulations. 
The coverage reports the fraction of simulations in which a 95\% nominal confidence interval centered around the plug-in regression coefficient $\widehat{\beta}$ and the debiased coefficient $\widehat{\beta}^{debiased}$ cover the target regression coefficient.
For each combination of left hand side variable $W_r$, large language model $m$ and prompting strategy $p$, we randomly sample $5,000$ Congressional bills and calculate the plug-in regression $\widehat{\beta}$ and bias-corrected coefficient $\widehat{\beta}^*$ with the policy topic as a covariate.
We vary the size of the validation sample over 2.5\%, 5\%, 10\%, 25\% and 50\%.
Results are averaged over $1,000$ simulations.
See Appendix \ref{section: LLM on RHS simulations}.
}
\label{table: congressional bills, LLM on RHS, summary statistics, GPT-4o, varying validation sample}
\end{table}

\begin{table}[htbp!]
\begin{subtable}[t]{0.475\textwidth}

\begin{tabular}{r|rrr}
\toprule
Validation Prop. & Median & 5\% & 95\%\\
\midrule
\multicolumn{1}{l|}{\textit{Normalized Bias}}\\
2.5\% & 0.031 & -1.615 & 1.884\\
5\% & 0.034 & -1.615 & 1.828\\
10\% & 0.035 & -1.604 & 1.901\\
25\% & 0.041 & -1.622 & 1.857\\
50\% & 0.039 & -1.596 & 1.845\\
\midrule
\multicolumn{1}{l|}{\textit{Coverage}}\\
2.5\% & 0.926 & 0.491 & 0.955\\
5\% & 0.927 & 0.521 & 0.953\\
10\% & 0.928 & 0.510 & 0.955\\
25\% & 0.930 & 0.505 & 0.953\\
50\% & 0.929 & 0.494 & 0.956\\
\bottomrule
\end{tabular}

\caption{Plug-in regression}
\end{subtable}
\hfill
\begin{subtable}[t]{0.475\textwidth}

\begin{tabular}{r|rrr}
\toprule
Validation Prop. & Median & 5\% & 95\%\\
\midrule
\multicolumn{1}{l|}{\textit{Normalized Bias}}\\
2.5\% & 0.005 & -0.049 & 0.065\\
5\% & 0.003 & -0.058 & 0.058\\
10\% & -0.004 & -0.052 & 0.059\\
25\% & -0.001 & -0.050 & 0.052\\
50\% & 0.003 & -0.054 & 0.055\\
\midrule
\multicolumn{1}{l|}{\textit{Coverage}}\\
2.5\% & 0.904 & 0.852 & 0.958\\
5\% & 0.932 & 0.906 & 0.955\\
10\% & 0.940 & 0.926 & 0.953\\
25\% & 0.947 & 0.935 & 0.958\\
50\% & 0.947 & 0.935 & 0.958\\
\bottomrule
\end{tabular}

\caption{Debiased regression}
\end{subtable}
\caption{Summary statistics for normalized bias and coverage using policy topic as a covariate for GPT-5-mini, varying the size of the validation sample.}
\floatfoot{\textit{Notes}: The normalized bias reports the average bias of the plug-in regression coefficient $\widehat{\beta}$ and the debiased coefficient $\widehat{\beta}^{debiased}$ for the target regression coefficient divided by their respective standard deviations across simulations. 
The coverage reports the fraction of simulations in which a 95\% nominal confidence interval centered around the plug-in regression coefficient $\widehat{\beta}$ and the debiased coefficient $\widehat{\beta}^{debiased}$ cover the target regression coefficient.
For each combination of left hand side variable $W_r$, large language model $m$ and prompting strategy $p$, we randomly sample $5,000$ Congressional bills and calculate the plug-in regression $\widehat{\beta}$ and bias-corrected coefficient $\widehat{\beta}^*$ with the policy topic as a covariate.
We vary the size of the validation sample over 2.5\%, 5\%, 10\%, 25\% and 50\%.
Results are averaged over $1,000$ simulations.
See Appendix \ref{section: LLM on RHS simulations}.
}
\label{table: congressional bills, LLM on RHS, summary statistics, GPT-5-mini, varying validation sample}
\end{table}

\begin{table}[htbp!]
\begin{subtable}[t]{0.475\textwidth}

\begin{tabular}{r|rrr}
\toprule
Validation Prop. & Median & 5\% & 95\%\\
\midrule
\multicolumn{1}{l|}{\textit{Normalized Bias}}\\
2.5\% & 0.118 & -1.631 & 2.112\\
5\% & 0.126 & -1.664 & 2.172\\
10\% & 0.119 & -1.635 & 2.201\\
25\% & 0.133 & -1.695 & 2.182\\
50\% & 0.158 & -1.638 & 2.177\\
\midrule
\multicolumn{1}{l|}{\textit{Coverage}}\\
2.5\% & 0.898 & 0.419 & 0.953\\
5\% & 0.900 & 0.405 & 0.950\\
10\% & 0.899 & 0.397 & 0.948\\
25\% & 0.899 & 0.383 & 0.949\\
50\% & 0.898 & 0.400 & 0.951\\
\bottomrule
\end{tabular}

\caption{Plug-in regression}
\end{subtable}
\hfill
\begin{subtable}[t]{0.475\textwidth}

\begin{tabular}{r|rrr}
\toprule
Validation Prop. & Median & 5\% & 95\%\\
\midrule
\multicolumn{1}{l|}{\textit{Normalized Bias}}\\
2.5\% & -0.003 & -0.052 & 0.057\\
5\% & -0.002 & -0.064 & 0.058\\
10\% & -0.001 & -0.047 & 0.053\\
25\% & -0.004 & -0.052 & 0.053\\
50\% & -0.003 & -0.057 & 0.052\\
\midrule
\multicolumn{1}{l|}{\textit{Coverage}}\\
2.5\% & 0.905 & 0.858 & 0.966\\
5\% & 0.934 & 0.908 & 0.958\\
10\% & 0.943 & 0.928 & 0.956\\
25\% & 0.946 & 0.935 & 0.957\\
50\% & 0.948 & 0.936 & 0.959\\
\bottomrule
\end{tabular}

\caption{Debiased regression}
\end{subtable}
\caption{Summary statistics for normalized bias and coverage using policy topic as a covariate for GPT-5-nano, varying the size of the validation sample.}
\floatfoot{\textit{Notes}: The normalized bias reports the average bias of the plug-in regression coefficient $\widehat{\beta}$ and the debiased coefficient $\widehat{\beta}^{debiased}$ for the target regression coefficient divided by their respective standard deviations across simulations. 
The coverage reports the fraction of simulations in which a 95\% nominal confidence interval centered around the plug-in regression coefficient $\widehat{\beta}$ and the debiased coefficient $\widehat{\beta}^{debiased}$ cover the target regression coefficient.
For each combination of left hand side variable $W_r$, large language model $m$ and prompting strategy $p$, we randomly sample $5,000$ Congressional bills and calculate the plug-in regression $\widehat{\beta}$ and bias-corrected coefficient $\widehat{\beta}^*$ with the policy topic as a covariate.
We vary the size of the validation sample over 2.5\%, 5\%, 10\%, 25\% and 50\%.
Results are averaged over $1,000$ simulations.
See Appendix \ref{section: LLM on RHS simulations}.
}
\label{table: congressional bills, LLM on RHS, summary statistics, GPT-5-nano, varying validation sample}
\end{table}

\begin{figure}[htbp!]
\includegraphics[width=\textwidth]{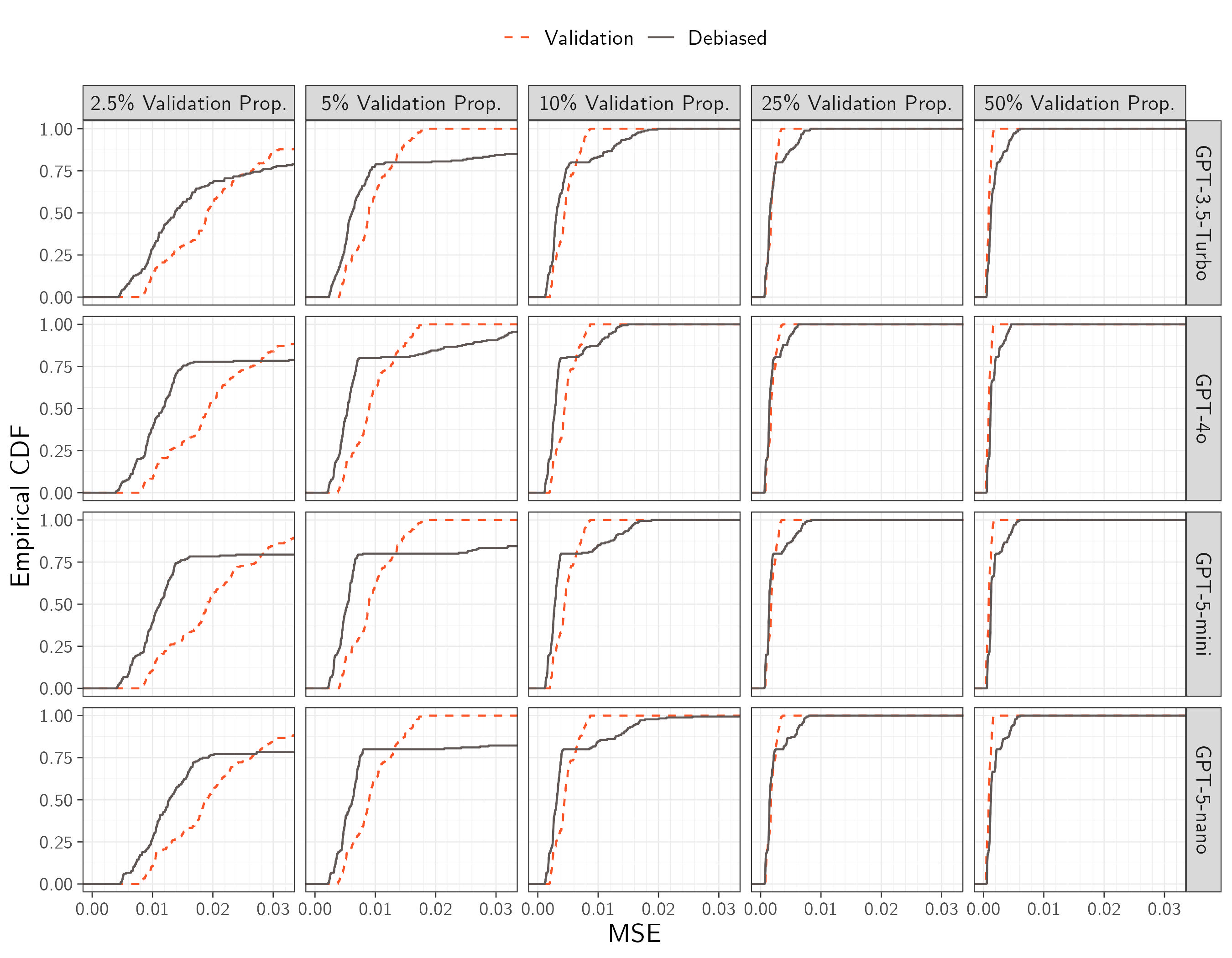}
\caption{Cumulative distribution function of mean square error for the bias-corrected estimator against validation-sample only estimator using policy topic as the validation sample size varies.}
\floatfoot{\textit{Notes}: For each combination of left hand side variable $W_r$, large language model $m$ and prompting strategy $p$, we randomly sample $5,000$ Congressional bills and calculate the plug-in regression $\widehat{\beta}$ and bias-corrected coefficient $\widehat{\beta}^*$ with the policy topic as a covariate.
We calculate the mean square error of $\widehat{\beta}^{debiased}$ and $\widehat{\beta}^*$ for the target regression $\beta^*$.
We vary the size of the validation sample over 2.5\%, 5\%, 10\%, 25\% and 50\%, and we average the results over $1,000$ simulations.
We summarize the distribution of average mean square error across regression specifications, choice of large language model and prompting strategies.
See Appendix \ref{section: LLM on RHS simulations}.}
\label{figure: congressional bills, LLM on RHS, MSE, all prop}
\end{figure}

\section{Prompts for Congressional Bills and Financial News Headlines}\label{section: prompts for empirical work}
\begin{figure}[htbp!]
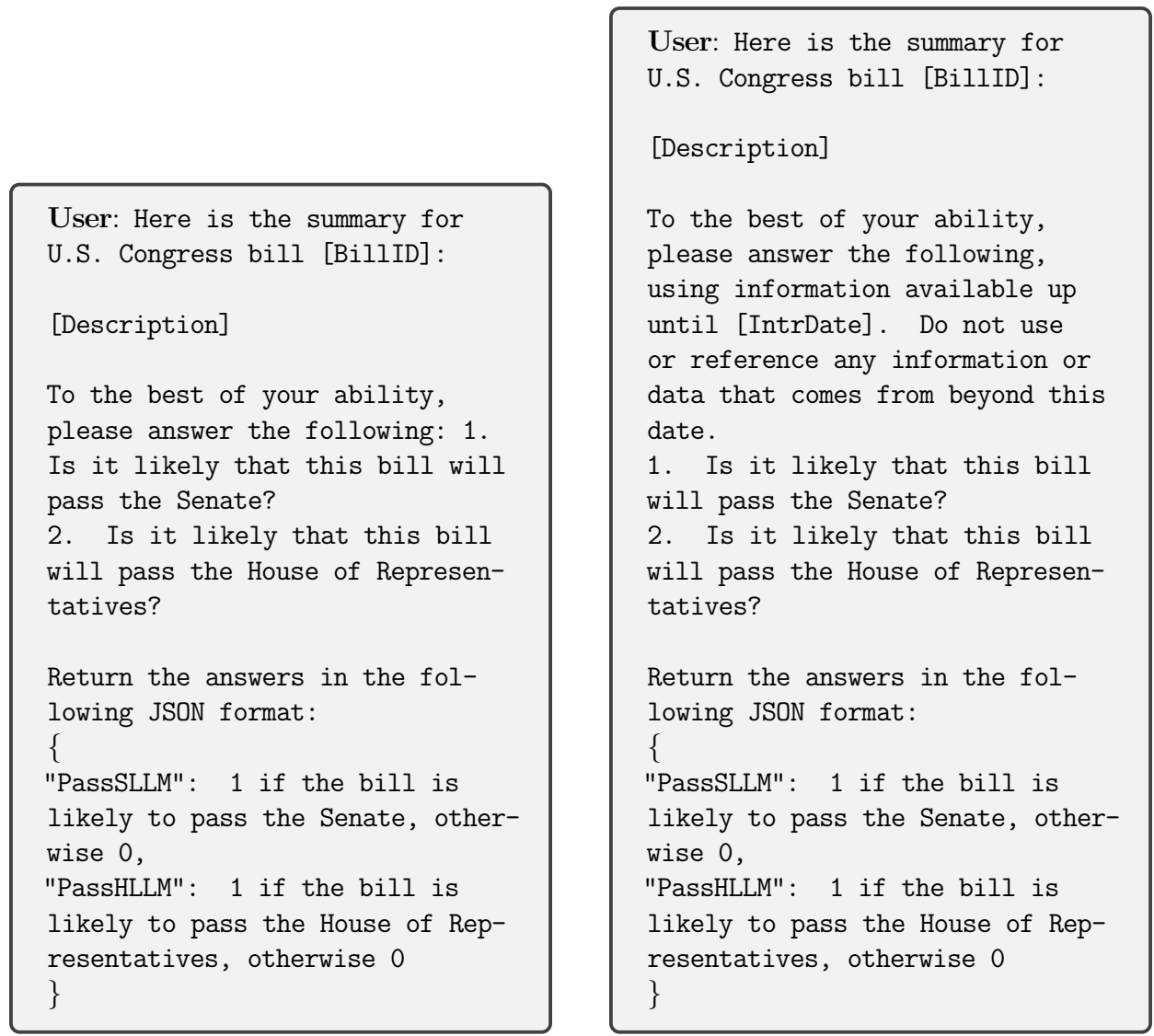

\centering
\begin{subfigure}[t]{0.475\textwidth}
    \centering
    \begin{tcolorbox}[colback=black!5!white,colframe=black!75!white] 
    \textbf{User}: \texttt{Here is the summary for U.S. Congress bill [BillID]:} \\\\
    \texttt{[Description]} \\\\
    \texttt{To the best of your ability, please answer the following:}
    \texttt{1. Is it likely that this bill will pass the Senate?} \\
    \texttt{2. Is it likely that this bill will pass the House of Representatives?} \\\\ 
    \texttt{Return the answers in the following JSON format:} \\
    \{ \\
    \texttt{"PassSLLM": 1 if the bill is likely to pass the Senate, otherwise 0,} \\
    \texttt{"PassHLLM": 1 if the bill is likely to pass the House of Representatives, otherwise 0} \\
    \}
    \end{tcolorbox}
    \caption{Base prompt}
    \label{subfig: congressional bills prediction, prompt without date restriction}
\end{subfigure}
    \hfill
\begin{subfigure}[t]{0.475\textwidth}
    \centering
    \begin{tcolorbox}[colback=black!5!white,colframe=black!75!white] 
    \textbf{User}: \texttt{Here is the summary for U.S. Congress bill [BillID]:} \\\\
    \texttt{[Description]} \\\\
    \texttt{To the best of your ability, please answer the following, using information available up until [IntrDate]. Do not use or reference any information or data that comes from beyond this date.} \\
    \texttt{1. Is it likely that this bill will pass the Senate?} \\
    \texttt{2. Is it likely that this bill will pass the House of Representatives?} \\\\
    \texttt{Return the answers in the following JSON format:} \\
    \{ \\
    \texttt{"PassSLLM": 1 if the bill is likely to pass the Senate, otherwise 0,} \\
    \texttt{"PassHLLM": 1 if the bill is likely to pass the House of Representatives, otherwise 0} \\
    \}
    \end{tcolorbox}
    \caption{Prompt with date restriction}
    \label{subfig: congressional bills prediction, prompt with date restriction}
\end{subfigure}
\caption{Prompts used for prediction based on large language models with Congressional legislation.}    
\floatfoot{\textit{Notes}: This figure documents the prompts used for the prediction exercise based on congressional legislation.
We prompt GPT-4o to predict whether 10,000 randomly selected congressional bills would pass the Senate or the House based on its text description. 
For each Congressional bill, we include its identifier [BillID], its text description [Description], and its introduction date [IntrDate] in the prompts. 
Figure (a) provides the base prompt, and Figure (b) provides the base prompt with the additional date restriction.
See Section \ref{section: evidence of training leakage} for further details.}
\label{fig: congressional bills prediction, prompts}
\end{figure}

\begin{figure}[htbp!]
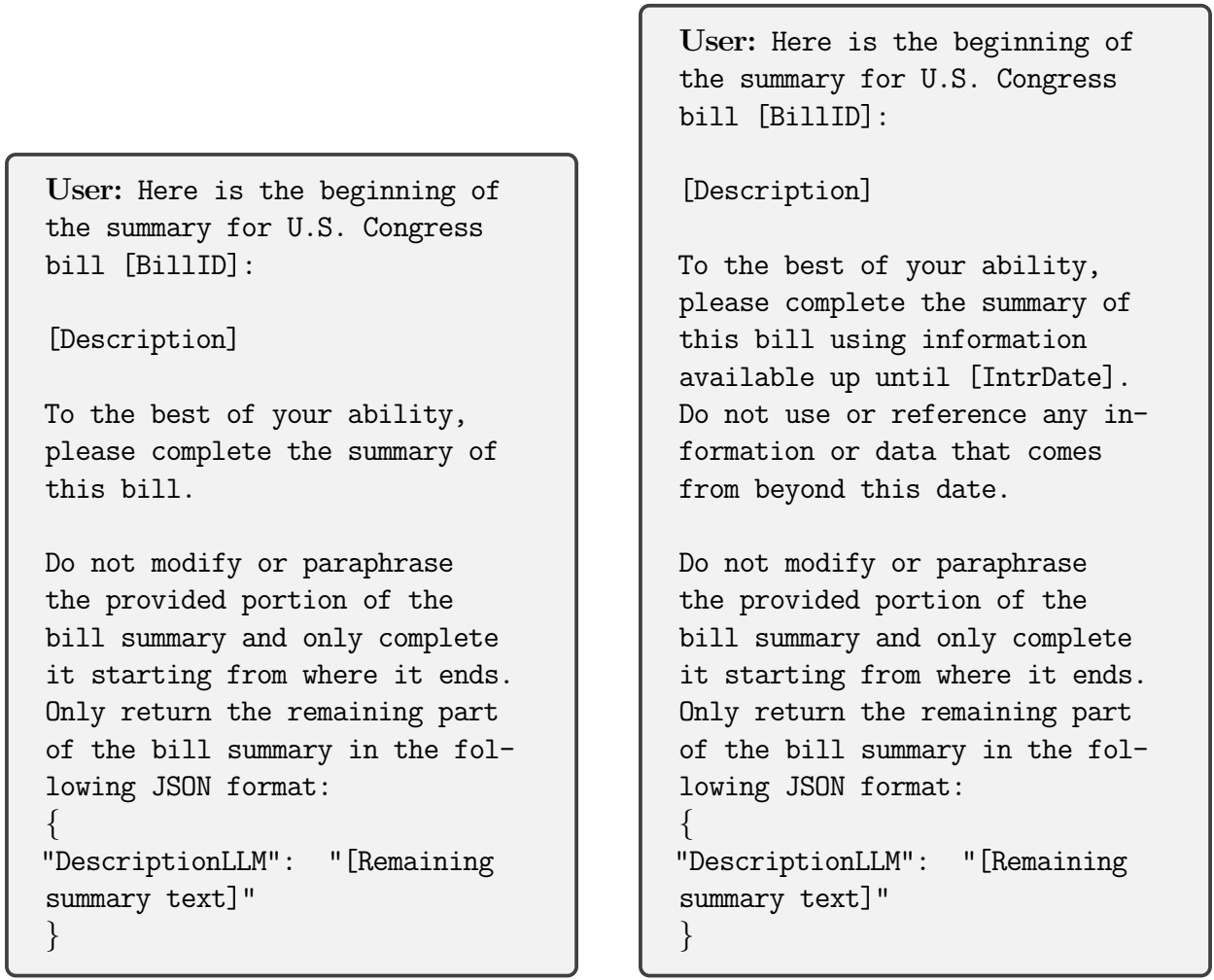

\centering
\begin{subfigure}[t]{0.475\textwidth}
    \centering
    \begin{tcolorbox}[colback=black!5!white,colframe=black!75!white] 
    \textbf{User:} \texttt{Here is the beginning of the summary for U.S. Congress bill [BillID]:} \\\\
    \texttt{[Description]} \\\\
    \texttt{To the best of your ability, please complete the summary of this bill.} \\\\
    \texttt{Do not modify or paraphrase the provided portion of the bill summary and only complete it starting from where it ends. Only return the remaining part of the bill summary in the following JSON format:} \\
    \{ \\
  \texttt{"DescriptionLLM": "[Remaining summary text]"} \\
    \}
    \end{tcolorbox}
    \caption{Base prompt}
    \label{subfig: congressional bills completion, prompt without date restriction}
\end{subfigure}
    \hfill
\begin{subfigure}[t]{0.475\textwidth}
    \centering
    \begin{tcolorbox}[colback=black!5!white,colframe=black!75!white] 
    \textbf{User:} \texttt{Here is the beginning of the summary for U.S. Congress bill [BillID]:} \\\\
    \texttt{[Description]} \\\\
    \texttt{To the best of your ability, please complete the summary of this bill using information available up until [IntrDate]. Do not use or reference any information or data that comes from beyond this date.} \\\\
    \texttt{Do not modify or paraphrase the provided portion of the bill summary and only complete it starting from where it ends. Only return the remaining part of the bill summary in the following JSON format:} \\
    \{ \\
  \texttt{"DescriptionLLM": "[Remaining summary text]"} \\
    \}
    \end{tcolorbox}
    \caption{Prompt with date restriction}
    \label{subfig: congressional bills completion, prompt with date restriction}
\end{subfigure}
\caption{Prompts used for text completion exercise based on large language models with Congressional legislation.}    
\floatfoot{\textit{Notes}: This figure documents the prompts used for the text completion exercise based on congressional legislation.
We prompt GPT-4o to complete the description of 10,000 randomly selected congressional bills based on a segment of its text. 
For each Congressional bill, we include its identifier [BillID], the beginning of its text description [Description], and its introduction date [IntrDate] in the prompts. 
Figure (a) provides the base prompt, and Figure (b) provides the base prompt with the additional date restriction.
See Section \ref{section: evidence of training leakage} for further details.}
\label{fig: congressional bills completion, prompts}
\end{figure}


\begin{figure}[htbp!]
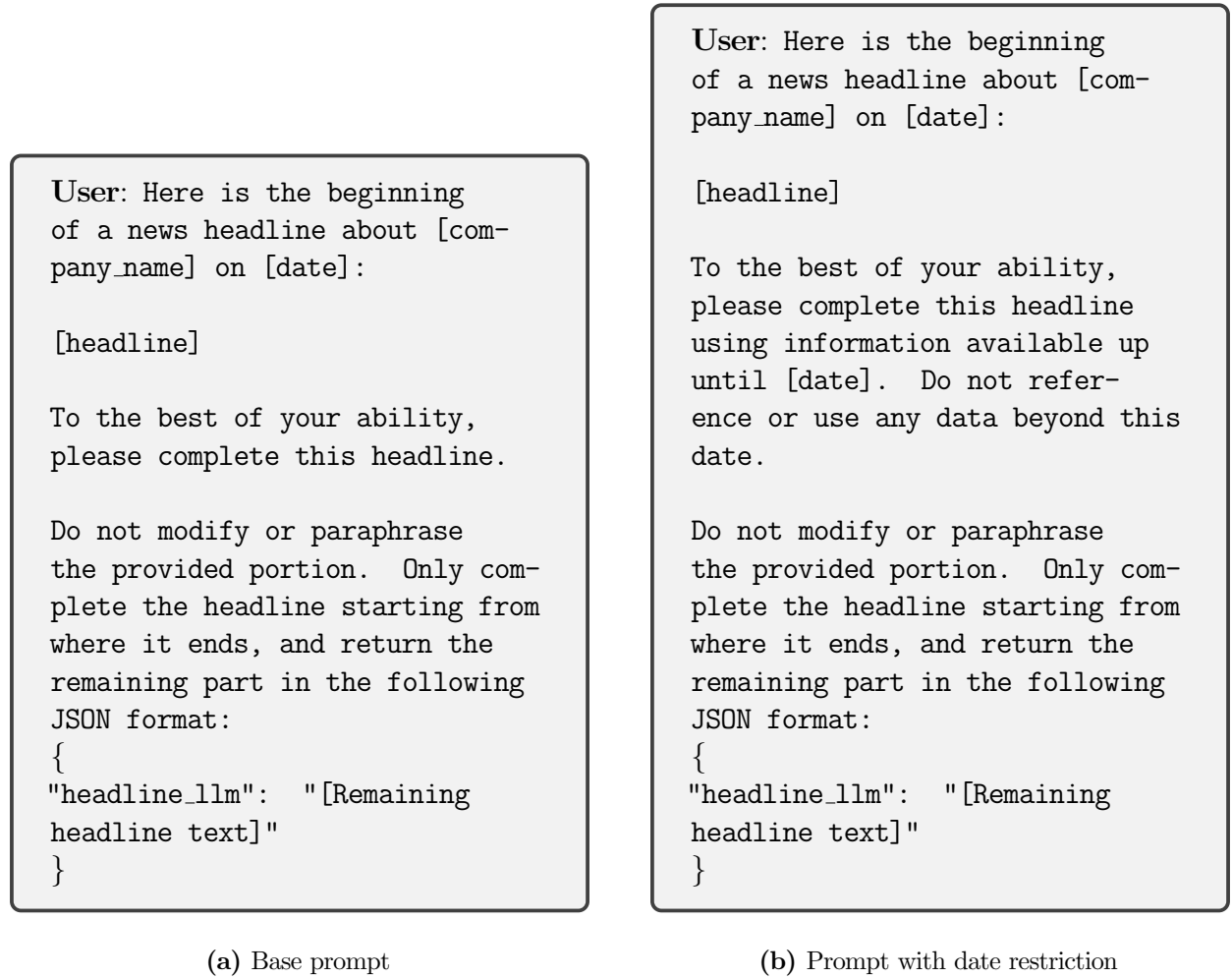

\centering
\begin{subfigure}[t]{0.475\textwidth}
    \centering
    \begin{tcolorbox}[colback=black!5!white,colframe=black!75!white] 
    \textbf{User}: \texttt{Here is the beginning of a news headline about [company\_name] on [date]:} \\\\
    \texttt{[headline]} \\\\
    \texttt{To the best of your ability, please complete this headline.} \\\\
    \texttt{Do not modify or paraphrase the provided portion. Only complete the headline starting from where it ends, and return the remaining part in the following JSON format:} \\
    \{ \\
    \texttt{"headline\_llm": "[Remaining headline text]"} \\
    \}
    \end{tcolorbox}
    \caption{Base prompt}
    \label{subfig: financial news headlines completion, prompt without date restriction}
\end{subfigure}
    \hfill
\begin{subfigure}[t]{0.475\textwidth}
    \centering
    \begin{tcolorbox}[colback=black!5!white,colframe=black!75!white] 
    \textbf{User}: \texttt{Here is the beginning of a news headline about [company\_name] on [date]:} \\\\
    \texttt{[headline]} \\\\
    \texttt{To the best of your ability, please complete this headline using information available up until [date]. Do not reference or use any data beyond this date.} \\\\
    \texttt{Do not modify or paraphrase the provided portion. Only complete the headline starting from where it ends, and return the remaining part in the following JSON format:} \\
    \{ \\
  \texttt{"headline\_llm": "[Remaining headline text]"} \\
    \}
    \end{tcolorbox}
    \caption{Prompt with date restriction}
    \label{subfig: financial news headlines completion, prompt with date restriction}
\end{subfigure}
\caption{Prompts used for text completion exercise based on large language models with financial news headlines.}    
\floatfoot{\textit{Notes}: This figure documents the prompts used for the text completion exercise based on financial news headlines.
We prompt GPT-4o to complete 10,000 randomly selected financial news headline based on a segment of its text. 
For each financial news headline, we include the name of the company it is about [company\_name], its publication date [date], and the beginning of its text [headline]. 
Figure (a) provides the base prompt, and Figure (b) provides the base prompt with the additional date restriction.
See Section \ref{section: evidence of training leakage} for further details.}
\label{fig: financial news headlines completion, prompts}
\end{figure}

\begin{figure}[htbp!]
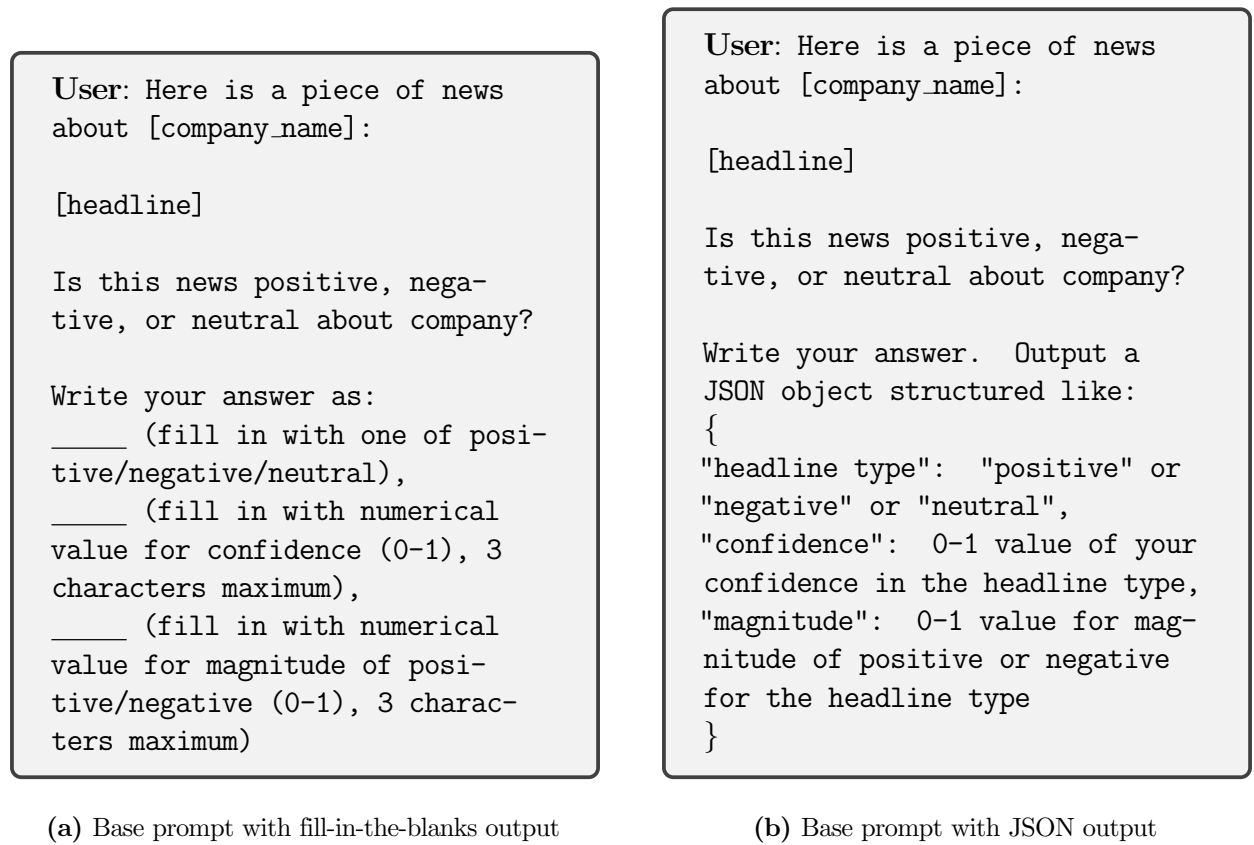

\centering
\begin{subfigure}[t]{0.475\textwidth}
    \centering
    \begin{tcolorbox}[colback=black!5!white,colframe=black!75!white] 
    \textbf{User}: \texttt{Here is a piece of news about [company\_name]:} \\\\
    \texttt{[headline]} \\\\
    \texttt{Is this news positive, negative, or neutral about company?} \\\\
    \texttt{Write your answer as:} \\
    \texttt{\underline{\hspace{1cm}} (fill in with one of positive/negative/neutral),} \\
    \texttt{\underline{\hspace{1cm}} (fill in with numerical value for confidence (0-1), 3 characters maximum),} \\
    \texttt{\underline{\hspace{1cm}} (fill in with numerical value for magnitude of positive/negative (0-1), 3 characters maximum)}
    \end{tcolorbox}
    \caption{Base prompt with fill-in-the-blanks output}
    \label{subfig: financial news headlines investor sentiment, base prompt with fill-in-the-blanks}
\end{subfigure}
    \hfill
\begin{subfigure}[t]{0.475\textwidth}
    \centering
    \begin{tcolorbox}[colback=black!5!white,colframe=black!75!white] 
    \textbf{User}: \texttt{Here is a piece of news about [company\_name]:} \\\\
    \texttt{[headline]} \\\\
    \texttt{Is this news positive, negative, or neutral about company?} \\\\
    \texttt{Write your answer. Output a JSON object structured like:} \\
    \{ \\
    \texttt{"headline type": "positive" or "negative" or "neutral",} \\
    \texttt{"confidence": 0-1 value of your confidence in the headline type,} \\    \texttt{"magnitude": 0-1 value for magnitude of positive or negative for the headline type} \\ 
    \}
    \end{tcolorbox}
    \caption{Base prompt with JSON output}
    \label{subfig: financial news headlines investor sentiment, base prompt with json}
\end{subfigure}
\caption{Base prompts for labeling financial news headlines with large language models.}    
\floatfoot{\textit{Notes}: This figure documents the base prompts used for labeling financial news headlines with large language models.
We prompt GPT-3.5-Turbo, GPT-4o, GPT-4o-mini, GPT-5-mini, and GPT-5-nano to label financial news headlines for whether they are positive, negative or neutral about the associated company.
For each financial news headline, we include the name of the company it is about [company\_name] and the text of the headline [headline]. Figure (a) provides the base prompt with fill-in-the-blanks output, and Figure (b) provides the base prompt with JSON output.
See Section \ref{section: evidence of measurement error} for further details.}
\label{figure: financial news headlines, positive/negative news, base prompts}
\end{figure}

\begin{figure}[htbp!]
\centering
\begin{subfigure}[t]{0.475\textwidth}
    \centering
    \begin{tcolorbox}[colback=black!5!white,colframe=black!75!white] 
    \textbf{User}: \texttt{You are a knowledgeable economic agent.}
    \end{tcolorbox}
    \caption{Economic agent persona}
    \label{subfig: financial news headlines, positive/negative news, economic agent persona}
\end{subfigure}
\hfill
\begin{subfigure}[t]{0.475\textwidth}
    \centering
    \begin{tcolorbox}[colback=black!5!white,colframe=black!75!white] 
    \textbf{User}: \texttt{Answer this question as if you are an expert in finance.}
    \end{tcolorbox}
    \caption{Finance expert persona}
    \label{subfig: financial news headlines, positive/negative news, finance expert persona}
\end{subfigure}
\begin{subfigure}[t]{0.475\textwidth}
    \centering
    \begin{tcolorbox}[colback=black!5!white,colframe=black!75!white] 
    \textbf{User}: \texttt{Answer this question as if you are an expert in the economy.}
    \end{tcolorbox}
    \caption{Economy expert persona}
    \label{subfig: financial news headlines, positive/negative news, economic expert persona}
\end{subfigure}
\hfill
\begin{subfigure}[t]{0.475\textwidth}
    \centering
    \begin{tcolorbox}[colback=black!5!white,colframe=black!75!white] 
    \textbf{User}: \texttt{Answer this question as if you were very knowledgeable about financial matters and in particular the stock market. So you are as knowledgeable as an analyst or trader at a very successful Wall Street Firm.}
    \end{tcolorbox}
    \caption{Successful trader persona}
    \label{subfig: financial news headlines, positive/negative news, trader persona}
\end{subfigure}
\caption{Persona modifications to the base prompt for labeling financial news headlines with large language models.}    
\floatfoot{\textit{Notes}: This figure documents the persona modifications to the base prompts for labeling financial news headlines with large language models. 
We prompt GPT-3.5-Turbo, GPT-4o, GPT-4o-mini, GPT-5-mini, and GPT-5-nano to label financial news headlines for whether they are positive, negative or neutral about the associated company.
Each persona modification is added to the beginning of the base prompt with JSON output (Panel (a) of Appendix Figure \ref{figure: financial news headlines, positive/negative news, base prompts}). 
See Section \ref{section: evidence of measurement error} for further details.}
\label{figure: financial news headlines, positive/negative news, persona modifications}
\end{figure}

\begin{figure}[htbp!]
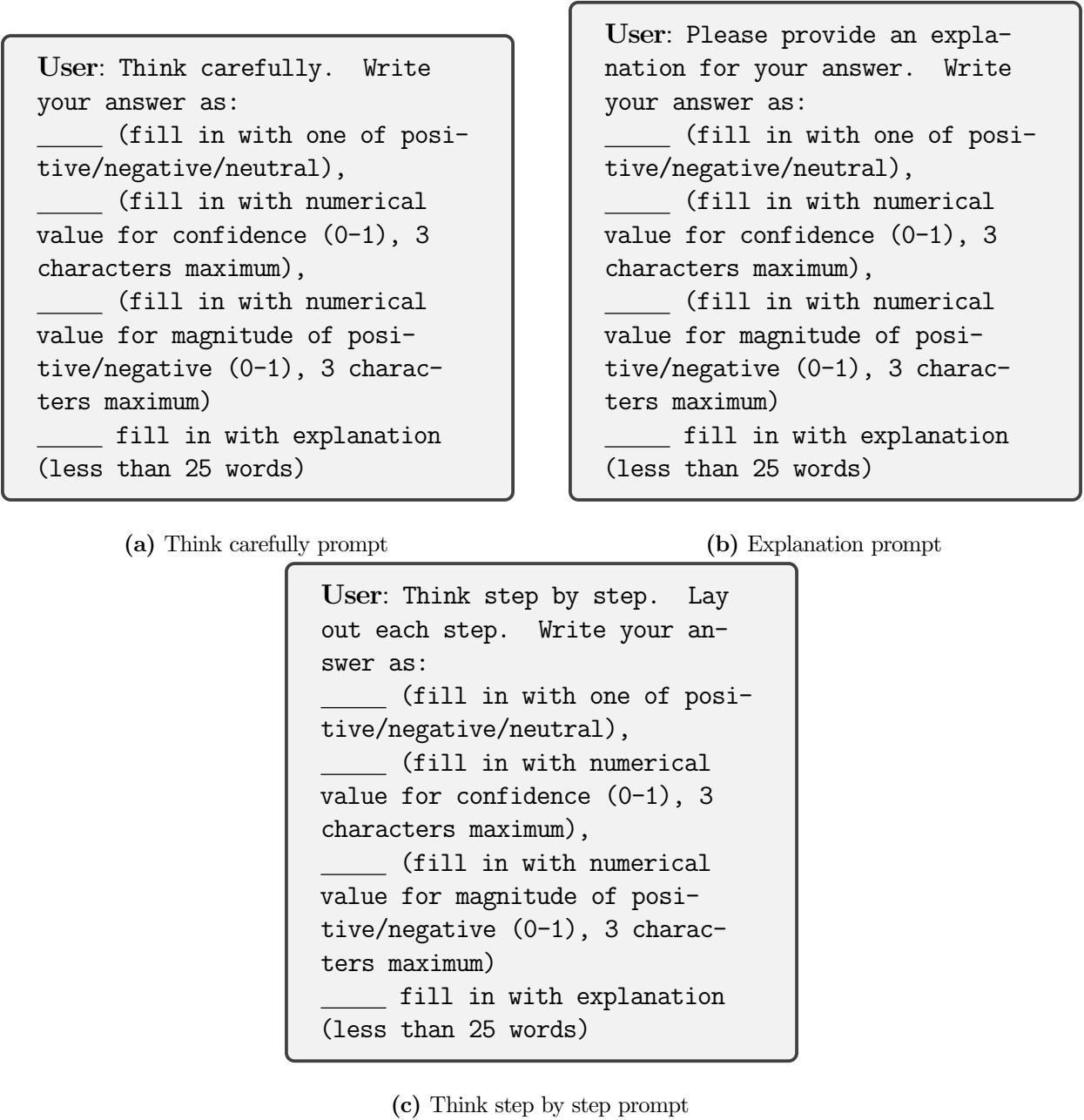

\centering
\begin{subfigure}[t]{0.475\textwidth}
    \centering
    \begin{tcolorbox}[colback=black!5!white,colframe=black!75!white] 
    \textbf{User}: \texttt{Think carefully. Write your answer as:} \\
    \texttt{\underline{\hspace{1cm}} (fill in with one of positive/negative/neutral),} \\
    \texttt{\underline{\hspace{1cm}} (fill in with numerical value for confidence (0-1), 3 characters maximum),} \\
    \texttt{\underline{\hspace{1cm}} (fill in with numerical value for magnitude of positive/negative (0-1), 3 characters maximum)} \\
    \texttt{\underline{\hspace{1cm}} fill in with explanation (less than 25 words)} 
    \end{tcolorbox}
    \caption{Think carefully prompt}
    \label{subfig: financial news headlines, positive/negative news, think carefully}
\end{subfigure}
\hfill
\begin{subfigure}[t]{0.475\textwidth}
    \centering
    \begin{tcolorbox}[colback=black!5!white,colframe=black!75!white] 
    \textbf{User}: \texttt{Please provide an explanation for your answer. Write your answer as:} \\
    \texttt{\underline{\hspace{1cm}} (fill in with one of positive/negative/neutral),} \\
    \texttt{\underline{\hspace{1cm}} (fill in with numerical value for confidence (0-1), 3 characters maximum),} \\
    \texttt{\underline{\hspace{1cm}} (fill in with numerical value for magnitude of positive/negative (0-1), 3 characters maximum)} \\
    \texttt{\underline{\hspace{1cm}} fill in with explanation (less than 25 words)} 
    \end{tcolorbox}
    \caption{Explanation prompt}
    \label{subfig: financial news headlines, positive/negative news, explanation}
\end{subfigure}
\begin{subfigure}[t]{0.475\textwidth}
    \centering
    \begin{tcolorbox}[colback=black!5!white,colframe=black!75!white] 
    \textbf{User}: \texttt{Think step by step. Lay out each step. Write your answer as:} \\
    \texttt{\underline{\hspace{1cm}} (fill in with one of positive/negative/neutral),} \\
    \texttt{\underline{\hspace{1cm}} (fill in with numerical value for confidence (0-1), 3 characters maximum),} \\
    \texttt{\underline{\hspace{1cm}} (fill in with numerical value for magnitude of positive/negative (0-1), 3 characters maximum)} \\
    \texttt{\underline{\hspace{1cm}} fill in with explanation (less than 25 words)} 
    \end{tcolorbox}
    \caption{Think step by step prompt}
    \label{subfig: financial news headlines, positive/negative news, step by step}
\end{subfigure}
\caption{Chain of thought modifications to the base prompt for labeling financial news headlines with large language models.} 
\floatfoot{\textit{Notes}: This figure documents the chain of thought modifications to the base prompts for labeling financial news headlines with large language models. 
We prompt GPT-3.5-Turbo, GPT-4o, GPT-4o-mini, GPT-5-mini, and GPT-5-nano to label financial news headlines for whether they are positive, negative or neutral about the associated company.
Each chain-of-though modification alters the base prompt with JSON output (Panel (a) of Appendix Figure \ref{figure: financial news headlines, positive/negative news, base prompts}).
See Section \ref{section: evidence of measurement error} for further details.}
\label{figure: financial news headlines, positive/negative news, chain of thought modifications}
\end{figure}

\begin{figure}[htbp!]
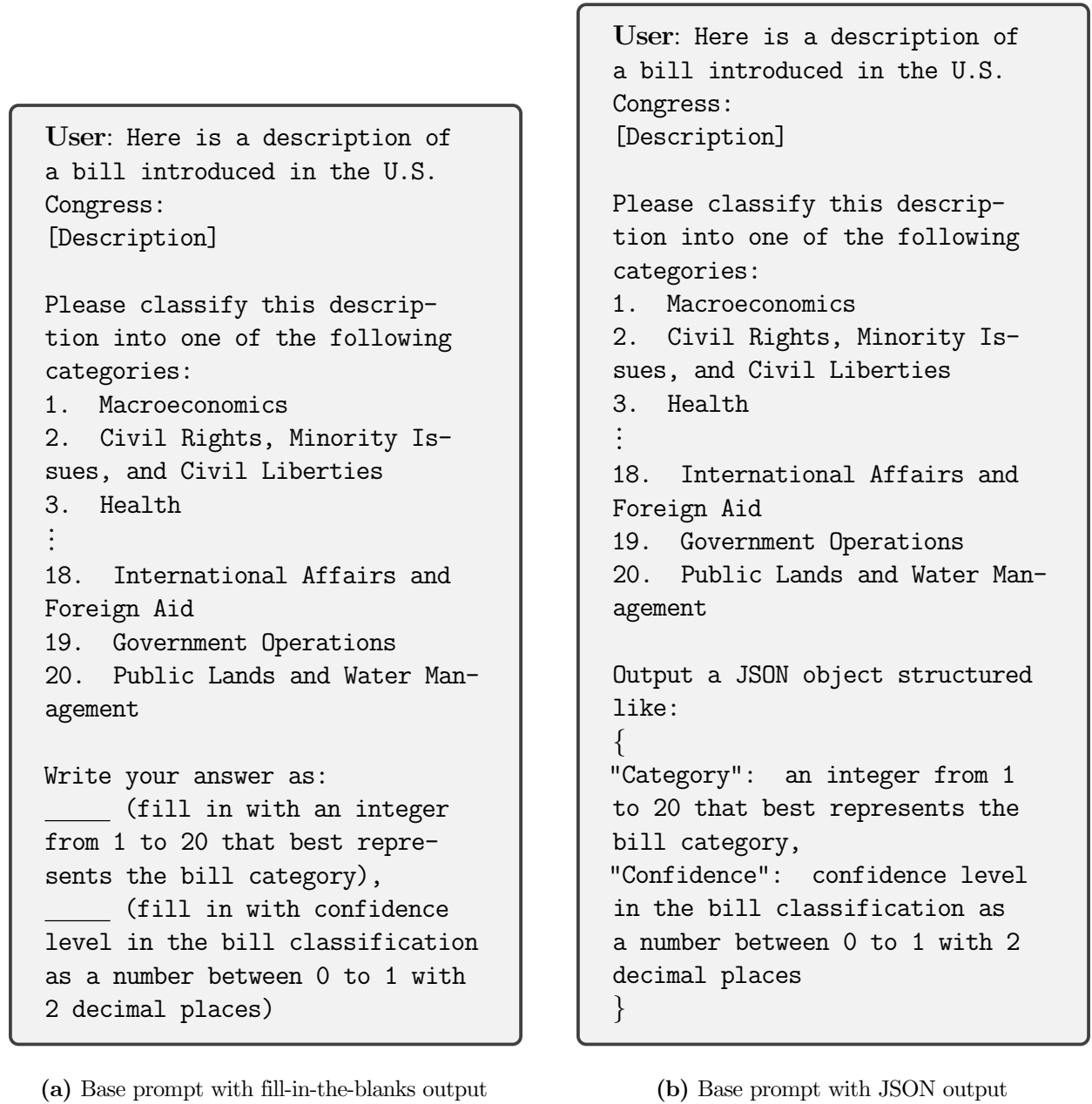

\centering
\begin{subfigure}[t]{0.475\textwidth}
    \centering
    \begin{tcolorbox}[colback=black!5!white,colframe=black!75!white] 
    \textbf{User}: \texttt{Here is a description of a bill introduced in the U.S. Congress:} \\
    \texttt{[Description]} \\\\
    \texttt{Please classify this description into one of the following categories:} \\
    \texttt{1. Macroeconomics} \\
    \texttt{2. Civil Rights, Minority Issues, and Civil Liberties} \\
    \texttt{3. Health} \\
    \texttt{$\vdots$} \\
    \texttt{18. International Affairs and Foreign Aid} \\
    \texttt{19. Government Operations} \\
    \texttt{20. Public Lands and Water Management} \\ \\
    \texttt{Write your answer as:} \\
    \texttt{\underline{\hspace{1cm}} (fill in with an integer from 1 to 20 that best represents the bill    category),} \\
    \texttt{\underline{\hspace{1cm}} (fill in with confidence level in the bill classification as a number between 0 to 1 with 2 decimal places)}
    \end{tcolorbox}
    \caption{Base prompt with fill-in-the-blanks output}
    \label{subfig: congressional bills, major topic labelling, base prompt with fill-in-the-blanks}
\end{subfigure}
    \hfill
\begin{subfigure}[t]{0.475\textwidth}
    \centering
    \begin{tcolorbox}[colback=black!5!white,colframe=black!75!white] 
    \textbf{User}: \texttt{Here is a description of a bill introduced in the U.S. Congress:} \\
    \texttt{[Description]} \\\\
    \texttt{Please classify this description into one of the following categories:} \\
    \texttt{1. Macroeconomics} \\
    \texttt{2. Civil Rights, Minority Issues, and Civil Liberties} \\
    \texttt{3. Health} \\
    \texttt{$\vdots$} \\
    \texttt{18. International Affairs and Foreign Aid} \\
    \texttt{19. Government Operations} \\
    \texttt{20. Public Lands and Water Management} \\ \\
    \texttt{Output a JSON object structured like:} \\
    \{ \\
    \texttt{"Category”: an integer from 1 to 20 that best represents the bill category,} \\
    \texttt{"Confidence”: confidence level in the bill classification as a number between 0 to 1 with 2 decimal places} \\
    \}
    \end{tcolorbox}
    \caption{Base prompt with JSON output}
    \label{subfig: congressional bills, major topic labelling, base prompt with json}
\end{subfigure}
\caption{Base prompts for labeling the policy topic with large language models on Congressional legislation.}    
\floatfoot{\textit{Notes}: This figure documents the base prompts used for labeling the policy with large language models on congressional legislation. 
We prompt GPT-3.5-turbo, GPT-4o, GPT-5-mini, and GPT-5-nano to label the descriptions of 10,000 randomly drawn Congressional bills for their major topic.
For each Congressional bill, we include the text of its description [description]. 
Figure (a) provides the base prompt with fill-in-the-blanks output, and Figure (b) provides the base prompt with JSON output.
See Section \ref{section: evidence of measurement error} for further details.}
\label{figure: congressional bills, major topic labelling, base prompts}
\end{figure}

\begin{figure}[htbp!]
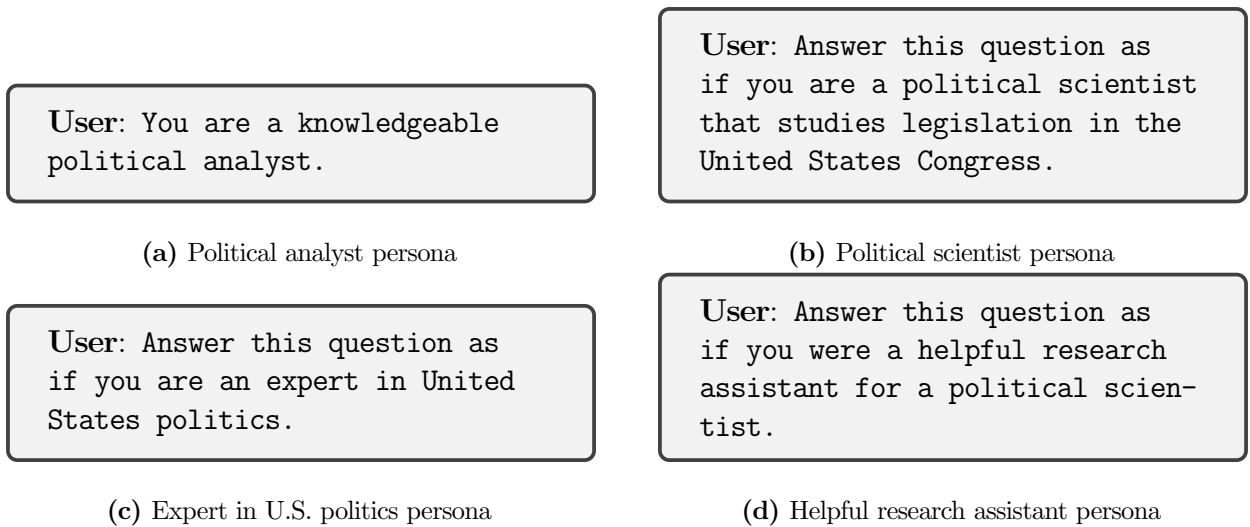

\centering
\begin{subfigure}[t]{0.475\textwidth}
    \centering
    \begin{tcolorbox}[colback=black!5!white,colframe=black!75!white] 
    \textbf{User}: \texttt{You are a knowledgeable political analyst.}
    \end{tcolorbox}
    \caption{Political analyst persona}
    \label{subfig: congressional bills, major topic labelling, political analyst persona}
\end{subfigure}
\hfill
\begin{subfigure}[t]{0.475\textwidth}
    \centering
    \begin{tcolorbox}[colback=black!5!white,colframe=black!75!white] 
    \textbf{User}: \texttt{Answer this question as if you are a political scientist that studies legislation in the United States Congress.}
    \end{tcolorbox}
    \caption{Political scientist persona}
    \label{subfig: congressional bills, major topic labelling, political scientists persona}
\end{subfigure}
\begin{subfigure}[t]{0.475\textwidth}
    \centering
    \begin{tcolorbox}[colback=black!5!white,colframe=black!75!white] 
    \textbf{User}: \texttt{Answer this question as if you are an expert in United States politics.}
    \end{tcolorbox}
    \caption{Expert in U.S. politics persona}
    \label{subfig: congressional bills, major topic labelling, politics expert persona}
\end{subfigure}
\hfill
\begin{subfigure}[t]{0.475\textwidth}
    \centering
    \begin{tcolorbox}[colback=black!5!white,colframe=black!75!white] 
    \textbf{User}: \texttt{Answer this question as if you were a helpful research assistant for a political scientist.}
    \end{tcolorbox}
    \caption{Helpful research assistant persona}
    \label{subfig: congressional bills, major topic labelling, helpful ra persona}
\end{subfigure}
\caption{Persona modifications to the base prompt for labeling the policy topic with large language models on Congressional legislation.}    
\floatfoot{\textit{Notes}: This figure documents the persona modifications to the base prompts for measuring the policy topic with large language models on Congressional legislation.
We prompt GPT-3.5-turbo, GPT-4o, GPT-5-mini, and GPT-5-nano to label the descriptions of 10,000 randomly drawn Congressional bills for their major topic.
Each persona modification is added to the beginning of the base prompt with JSON output (Panel (b) of Appendix Figure \ref{figure: congressional bills, major topic labelling, base prompts}). 
See Section \ref{section: evidence of measurement error} for further details.}
\label{figure: congressional bills, major topic labelling, persona modifications}
\end{figure}

\begin{figure}[htbp!]
\centering
\begin{subfigure}[t]{0.475\textwidth}
    \centering
    \begin{tcolorbox}[colback=black!5!white,colframe=black!75!white] 
    \textbf{User}: \texttt{Think carefully. Output a JSON object structured like:} \\
    \{ \\
    \texttt{"Category”: an integer from 1 to 20 that best represents the bill category,} \\
    \texttt{"Confidence”: confidence level in the bill classification as a number between 0 to 1 with 2 decimal places,} \\
    \texttt{“Explanation”: a one-sentence explanation of your bill category answer} \\
    \}
    \end{tcolorbox}
    \caption{Think carefully prompt}
    \label{subfig: congressional bills, major topic labelling, think carefully}
\end{subfigure}
\hfill
\begin{subfigure}[t]{0.475\textwidth}
    \centering
    \begin{tcolorbox}[colback=black!5!white,colframe=black!75!white] 
    \textbf{User}: \texttt{Please provide an explanation for your answer. WOutput a JSON object structured like:} \\
    \{ \\
    \texttt{"Category”: an integer from 1 to 20 that best represents the bill category,} \\
    \texttt{"Confidence”: confidence level in the bill classification as a number between 0 to 1 with 2 decimal places,} \\
    \texttt{“Explanation”: a one-sentence explanation of your bill category answer} \\
    \}
    \end{tcolorbox}
    \caption{Explanation prompt}
    \label{subfig: congressional bills, major topic labelling, explanation}
\end{subfigure}
\begin{subfigure}[t]{0.475\textwidth}
    \centering
    \begin{tcolorbox}[colback=black!5!white,colframe=black!75!white] 
    \textbf{User}: \texttt{Think step by step. Lay out each step. Output a JSON object structured like:} \\
    \{ \\
    \texttt{"Category”: an integer from 1 to 20 that best represents the bill category,} \\
    \texttt{"Confidence”: confidence level in the bill classification as a number between 0 to 1 with 2 decimal places,} \\
    \texttt{“Explanation”: a one-sentence explanation of your bill category answer} \\
    \}
    \end{tcolorbox}
    \caption{Think step by step prompt}
    \label{subfig: congressional bills, major topic labelling, step by step}
\end{subfigure}
\caption{Chain of thought modifications to the base prompt for labeling the policy topic with large language models on Congressional legislation.} 
\floatfoot{\textit{Notes}: This figure documents the chain of thought modifications to the base prompts for labeling the policy topic with large language models on Congressional legislation. 
We prompt GPT-3.5-turbo, GPT-4o, GPT-5-mini, and GPT-5-nano to label the descriptions of 10,000 randomly drawn Congressional bills for their major topic.
Each chain-of-though modification alters the base prompt with JSON output (Panel (b) of Appendix Figure \ref{figure: congressional bills, major topic labelling, base prompts}).
See Section \ref{section: evidence of measurement error} for further details.}
\label{figure: congressional bills, major topic labelling, chain of thought modifications}
\end{figure}

\end{document}